\documentclass[11pt]{article}

\usepackage[authoryear,round]{natbib}
\bibpunct{(}{)}{;}{a}{}{,}

\let\cite\citep

\usepackage{amsmath}
\usepackage{amssymb}
\usepackage{amsthm}
\usepackage{hyperref}

\providecommand{\doi}[1]{%
  \url{https://doi.org/#1}%
}

\usepackage{graphicx}
\usepackage[margin=1in]{geometry}
\usepackage{authblk}
\usepackage{xcolor}
\usepackage{booktabs}
\usepackage{tabularx}
\usepackage{array}
\usepackage{tikz}
\usepackage{comment}
\usetikzlibrary{arrows.meta,positioning}
\newtheorem{theorem}{Theorem}[section]
\newtheorem{lemma}[theorem]{Lemma}
\newtheorem{proposition}[theorem]{Proposition}

\newtheorem{corollary}[theorem]{Corollary}
\theoremstyle{remark}\newtheorem{remark}[theorem]{Remark}

\title{Milk-removal feedback shapes lactation sustainability and daily output in a model of human lactation}

\author[1]{Angelika Manhart\thanks{Corresponding author: \texttt{angelika.manhart@univie.ac.at}, ORCID: 0000-0002-0303-8114}}
\author[1]{Tobias M\"uller}

\affil[1]{Faculty of Mathematics, University of Vienna, Austria}

\date{}

\begin{document}

\maketitle

%\tableofcontents
%\newpage

\begin{abstract}
Milk production during lactation is dynamically regulated by milk removal, yet quantitative models linking feeding or pumping schedules to subsequent production remain scarce. We develop a minimal model describing milk accumulation and production between removal sessions, and the stimulation of subsequent production through milk removal.

We compare different assumptions about how stimulation depends on the amount of milk removed or the duration of removal, and consider two removal protocols: prescribing either the removed amount or the removal duration. When stimulation depends on the removed amount, declining production can reduce subsequent stimulation and reinforce the decline. This leads to a critical wait interval beyond which no repeated constant schedule can sustain lactation. If stimulation is inefficient for small removal amounts but more efficient at intermediate amounts, the model can exhibit bistability: depending on the initial conditions, milk production either persists or ceases. In contrast, duration-dependent stimulation does not permit a zero-production state under continued removal, but can generate non-constant long-term dynamics under certain conditions.

We further show that maximizing milk removal per session need not maximize daily output, because increased removal competes with session frequency. The framework provides a basis for testing alternative feedback hypotheses against longitudinal data and for optimizing feeding and pumping schedules.
\end{abstract}



\section{Introduction}

%The goal of paper is to gain an in-depth understanding of the long-term behaviour of the production-removal-stimulation cycle of lactating women under a regular wait/removal schedules. In particular we want to understand how the the establishment and maintenance of lactation, as well as the obtained milk output depend on biological and scheduling parameters and initial conditions. Further, we will discuss the impact of the stimulation mechanism as well as the shape of the stimulation function.

\paragraph{Why we need a quantitative understanding of lactation.} Human lactation affects a large proportion of mothers and infants, yet difficulties in establishing or maintaining an adequate milk supply are common and are a frequent source of concern during breastfeeding and pumping \cite{Gatti2008}. Human milk provides appropriate nutrition as well as immune and other bioactive components to the infant, making the successful establishment and maintenance of lactation clinically and practically important. This is especially true for prematurely born babies \cite{Haiden2016}. At the same time, the regulation of milk production is inherently dynamic: milk removal alters subsequent production, but much of the practical guidance available to mothers remains qualitative. A quantitative mechanistic framework could therefore help translate general principles such as frequent and effective milk removal into more precise predictions about how feeding or pumping schedules influence milk supply. Ultimately, such models could be used to address practical questions including how lactation is best established, how supply can be maintained or increased under time and scheduling constraints, how pumping schedules can be optimized, and how lactation can be reduced or stopped in a controlled manner. Before making such predictions, however, it is important to understand which qualitative behaviours arise robustly from the underlying feedback structure and how they depend on specific assumptions about how milk removal stimulates subsequent production.

\paragraph{Removing milk stimulates milk production.} Human milk is synthesized and secreted by mammary epithelial cells lining the alveoli of the lactating breast. During breastfeeding or pumping, milk is transported from the alveoli through the ductal system and removed from the breast. Once lactation is established, milk production is strongly regulated by milk removal, allowing supply to adapt to demand \cite{Truchet2017,Kent2012}. In particular, short-term measurements in lactating women indicate that the subsequent rate of milk synthesis is associated with the degree to which the breast is emptied during milk removal \cite{Daly1996}. Several mechanisms have been implicated in this local feedback, including mechanical stretching of the mammary epithelium or local biochemical feedback signals within the breast \cite{Weaver2016,Truchet2017,Peaker1996}. If milk is not removed for sufficiently long periods, milk synthesis declines and the mammary gland eventually enters involution \cite{Truchet2017}, the process by which the lactating mammary gland returns toward its non-lactating state.

\medskip
In addition to the \emph{local} feedback, milk removal also stimulates lactation through neuroendocrine pathways, whereby nipple stimulation triggers neural signals that induce the release of hormones including oxytocin and prolactin \cite{McNeilly1983}. Oxytocin induces contraction of myoepithelial cells surrounding the alveoli and thereby drives milk ejection, whereas prolactin supports milk synthesis and secretory activity of mammary epithelial cells \cite{Truchet2017}. However, the relationship between stimulation and subsequent milk production is not straightforward: for example, although prolactin is released in response to nipple stimulation, circulating prolactin concentrations during established lactation are not directly correlated with the short-term rate of milk synthesis \cite{Cox1996}. Thus, the physiological feedback from a milk-removal event to subsequent production likely combines neuroendocrine stimulation with local effects associated with the amount of milk removed. In our model we therefore describe this feedback phenomenologically and consider separately stimulation that depends on the duration of milk removal and stimulation that depends on the amount removed.

\paragraph{On the state of lactation modelling.} Despite extensive experimental work on human lactation, its dynamic regulation has received relatively little attention from mathematical modelling. Mechanistic models of the human lactating breast have primarily focused on the biomechanics and fluid dynamics of milk removal, including transport through the mammary ductal system and the mechanical processes involved in milk removal by the infant \cite{Mortazavi2015,Mortazavi2017,Azarnoosh2020}. A notable exception is the MSc thesis of \citet{Faulkner2019}, which developed ordinary differential equation models coupling milk production and removal with hormonal, neural and local regulatory mechanisms, and used them to investigate the effect of feeding frequency. In other mammalian systems, particularly dairy cattle, mechanistic modelling has a longer history, ranging from models of mammary metabolism and secretory-cell dynamics to models explicitly linking udder filling and milking frequency to milk yield and mammary regression \cite{Dijkstra1997,Vetharaniam2003}. More recently, data-driven approaches have also been applied to human lactation, for example to predict breastfeeding outcomes or identify low milk supply from milk composition and maternal and infant characteristics \cite{AgudeloPerez2024,Jin2025}. However, these approaches do not describe the dynamic feedback between individual milk-removal events and subsequent milk production. %Here, we develop a deliberately minimal mechanistic model of this feedback, with the aim of identifying general principles governing the maintenance and output of lactation under different milk-removal schedules.

\paragraph{This work.} In Sec.~\ref{sec:model}, we develop a minimal mechanistic model of the feedback between milk production, milk removal, and the stimulation of subsequent milk production. The model has a hybrid continuous--discrete structure: milk accumulation and changes in the production rate are described continuously between removal sessions, while breastfeeding or pumping sessions are represented as discrete removal and stimulation events (see Rem.~\ref{rem:impulsive} on impulsive systems). We use this framework to investigate several alternative assumptions about this feedback. First, we ask whether the stimulation of subsequent production depends primarily on the \emph{amount} of milk removed or on the \emph{duration} of the removal session. Second, we consider different possible \emph{shapes} of the stimulation response: in particular, whether even small amounts of milk removal efficiently stimulate production, or whether stimulation is inefficient for small removals and becomes most efficient only at intermediate amounts. Finally, we compare two removal protocols: \emph{set-target} removal, in which removal continues until a prescribed amount of milk has been obtained or the breast is emptied, and \emph{set-duration} removal, in which milk is removed for a prescribed period of time and the amount obtained depends on the milk available. In Sec.~\ref{sec:analysis} we study the resulting models in depth, focusing on long-term behaviour. In particular we determine under which conditions lactation can be sustained, and when the system instead progresses towards zero milk production. We will discuss how these outcomes depend on the initial state, biological parameters, and the removal schedule. In Sec.~\ref{sec:daily_output} we further investigate how these factors determine daily milk output and the trade-offs involved in increasing milk production. Finally, Secs.~\ref{sec:compare} and \ref{sec:discussion} summarize and discuss the findings. The \emph{Interactive Lactation Explorer} (SI: Online Resource 1) allows readers to vary the model parameters and schedules and to follow the resulting dynamics in phase space and over time.

\section{Model}
\label{sec:model}

% ---------------------------------------------------------------- FIGURE ---

\paragraph{Set-up.} The model describes the feedback loop of milk production stimulation and milk removal for one breast. Here \emph{milk removal} can refer to any mode of removing milk via the nipple, which can include breastfeeding, hand expressing or using a breast pump. Individual lobes are not resolved, rather we model the total amount of milk present in the breast $m$ (in ml), and the underlying milk production rate, $p$ (in ml per hour). We refer to the state $(m,p)=(0,0)$ as the \emph{involution state}, representing a state with no stored milk and zero milk-production rate within the present model. This terminology is used as a modelling shorthand: the model does not resolve the cellular and tissue-remodelling processes of mammary-gland involution.

\begin{figure}[t]
  \centering
  \includegraphics[width=\textwidth]{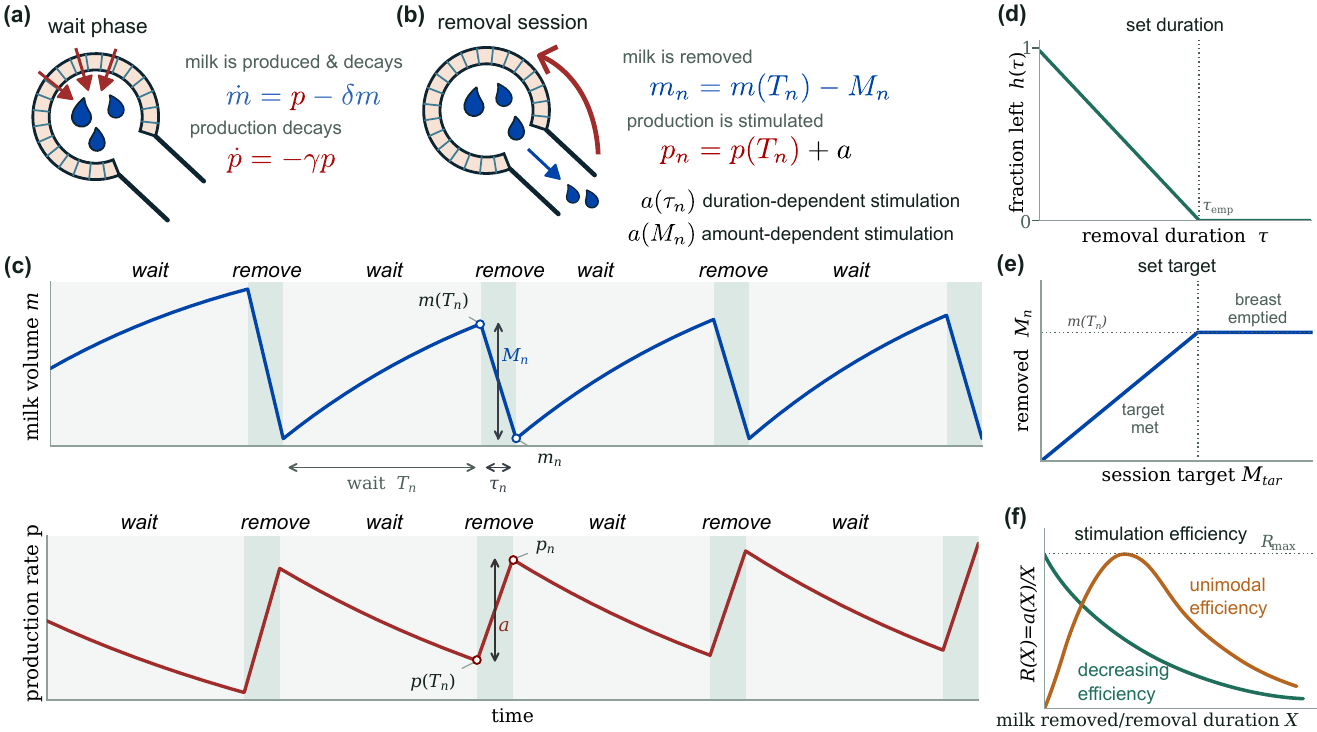}
  \caption{Model. (a),(b) Schematics of the two phases of a cycle. During the wait phase milk is produced and secreted into the lumen of the alveolus. At a removal session milk drains through the duct and the removal stimulates future production.
  (c) Development of amount of milk in the breast ($m$, upper panel) and production rate ($p$ lower panel) during wait and removal phases. Note that we model removal sessions as discrete (short) events, while in the figure removal durations are extended for graphical clarity. (d) For a set-duration removal protocol, the fraction of milk left behind is $h(\tau)=\mathrm{max}\{1-\nu\tau,0\}$. (e) For a set-target protocol, the removed volume is $\min\{M_\mathrm{tar},m(T_n)\}$. (f) The two stimulation efficiency classes considered.}
  \label{fig:model}
\end{figure}

\medskip
Time is divided into alternating \emph{wait} phases, during which no milk is removed and \emph{removal} sessions, at which milk is removed and the production rate is updated in response, see Fig.~\ref{fig:model}. \emph{Wait} intervals are typically much longer (several hours) than \emph{removal} durations (tens of minutes). We therefore describe the wait phase in continuous time using ordinary differential equations, Fig.~\ref{fig:model}(a). While we account for the fact that removal sessions have a finite removal duration contributing to elapsed schedule time, we do not resolve production or decay during the removal session itself; instead, its effect is represented by discrete-in-time events, Fig.~\ref{fig:model}(b). We write $m_n$ and $p_n$ for the milk volume and production rate immediately after the $n$-th removal session, with $(m_0,p_0)$ denoting the prescribed initial state. For $n\ge 1$, $T_n$ denotes wait interval, which is the length of the wait phase before the $n$-th removal session.
%and $\tau_n$ the removal session's duration.

\begin{remark}[Relation to impulsive systems]\label{rem:impulsive}
Mathematically, the model belongs to the class of impulsive differential equations \cite{Lakshmikantham1989}: the state evolves according to an ordinary differential equation between sessions and jumps instantaneously at each session. Such systems have been used, for example, to describe pulse vaccination, drug dosing and periodic harvesting, and are typically analysed through the map obtained by composing the flow over one period with the jump. In the language of impulsive harvesting \cite{Clark1990, Zhang2003, Lawson2023}, set-duration and set-target removal correspond to proportional and constant-quota harvesting, respectively. Unlike in standard harvesting models, however, removal here also stimulates subsequent production.
\end{remark}

\paragraph{Removal schedules.}
A removal schedule specifies the prescribed timing and removal protocol
over successive removal cycles (see details in Sec.~\ref{ssec:model_amount}). Under a \emph{set-duration protocol}, a schedule is given by the sequence
\[
    (T_n,\tau_n),\qquad n=1,2,\hdots 
\]
where $T_n$ is the wait interval preceding the $n$-th removal session and
$\tau_n$ is its prescribed duration. Under a
\emph{set-target protocol}, a schedule is given by the sequence
\[
    (T_n,M_{\mathrm{tar},n}),\qquad n=1,2,\hdots 
\]
where $M_{\mathrm{tar},n}$ is the prescribed target amount for the $n$-th
removal session. Throughout this work we consider \emph{regular schedules}, meaning that the same schedule is repeated after every removal session:
$$
(T_n,\tau_n)\equiv (T,\tau)\quad \text{or} \quad (T_n,M_{\mathrm{tar},n})\equiv (T,M_{\mathrm{tar}}).
$$
Note that the state variables \(m_n,p_n\), the amount actually removed \(M_n\), and, under set-target removal, the resulting session duration \(\tau_n\), may nevertheless vary between sessions.

\subsection{Wait phase}

Consider a wait interval of length $T>0$ prior to the $n$-th removal session. During the time $t\in[0,T]$ milk is produced and accumulates in the breast at rate $p(t)$ and decays (e.g. through reabsorption) at rate $\delta>0$, while, in the absence of stimulation, the production rate itself decays at rate $\gamma$:
\begin{align}
\label{eqn:waitODE}
\dot m &= p - \delta m, \qquad \dot p = -\gamma p,
\end{align}
for $t \in [0,T]$. The initial conditions are given by the amount of milk in the breast, and the production rate directly after the previous removal event:
\begin{equation}
\label{eqn:waitODE_IC}
m(0) = m_{n-1}, \qquad p(0) = p_{n-1}.
\end{equation}
Solving this system forward to $t = T$ gives $m(T;m_{n-1},p_{n-1})$ and $p(T;p_{n-1})$, the state of the breast immediately before the next removal session. In the following we define $x_n:=m(T;m_{n-1},p_{n-1})$ to simplify notation.

\begin{remark}
Here we assume that the production rate simply follows exponential decay. However, it has been suggested that there is a negative feedback between the amount of milk in the breast (or breast fullness) and the production rate \cite{Peaker1996}. A more general model would therefore have the form $\dot p=f(p,m)$ where $f$ is a decreasing function of $m$.
\end{remark}

\subsection{Removal session}

At a removal session, a volume $M_n$ (in ml) with $0\leq M_n\leq x_n$ of milk is removed from the breast. This removal stimulates future production via an increment $a$ (in ml per hour), which can depend on the amount of removed milk $M_n$ and/or the removal duration $\tau_n$. We will discuss the first option in Sec.~\ref{sec:amount_stim_target}, \ref{sec:amount_stim_duration} and the second option in Sec.~\ref{sec:dur_stim_duration}, \ref{sec:dur_stim_target},
\begin{align}
m_{n} &= x_n - M_n, \\
p_{n} &= p(T;p_{n-1}) + a(M_n,\tau_n).
\end{align}

The removal session is then followed by a new wait phase, starting from $m_n$, $p_n$.

\subsection{Modelling the removed amount of milk $M_n$}
\label{ssec:model_amount}
If $x$ is the amount of milk in the breast directly prior to removal and $\tau>0$ (in hours) is the length of the removal session, we model the amount of milk removed by
\begin{equation}
\label{eqn:removal_law}
M(x,\tau) = x \left( 1 -h(\tau) \right),
\end{equation}
where $h(\tau)\in[0,1]$ the fraction of milk remaining in the breast at the end of a removal session of length $\tau$. We use the linear law, see Fig.~\ref{fig:model}(d),
$$ h(\tau)=\mathrm{max}\{1-\nu\tau,0\}.$$
The removal rate parameter is denoted by $\nu>0$ (per hour), where $\nu$ can vary with infant age, pump type, pump setting, etc. Complete emptying occurs at $\tau=\frac{1}{\nu}=:\tau_\mathrm{emp}$. One alternative would be to use the exponential removal law \(h(\tau)=e^{-\nu\tau}\); we do not consider this alternative further, since it cannot describe complete emptying of the breast.\newline

We consider two possible \emph{removal protocols} of how to decide when to stop a removal session and to determine the volume $M_n$ removed: In the first case, the removal duration $\tau$ is set, and the pumping or feeding is stopped after that time. In the second case, the target amount of removed milk is set and the pumping or feeding is stopped when the target is met, or when the breast is emptied.

\paragraph{Set-duration removals.} The schedule is given by $(T,\tau)$, i.e. in addition to the wait interval $T$ we set the removal duration $\tau$. Then we obtain
\begin{equation*}
%\label{eqn:removal_law}
M_n = x_n \left( 1 -h(\tau) \right).
\end{equation*}

\paragraph{Set-target removals.} Here the schedule is given by $(T,M_\mathrm{tar})$, i.e. in addition to the length of the wait interval, we prescribe a target amount of removed milk $M_{\mathrm{tar}}$ (in ml). This can represent infant demand or a pumping goal. The volume actually removed is capped by the milk available in the breast, see Fig.~\ref{fig:model}(e):
\begin{equation}
M_n = \min\{M_{\mathrm{tar}}, \, x_n\}.
\end{equation}
In this case the associated removal duration $\tau_n$ (which now depends on the session index $n$) is defined using \eqref{eqn:removal_law} by
\begin{align*}
  \tau_n=\begin{cases}
    \frac{M_\mathrm{tar}}{x_n}\tau_\mathrm{emp}, & x_n>M_\mathrm{tar}\\
    \tau_\mathrm{emp}, & x_n\leq M_\mathrm{tar}.
  \end{cases}
\end{align*}

\subsection{Modelling the stimulation function $a(M,\tau)$}
\label{sec:stim}

The stimulation function $a$ describes how strongly the removal of milk up-regulates
subsequent production. In the following we will focus on two scenarios: 1. Stimulation primarily depends on the amount of milk removed, $a=a(M)$, see Sec.~\ref{sec:amount_stim_target} and \ref{sec:amount_stim_duration}. 2. Stimulation primarily depends on the duration of milk removal, $a=a(\tau)$, see Sec.~\ref{sec:dur_stim_target} and \ref{sec:dur_stim_duration}. We do not consider stimulation functions that depend on $M$ and $\tau$ simultaneously.\newline

We assume that $a\in\mathcal{C}^1([0,\infty))$, $a(0)=0$, with $a$ strictly increasing, bounded and with limiting value $a_{\mathrm{max}}>0$ for large inputs. We also introduce the associated \emph{stimulation efficiencies} $R(X)$, where $X$ can be $M$ or $\tau$,
\[
  R(X)=\frac{a(X)}{X},\qquad X>0,
\]
we extend the corresponding efficiency continuously to zero by $a'(0)$.\newline

The analysis below considers two structurally different \emph{efficiency classes}, see Fig.~\ref{fig:model}(f). In a \emph{decreasing-efficiency}
response, $R$ is largest for very small arguments and is monotonically decreasing. In a
\emph{unimodal-efficiency} response,  $R$ is small near zero, rises to a maximum at an intermediate value, and then decreases again. These properties, rather than the particular functional form for $a$, determine the qualitative dynamics.

%----------

%-------------

\begin{table}[t]
\centering
\caption{State variables, schedule quantities, model parameters, and derived quantities.}
\label{tab:variables}
\begin{tabularx}{\textwidth}{@{} l l X @{}}
\toprule
\textbf{Symbol} & \textbf{Units} & \textbf{Meaning} \\
\midrule

\multicolumn{3}{@{}l}{\textit{State variables and schedule quantities}} \\[1mm]

$m$ 
& ml 
& Amount of milk in the breast. \\

$p$ 
& ml\,h$^{-1}$ 
& Milk production rate. \\

$T$ 
& h 
& Wait interval between consecutive removal sessions. \\

$\tau$ 
& h 
& Duration of a removal session. \\

$M$ 
& ml 
& Amount of milk removed during a removal session. \\

$M_{\rm tar}$ 
& ml 
& Prescribed amount to be removed under a set-target protocol. \\

\addlinespace
\multicolumn{3}{@{}l}{\textit{Model parameters and response functions}} \\[1mm]

$\delta$ 
& h$^{-1}$ 
& Rate of milk loss from the breast during the wait phase. \\

$\gamma$ 
& h$^{-1}$ 
& Decay rate of the milk production rate in the absence of stimulation. \\

$\nu, \quad \tau_\mathrm{emp}=\frac{1}{\nu}$ 
& h$^{-1}$, h 
& $\nu$: Removal-rate parameter. $\tau_\mathrm{emp}$: Emptying duration.\\

$h(\tau)$ 
& 1 
& Fraction of the milk available after a removal session of duration $\tau$. \\

$a(M)$, $a(\tau)$ 
& ml\,h$^{-1}$ 
& Increase in production rate induced by removing an amount $M$ of milk/a removal session of duration $\tau$. \\

$a_{\mathrm{max}}$ 
& ml\,h$^{-1}$ 
& Saturation value of the stimulation response. \\

$\mu_M$, $\mu_\tau$ 
& ml, h 
& Characteristic amount scale of the amount/duration-dependent stimulation response. \\

\addlinespace
\multicolumn{3}{@{}l}{\textit{Frequently used derived quantities}} \\[1mm]

$d(T)=e^{-\delta T}$ 
& 1 
& Fraction of stored milk remaining after a wait interval $T$ in the absence of new production. \\

$g(T)=e^{-\gamma T}$ 
& 1
& Fraction of the production rate remaining after a wait interval $T$. \\

$G(T)$ 
& h 
& Factor describing the contribution of milk production during a wait interval $T$ to the milk available before removal. \\

$x=md(T)+pG(T)$
& ml
& Amount of milk available immediately before a removal session. \\

\bottomrule
\end{tabularx}

\end{table}

%----------

\paragraph{Examples.}
For illustrations and closed-form examples, we use two simple saturating
responses. Let $X\in\{M,\tau\}$ for amount-dependent and duration-dependent stimulation respectively and let $\mu_X$ have the same units as $X$.\newline

The \emph{Michaelis--Menten} response (also called a \emph{Type-II functional response})
\begin{equation}
\label{eqn:mich_ment}
    a(X)
    =
    a_{\mathrm{max}}\frac{X}{X+\mu_X}
\end{equation}
is concave and has strictly decreasing efficiency $a(X)/X=a_{\mathrm{max}}/(X+\mu_X)$.\newline

The \emph{Hill} response (also called a \emph{Type-III functional response})
\begin{equation}
\label{eqn:hill}
    a(X)
    =
    a_{\mathrm{max}}
    \frac{X^2}{X^2+\mu_X^2}
\end{equation}
has a sigmoidal form and has unimodal efficiency, with its maximum at $X=\mu_X$.\newline

The explicit forms are used only for examples and
figures; the analytical results below are formulated in terms of the
shape of the corresponding efficiency function.

\subsection{Full Model}

An advantage of the model formulation is that the solution to \eqref{eqn:waitODE},\eqref{eqn:waitODE_IC} can be determined explicitly using the standard variations-of-constants method for solving differential equations. One obtains for a general time $t\in[0,T]$ 
$$
m(t;m_{n-1},p_{n-1} )=m_{n-1} d(t) +p_{n-1} G(t),\qquad p(t,p_{n-1})=p_{n-1} g(t)
$$
where
$$
g(t)=e^{-\gamma t},\qquad d(t)=e^{-\delta t},\qquad G(t)=\frac{e^{-\gamma t}-e^{-\delta t}}{\delta-\gamma}.
$$
Note that for $\delta=\gamma$ we have $G(t)=te^{-\gamma t}$. We see that the milk in the breast just prior to removal $x_n=m(T;m_{n-1},p_{n-1} )$ has two contributions: The first term describes the decay of the left-over milk from the previous removal session. The second term describes the new milk produced since the last session.\newline

For a regular schedule, the wait and removal phases therefore define
an autonomous discrete dynamical system
\[
  \begin{pmatrix}m_n\\p_n\end{pmatrix}
  =
  F
  \begin{pmatrix}m_{n-1}\\p_{n-1}\end{pmatrix}
  .
\]

where, writing
\[
    g=g(T),\qquad d=d(T),\qquad G=G(T),
    \qquad
    x_n=d\,m_{n-1}+G\,p_{n-1},
\]
one cycle is given by
\[
    m_n=x_n-M_n,
    \qquad
    p_n=g\,p_{n-1}+a(M_n,\tau_n).
\]
The quantities $M_n$ and $\tau_n$ are determined by the chosen removal
protocol as described above.

Tab.~\ref{tab:variables} summarizes the main model variables.

\section{Model Analysis}
\label{sec:analysis}

The main goal is to understand how system long-term behaviour depends on the stimulation mechanism (amount-dependent or duration-dependent), the removal protocol (set-target or set-duration) and the efficiency class (decreasing efficiency or unimodal efficiency). In order to make the main narrative and insights easier to follow, the main text focuses on results and interpretation. Technical details and proofs are collected in the Appendix. For the same reason, non-generic boundary cases are generally excluded in the statements. We encourage readers to follow the analysis below with the \emph{Interactive Lactation Explorer} (Online Resource 1), in which phase planes, orbits and schedule spaces can be
recomputed for any parameter choice. 

\medskip
We define the key quantity
\begin{equation}
\label{eqn:Gamma}
    \Gamma=\Gamma(T):=\frac{1-g}{G}.
\end{equation}

The quantity $\Gamma(T)$ will repeatedly arise as the stimulation efficiency required to balance the decay of the production rate over
one removal cycle.\newline

Here is an overview over the analysis sections of the four discussed cases:
\begin{center}
\begin{tabular}{lcc}
\toprule
& \textbf{Set-target protocol} & \textbf{Set-duration protocol} \\
\midrule
\textbf{Amount-dependent stimulation}
& Sec.~\ref{sec:amount_stim_target} (Proofs in App.~\ref{app:amount_stim_target}) & Sec.~\ref{sec:amount_stim_duration} (Proofs in App.~\ref{app:amount_stim_duration})\\
\textbf{Duration-dependent stimulation}
& Sec.~\ref{sec:dur_stim_target} (Proofs in App.~\ref{app:dur_stim_target}) & Sec.~\ref{sec:dur_stim_duration} (Proofs in App.~\ref{app:dur_stim_duration})\\
\bottomrule
\end{tabular}
\end{center}
Then in Sec.~\ref{sec:protocol_correspondence} we discuss in more detail how protocol choice affects the long-term dynamics. We start by analysing amount-dependent stimulation under a set-target removal protocol.

\subsection{Amount-dependent stimulation, set-target protocol}
\label{sec:amount_stim_target}

\paragraph{Preliminaries.}

Throughout this section we assume that production stimulation depends on the amount of milk removed during removal sessions, i.e. $a=a(M)$ with $a:[0,\infty)\to[0,a_\mathrm{max})$ is $C^1$, strictly increasing, bounded, satisfies $a(0)=0$, and has finite saturation level $a_\mathrm{max}$.

\medskip
Further the schedule $(T,M_\mathrm{tar})$ is fixed with $T>0$ and $M_\mathrm{tar}>0$, and we use a set-target removal protocol: The amount of removed milk is the minimum of the session target $M_\mathrm{tar}$ and the available milk.

\medskip
We abbreviate
\begin{equation}
\label{eq:abbrev}
  a_\mathrm{tar}=a(M_\mathrm{tar}),\quad R_\mathrm{tar}=R(M_\mathrm{tar}),
\end{equation}
where $R(M)=a(M)/M$ is the stimulation efficiency as defined and discussed in Sec.~\ref{sec:stim}.
The state space is $X=[0,\infty)^2$ and the dynamics is generated by the map $F:X\to X$,
\begin{equation}\label{eq:Fmap}
  F(m,p)=\Big(\max\{x-M_\mathrm{tar},\,0\},\;\;p\,g+a\big(\min\{M_\mathrm{tar},\,x\}\big)\Big),
\end{equation}
where $x=md+pG$ is the amount of milk in the breast just prior to removal. Lem.~\ref{lem:monotone} and Lem.~\ref{lem:bounded} in App.~\ref{app:amount_stim_target} establish basic monotonicity and boundedness results.\newline

\begin{figure}[t]
  \centering
  \includegraphics[width=\textwidth]{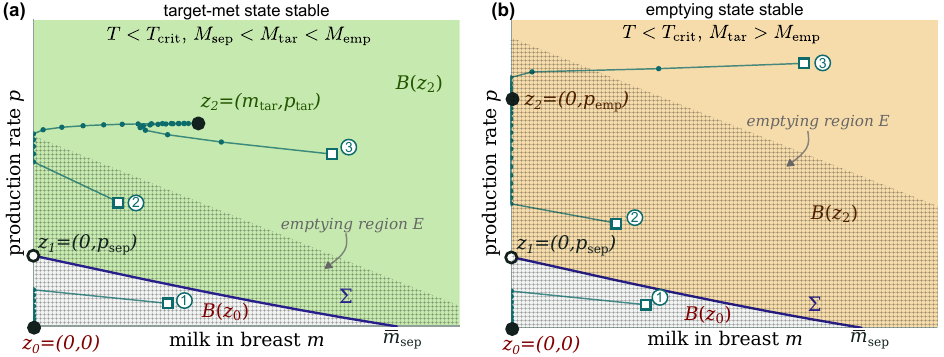}
  \caption{Phase planes for the set-target map \eqref{eq:Fmap} with a unimodal-efficiency response, illustrated with the Hill example \eqref{eqn:hill}. The grey lattice pattern indicates the emptying region $E$, filled circles denote the stable states $z_0$ and $z_2$, the open circle the unstable state $z_1$. The separatrix $\Sigma$ (blue) runs from $(0,p_\mathrm{sep})$ to $(\bar m_\mathrm{sep},0)$ and separates the two basins $B(z_0)$ (grey) and $B(z_2)$ (green for the target-met state, orange for the emptying state). In (a) the attractor $z_2$
    lies in the interior, in (b) it lies on the $p$-axis. Orbits (1)-(3) show example behaviours.}
  \label{fig:amtStim_setTar_uni_phase}
\end{figure}

We call $E=\{z\in X:\;x(z)\le M_\mathrm{tar}\}$ the \emph{emptying region}: on $E$ all milk available immediately before the session is removed. If $x<M_\mathrm{tar}$ the target cannot be met, while on the boundary $x=M_\mathrm{tar}$ the target is met exactly and the breast is emptied. We call $E^c:=X\setminus E$ the \emph{target-met region}: Here the target is met with milk left behind. See also Fig.~\ref{fig:amtStim_setTar_dec_phase} and Fig.~\ref{fig:amtStim_setTar_uni_phase}.

\paragraph{Fixed points in the target-met region $E^c$.} On $E^c$ the map $F$ reduces to the affine map
\begin{equation}\label{eq:affine}
  F(m,p)=\big(x-M_\mathrm{tar},\;p\,g+a(M_\mathrm{tar})\big).
\end{equation}
Solving the fixed point equation $F(z)=z$ gives a unique fixed point $z_\mathrm{tar}=(m_\mathrm{tar},p_\mathrm{tar})$, the \emph{target-met fixed point}, given by
\begin{equation}\label{eqn:target_met}
  p_\mathrm{tar}=\frac{a(M_\mathrm{tar})}{1-g}=\frac{M_\mathrm{tar}R_\mathrm{tar}}{G\Gamma},\qquad
  m_\mathrm{tar}=\frac{p_\mathrm{tar}G-M_\mathrm{tar}}{1-d}=\frac{M_\mathrm{tar}}{1-d}
\left(
\frac{R_\mathrm{tar}}{\Gamma}-1
\right).
\end{equation}
For $z_\mathrm{tar}$ to lie in $E^c$ it has to fulfill $z_\mathrm{tar}\in X$ and $x(z_\mathrm{tar})>M_\mathrm{tar}$. It is easy to see that this is equivalent to
\begin{equation}
\label{eqn:R_Ga}
R_\mathrm{tar}>\Gamma(T).
\end{equation}
%Inspecting \eqref{eqn:target_met} we see that this also implies $m_\mathrm{tar}> 0$. 

\paragraph{Fixed points in the emptying region $E$.}
On $E$ the map $F$ becomes
\begin{equation*}
  F(m,p)=\big(0,\;p\,g+a\big(x(m,p)\big)\big).
\end{equation*}
We see that the first component of $F$ vanishes, so every point of $E$ is mapped onto the $p$-axis. While an orbit lies on the \(p\)-axis within the emptying region, its dynamics reduces to the scalar map
$$ \Phi(p)=gp+a(Gp), $$
with
$$ \Phi(p)-p=pG\bigl(R(Gp)-\Gamma\bigr).$$
Hence the sign of \(R-\Gamma\) determines whether production increases or decreases along the \(p\)-axis. The resulting scalar dynamics for the decreasing- and unimodal-efficiency cases is used in the proofs of the global results and is detailed in App.~\ref{app:amount_stim_target} in Lem.~\ref{lem:axisconc} and Lem.~\ref{lem:axis}.\newline

Fixed points in $E$ all have $m=0$ and are given by $z_0=(0,0)$ and $(0,p)$ with $p>0$ a solution to
\begin{equation}
\label{eqn:M_emp}
R(pG)=\Gamma(T).
\end{equation}
For the resulting fixed point to lie in $E$, we need the associated volume of removed milk $x=pG$ to fulfil
\begin{equation}
\label{eqn:M_emp_exist}
pG\leq M_\mathrm{tar}.
\end{equation}

\paragraph{Stimulation efficiency vs inter-session production loss.}
Eqs.~\eqref{eqn:R_Ga} and \eqref{eqn:M_emp} show that the key to understanding fixed point existence is the relation of two central quantities: $\Gamma(T)$, and the stimulation efficiency $R(M)$. Straight-forward calculations show that $\Gamma(T)$ is strictly increasing with $\lim_{T\to0^+}\Gamma(T)=\gamma$ and $\lim_{T\to\infty}\Gamma(T)=\infty$. For $R$, we define
\begin{equation}
  \label{eqn:Rmax}
  R_\mathrm{max}=\sup_{M>0}R(M).
\end{equation}

\begin{figure}[t]
  \centering
  \includegraphics[width=\textwidth]{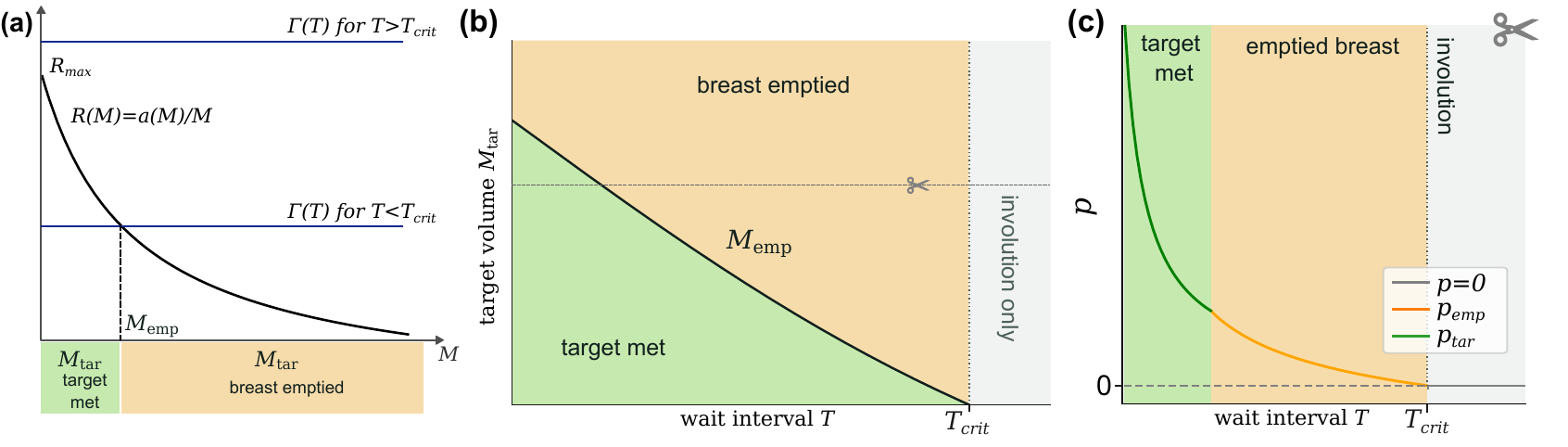}
  \caption{Decreasing-efficiency stimulation, illustrated with the Michaelis--Menten response in \eqref{eqn:mich_ment} with $\gamma<R_\mathrm{max}$. (a) The stimulation efficiency $R(M)$ and its intersection with $\Gamma(T)$ for $T>T_\mathrm{crit}$ (no intersection) and $T<T_\mathrm{crit}$ (one intersection). (b) The corresponding schedule space $(T,M_\mathrm{tar})$. For $T<T_\mathrm{crit}$ the maximal achievable target curve $M_\mathrm{emp}(T)$, defined in \eqref{eq:windowconc}, separates the schedule space: for $M_\mathrm{tar}>M_\mathrm{emp}$ the target is not met at the attractor and the breast is always emptied (orange), for $M_\mathrm{tar}<M_\mathrm{emp}$ the target is met with milk left behind (green), and for $T>T_\mathrm{crit}$ the involution state is globally attracting (grey). The horizontal dashed-grey line indicates the slice shown in (c). (c) Slice through schedule space for a fixed value of $M_\mathrm{tar}$. Shown are the steady state values of $p$ as functions of $T$ with solid lines indicating stable fixed points and dashed lines unstable ones.}
  \label{fig:amtStim_setTar_dec_schedule}
\end{figure}

\begin{figure}[t]
  \centering
  \includegraphics[width=\textwidth]{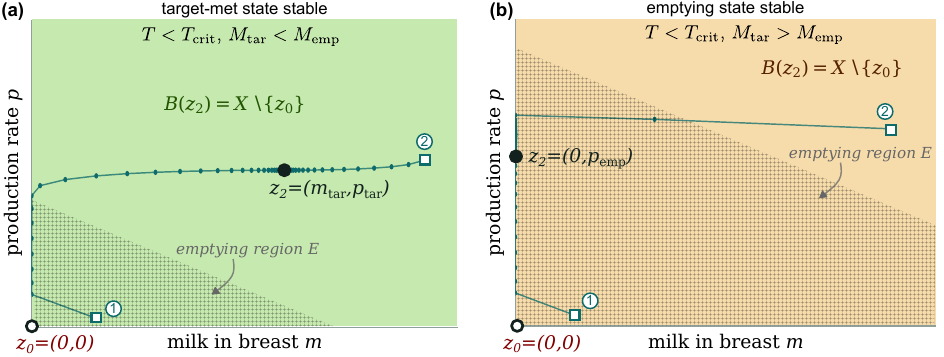}
  \caption{Phase planes for the set-target map \eqref{eq:Fmap} with a decreasing-efficiency
    response, illustrated with the Michaelis--Menten example of Fig.~\ref{fig:amtStim_setTar_dec_schedule}. The grey lattice pattern marks the emptying region $E$.
    The filled circle is the stable state $z_2$; the open circle at the origin is the unstable
    involution state $z_0$. In (a) $z_2$ is a target-met state in $E^c$, in (b) $z_2$ is an emptying state on the $p$-axis. Orbits (1),(2) show example behaviours.}
  \label{fig:amtStim_setTar_dec_phase}
\end{figure}

\begin{figure}[t]
  \centering
  \includegraphics[width=\textwidth]{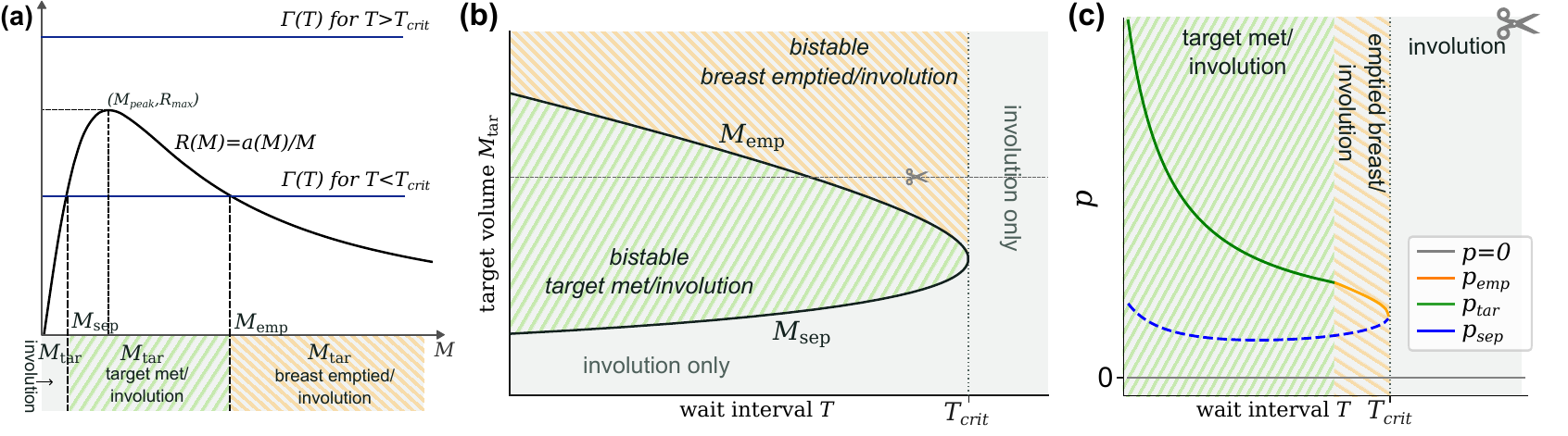}
  \caption{Unimodal-efficiency stimulation, illustrated with sigmoidal Hill response \eqref{eqn:hill} for $\gamma<R_\mathrm{max}$. (a) The stimulation efficiency $R(M)$ and its intersection with $\Gamma(T)$ for $T>T_\mathrm{crit}$ (no intersection) and $T<T_\mathrm{crit}$ (two intersections). (b) The corresponding schedule space $(T,M_\mathrm{tar})$. A globally attracting involution state is shown in grey, bistable regions are shown in two colours: green/grey whenever the target-met and the involution state are both stable, and orange/grey whenever the emptying state and the involution state are both stable. The horizontal dashed-grey line indicates the slice shown in (c). (c) Slice through schedule space for a fixed value of $M_\mathrm{tar}$. Shown are the steady state values of $p$ as functions of $T$. Solid lines indicate stable states, dashed lines unstable ones.}
  \label{fig:amtStim_setTar_uni_schedule}
\end{figure}

\noindent Therefore, if $\gamma<R_\mathrm{max}$, there exists a unique $T_\mathrm{crit}>0$ defined by 
\begin{equation}
\label{eqn:Tcrit}
\Gamma(T_\mathrm{crit})=R_\mathrm{max}.
\end{equation}
For $T<T_\mathrm{crit}$ there exists at least one solution $p>0$ to \eqref{eqn:M_emp} and a range of $M_\mathrm{tar}$ values that fulfil \eqref{eqn:R_Ga}, see also Fig.~\ref{fig:amtStim_setTar_dec_schedule}(a) and Fig.~\ref{fig:amtStim_setTar_uni_schedule}(a). If $\gamma\geq R_\mathrm{max}$ then $\Gamma(T)>R_\mathrm{max}$ for all $T>0$ and both \eqref{eqn:R_Ga} and \eqref{eqn:M_emp} can never be fulfilled. In this case $z_0=(0,0)$ is the only fixed point and we define $T_\mathrm{crit}=0$.\newline

Thus $T_\mathrm{crit}$ represents the critical wait threshold below which sustained lactation can in principle be achieved. For $T_\mathrm{crit}=0$ no positive wait interval can sustain lactation.
If $0<T<T_\mathrm{crit}$, there exists a range of removed volumes for which stimulation exceeds inter-session production loss. These considerations also show that $\Gamma(T)$ can be interpreted as the wait-interval dependent stimulation efficiency required to balance between-session production loss. Next we will discuss how the existence of the fixed points depend on the shape of $R$.

\paragraph{Decreasing stimulation efficiency.}

% ---------------------------------------------------------------- FIGURE ---

We first consider responses for which stimulation efficiency $R(M)=a(M)/M$ is largest at small removed volumes and decreases thereafter, see also Sec.~\ref{sec:stim}. We assume $R(M)$ is defined for $M\geq 0$ and fulfils $R'(M)<0$ for $M>0$, and $R(M)\to 0$ as $M\to\infty$. Then $R_\mathrm{max}=R(0)=a'(0)$, see Fig.~\ref{fig:amtStim_setTar_dec_schedule}(a).

\medskip
We see that for $T<T_\mathrm{crit}$, there exists a unique $M_\mathrm{emp}(T)>0$, such that 
\begin{equation}\label{eq:windowconc}
  R(M_\mathrm{emp}(T))=\Gamma(T),
  \qquad
  R(M)>\Gamma(T)\quad\Longleftrightarrow\quad 0<M<M_\mathrm{emp}(T).
\end{equation}
We write $p_\mathrm{emp}=M_\mathrm{emp}/G$, see \eqref{eqn:M_emp}, and call $(0,p_\mathrm{emp})$ the \emph{emptying fixed point}.
Holding $T$ fixed, we can now inspect how the existence of fixed points depend on the second scheduling parameter $M_\mathrm{tar}$: If $M_\mathrm{emp}< M_\mathrm{tar}$ the emptying fixed point exists/lies in the emptying region, see \eqref{eqn:M_emp_exist}. If $M_\mathrm{emp}> M_\mathrm{tar}$ the emptying fixed point does not exist, but the target-met fixed points exists, see \eqref{eqn:R_Ga}. This is summarized in Fig.~\ref{fig:amtStim_setTar_dec_schedule}(a),(b).

%For fixed points $(0,p_\mathrm{emp})$ in the emptying region, we find that if $T<T_\mathrm{crit}$, there is a unique $M_\mathrm{emp}>0$ such that (compare \eqref{eqn:M_emp} and see Fig.~\ref{fig:amtStim_setTar_dec_schedule}(a))

 %If $T>T_\mathrm{crit}$, $R(M)<\Gamma$ for every $M>0$ and no
%positive root exists. Since $\Gamma$ depends on $T$, so does $M_\mathrm{emp}$ and we sometimes write $M_\mathrm{emp}(T)$ to emphasise this.
%\newline

\paragraph{Example: concave Michaelis--Menten response.}
For the Michaelis--Menten response from \eqref{eqn:mich_ment}, i.e.
\begin{equation*}
  a(M)=a_\mathrm{max}\frac{M}{M+\mu_M},
\end{equation*}
one has $R(M)=a_\mathrm{max}/(M+\mu_M)$ and $R_\mathrm{max}=a_\mathrm{max}/\mu_M$. Hence for $T<T_\mathrm{crit}$ we find, see Fig.~\ref{fig:amtStim_setTar_dec_schedule}(b),(c),
\begin{equation}\label{eq:rootconc}
  p_\mathrm{emp}=\frac{\mu_M}{G}(\kappa-1),
  \qquad
  M_\mathrm{emp}=\mu_M(\kappa-1),\qquad \kappa:=\frac{R_\mathrm{max}}{\Gamma}.
\end{equation}
This example of a decreasing-efficiency response is used in the figures, but the theorem below does not depend on this particular choice.\newline

The following theorem establishes how system long-term behaviour depends on the scheduling parameters $(T,M_\mathrm{tar})$:

\begin{theorem}\label{thm:amtStim_tar_dec}
Let $R(M)=a(M)/M$ be a decreasing-efficiency response as above.
\begin{enumerate}
  \item[(i)] If $T> T_\mathrm{crit}$, then $z_0=(0,0)$ is the unique fixed point of $F$ in $X$ and it is
    globally asymptotically stable.
  \item[(ii)] If $T<T_\mathrm{crit}$ and $M_\mathrm{tar}\neq M_\mathrm{emp}$, then $F$ has exactly two fixed
    points, namely $z_0$ and
    \[
      z_2=\big(m_\mathrm{tar},p_\mathrm{tar}\big)\ \text{ with }m_\mathrm{tar}>0
      \quad\text{if }M_\mathrm{tar}<M_\mathrm{emp},
      \qquad
      z_2=\big(0,p_\mathrm{emp}\big)
      \quad\text{if }M_\mathrm{tar}>M_\mathrm{emp}.
    \]
    The state $z_0$ is unstable, and $z_2$ attracts every other initial condition, i.e. the basin of attraction of $z_2$,
    $B(z_2)=X\setminus\{z_0\}$.
\end{enumerate}
\end{theorem}

For proof see App.~\ref{app:amount_stim_target}, for visualisations see Fig.~\ref{fig:amtStim_setTar_dec_schedule} and Fig.~\ref{fig:amtStim_setTar_dec_phase}.

%-------------

\paragraph{Interpretation.}

Thm.~\ref{thm:amtStim_tar_dec} gives the following conclusion: whenever stimulation efficiency
$R(M)=a(M)/M$ decreases with the removed volume, the schedule alone decides the long-term outcome. If $\gamma\ge R_\mathrm{max}$ sustained lactation is impossible for any schedule $(T,M_\mathrm{tar})$ and involution occurs. In this case even the maximal stimulation efficiency cannot overcome the minimal production loss. If $\gamma<R_\mathrm{max}$, there exists a critical wait interval $T_\mathrm{crit}>0$. For wait intervals with $T>T_\mathrm{crit}$, again lactation cannot be sustained, since stimulation cannot compensate for the intersession loss. However for wait intervals with $T<T_\mathrm{crit}$, the system converges to a sustained lactation state from every nonzero starting state.

\medskip
For such sufficiently short wait intervals $T<T_\mathrm{crit}$ the sustained state can have two different forms depending on whether the session target $M_\mathrm{tar}$ lies below or above the maximal achievable target $M_\mathrm{emp}(T)$. If $M_\mathrm{tar}<M_\mathrm{emp}$ the session target is met with milk left behind. If $M_\mathrm{tar}>M_\mathrm{emp}$, every session empties the available milk, and the target is not met. The shorter the intersession wait time $T$, the larger the maximal achievable target $M_\mathrm{emp}$. These regimes are summarized in Fig.~\ref{fig:amtStim_setTar_dec_schedule}(b),(c).

\medskip
Although the initial milk amount and production rate do not affect the long-term outcome, they can affect the transient route to the attractor. For example, Fig.~\ref{fig:amtStim_setTar_dec_phase}(a) shows an orbit that initially undergoes complete emptying before production increases sufficiently for the session target to be met. The unimodal response considered next shows that a different efficiency profile can instead make the long-term outcome itself initial-state dependent.

%==========================================================================

\paragraph{Unimodal stimulation efficiency.}

% ---------------------------------------------------------------- FIGURE ---

Next we consider responses for which stimulation efficiency is weak for very small removals,
maximal at an intermediate volume, and decreases again for large removals. Specifically, assume
that $R(M)=a(M)/M$ extends continuously to $M=0$ with $R(0)=0$, has a unique maximum at $M=M_\mathrm{peak}>0$, satisfies $R'(M)>0$ for $0<M<M_\mathrm{peak}$ and $R'(M)<0$ for $M>M_\mathrm{peak}$, and tends to zero as $M\to\infty$. Then $R_\mathrm{max}=R(M_\mathrm{peak})$, see Fig.~\ref{fig:amtStim_setTar_uni_schedule}(a).\newline

For $0<T<T_\mathrm{crit}$ $R(M)=\Gamma$ (compare \eqref{eqn:M_emp} and see Fig.~\ref{fig:amtStim_setTar_uni_schedule}(a)) has exactly two positive solutions $M_\mathrm{sep}$ and $M_\mathrm{emp}$ with
\[
  M_\mathrm{sep}<M_\mathrm{peak}<M_\mathrm{emp},
  \qquad
  p_\mathrm{sep}=M_\mathrm{sep}/G,\quad p_\mathrm{emp}=M_\mathrm{emp}/G,
\]
and
\begin{equation}\label{eq:window}
  R(M)>\Gamma
  \qquad\Longleftrightarrow\qquad
  M_\mathrm{sep}<M<M_\mathrm{emp}.
\end{equation}
Holding $T$ fixed, we can now inspect how the existence of fixed points depend on the second scheduling parameter $M_\mathrm{tar}$: The fixed point $z_0=(0,0)$ always exists. We see that for $M_\mathrm{sep}<M_\mathrm{tar}<M_\mathrm{emp}$ the target-met fixed point exists, see \eqref{eqn:R_Ga}. Additionally the fixed point $(0,p_\mathrm{sep})$ in the emptying region exists, see \eqref{eqn:M_emp_exist}. For $M_\mathrm{emp}<M_\mathrm{tar}$ the target-met fixed point does not exist, but the two fixed points in the emptying region exist, $(0,p_\mathrm{sep})$ and $(0,p_\mathrm{emp})$, see \eqref{eqn:M_emp_exist}. If $M_\mathrm{tar}<M_\mathrm{sep}$ only $z_0=(0,0)$ exists. This is summarized in Fig.~\ref{fig:amtStim_setTar_uni_schedule}(a),(b).

% ---------------------------------------------------------------- FIGURE ---

\paragraph{Example: Sigmoidal Hill response.}
For the Hill response \eqref{eqn:hill}, i.e.
\[
  a(M)=a_\mathrm{max}\frac{M^2}{M^2+\mu_M^2},
\]
one has unimodal efficiency with $M_\mathrm{peak}=\mu_M$ and
$R_\mathrm{max}=a_\mathrm{max}/(2\mu_M)$. For $T<T_\mathrm{crit}$, and defining $\kappa:=R_\mathrm{max}/\Gamma$, the two thresholds are explicit, see Fig.~\ref{fig:amtStim_setTar_uni_schedule}(b),(c):
\begin{equation}\label{eq:roots}
  p_\mathrm{sep}=\frac{\mu_M}{G}\left(\kappa-\sqrt{\kappa^2-1}\right),
  \qquad
  p_\mathrm{emp}=\frac{\mu_M}{G}\left(\kappa+\sqrt{\kappa^2-1}\right),
\end{equation}
\[
  M_\mathrm{sep}=\mu_M\left(\kappa-\sqrt{\kappa^2-1}\right),
  \qquad
  M_\mathrm{emp}=\mu_M\left(\kappa+\sqrt{\kappa^2-1}\right).
\]
These closed forms are specific to the example; the global results below use only the shape of $R$.\newline

The following definitions are useful for formulating the next theorem: The state space $X=[0,\infty)^2$ is equipped with the component-wise partial order: $(m,p)\le(\tilde m,\tilde p)$
iff $m\le\tilde m$ and $p\le\tilde p$.
A set $S\subseteq X$ is a \emph{lower set} if
$z\le w\in S$ implies $z\in S$, and an \emph{upper set} if $z\ge w\in S$ implies $z\in S$.\newline

The following theorem and corollary describe how the long-term dynamics depend on the scheduling parameters $(T,M_\mathrm{tar})$ and the initial conditions.

\begin{theorem}\label{thm:amtStim_tar_uni}
Let $R=a(M)/M$ be a unimodal-efficiency response as above. Assume $0<T<T_\mathrm{crit}$, $M_\mathrm{tar}>M_\mathrm{sep}$ and
$M_\mathrm{tar}\neq M_\mathrm{emp}$. Then the following hold.

\smallskip
\textbf{(i) Fixed points.} $F$ has exactly three fixed points in $X$:
the involution state $z_0=(0,0)$; the state
$z_1=(0,p_\mathrm{sep})$; and a third state
\[
\begin{cases}
  z_2=\big(m_\mathrm{tar},\,p_\mathrm{tar}\big)
  \ \text{ with } m_\mathrm{tar}>0
  \quad&\text{if } M_\mathrm{tar}<M_\mathrm{emp},
  \\
  z_2=\big(0,\,p_\mathrm{emp}\big)
  \quad&\text{if } M_\mathrm{tar}>M_\mathrm{emp}.
\end{cases}
\]
The states $z_0$ and $z_2$ are locally asymptotically stable, and
$z_1$ is unstable.

\smallskip
\textbf{(ii) Global behaviour.} Writing $B(z)$ for the basin of
attraction of a fixed point $z$, we can write the state space in terms of the
disjoint union
\[
  X=B(z_0)\ \sqcup\ \Sigma\ \sqcup\ B(z_2),
  \qquad \Sigma=W^s(z_1),
\]
where $W^s(z_1)$ is the stable set of the unstable fixed point $z_1$.
Thus every orbit converges to one of the three fixed points.
The set $B(z_0)$ is an open lower set and $B(z_2)$ is an open upper set.

\smallskip
\textbf{(iii) Separatrix.} The separatrix $\Sigma$ is the graph of a
continuous, strictly decreasing function
$\sigma:[0,\bar m_\mathrm{sep}]\to[0,p_\mathrm{sep}]$ with
$\sigma(0)=p_\mathrm{sep}$ and $\sigma(\bar m_\mathrm{sep})=0$.
Its explicit construction is detailed in App.~\ref{app:amount_stim_target}.

\end{theorem}

\begin{corollary}\label{cor:regime1}
If $T>T_\mathrm{crit}$, or if $0<T<T_\mathrm{crit}$ and $M_\mathrm{tar}<M_\mathrm{sep}$, then $z_0$ is the unique fixed point of $F$ in $X$, and it is globally asymptotically stable.
\end{corollary}

For proofs see App.~\ref{app:amount_stim_target}, for visualisations see Fig.~\ref{fig:amtStim_setTar_uni_schedule} and Fig.~\ref{fig:amtStim_setTar_uni_phase}.

% ================================================================ INTERPRETATION
\subsubsection*{Interpretation}

The critical-wait threshold $T_\mathrm{crit}$ and the distinction between target-met and emptying lactation states are unchanged from the decreasing-efficiency case. However, the unimodal efficiency response introduces two additional features:

\medskip
First, for $T<T_\mathrm{crit}$ in addition to the maximal achievable target amount $M_\mathrm{emp}$ (beyond which stable lactation is possible, but the session target cannot be met), there is also a lower threshold $M_\mathrm{sep}$. For session targets $M_\mathrm{tar}$ below this value, the amount removed does not provide sufficient stimulation to maintain production, which leads to involution. Together these thresholds define a range of session targets that can be sustainably met $M_\mathrm{sep}(T)<M_\mathrm{tar}<M_\mathrm{emp}(T)$, see Fig.~\ref{fig:amtStim_setTar_uni_schedule}(b).

\medskip
Second, Theorem~\ref{thm:amtStim_tar_uni} says that, for a unimodal stimulation efficiency, whether stable lactation can be achieved is not determined by the schedule alone. For  $M_\mathrm{tar}>M_\mathrm{sep}$ the schedule admits two outcomes, involution and stable lactation, and which one is realised depends on initial milk in the breast and the initial production rate. The model therefore separates two questions: what state a schedule can sustain, and whether that state can be reached from the current condition.

\medskip
Whenever a positive lactation state exists, the $(m,p)$ state space is divided into two regions representing the possible initial states that lead to involution or sustained lactation, which is shown in Fig.~\ref{fig:amtStim_setTar_uni_phase}. The dividing separatrix is a decreasing curve, quantifying the trade-off between initially stored milk and initial production rate: Starting from an empty breast, production must exceed $p_\mathrm{sep}$; starting from zero production, stored volume must exceed $\bar m_\mathrm{sep}$. Between these $\Sigma$ interpolates monotonically.

\subsection{Amount-dependent stimulation, set-duration protocol}
\label{sec:amount_stim_duration}

We continue to assume that stimulation depends on the amount of milk removed, but now prescribe the removal duration \(\tau>0\) rather than a target amount. Writing \(h=h(\tau)\) for the fraction of milk remaining after the session and \(x=md+pG\) for the available milk immediately before removal, the removed amount is
$$ M=x(1-h). $$
We use the linear removal law introduced in Sec.~\ref{sec:model}, \(h(\tau)=\max\{1-\nu\tau,0\}\) and define $\tau_\mathrm{emp}=1/\nu$. Removal durations \(\tau\ge \tau_\mathrm{emp}\) produce no additional removal or stimulation, so we analyse \(0<\tau<\tau_\mathrm{emp}\) and treat \(\tau=\tau_\mathrm{emp}\) as the complete-emptying endpoint. For $\tau\ge\tau_{\mathrm{emp}}$ the map reduces to $F(m,p)=\bigl(0,\,pg+a(x)\bigr)$, which we have already analysed as the emptying scenario in Sec.~\ref{sec:amount_stim_target}.\newline

Hence, if $0<\tau<\tau_\mathrm{emp}$ we have $0<h<1$, and the dynamics is generated by
\begin{equation}\label{eq:Fdur}
  F(m,p)=\Big(x\,h,\;\;p\,g+a\big(x(1-h)\big)\Big).
\end{equation}

A fixed point $(m_*,p_*)$ will satisfy
\begin{equation}
\label{eqn:dursteady}
m_*=\frac{p_*Gh}{1-dh},\qquad p_*(1-g)=a\left(p_* G r\right)  
\end{equation}
with $r$ defined as
\begin{equation}
\label{eqn:r}
r=r(T,\tau):=\frac{1-h(\tau)}{1-h(\tau)d(T)} \in (0,1).  
\end{equation}
Since $a(0)=0$, there always exists the zero fixed point $z_0=(0,0)$.% and every non-trivial fixed point will be interior.

%------------

% ---------------------------------------------------------------- FIGURE ---
\begin{figure}[t]
  \centering
  \includegraphics[width=\textwidth]{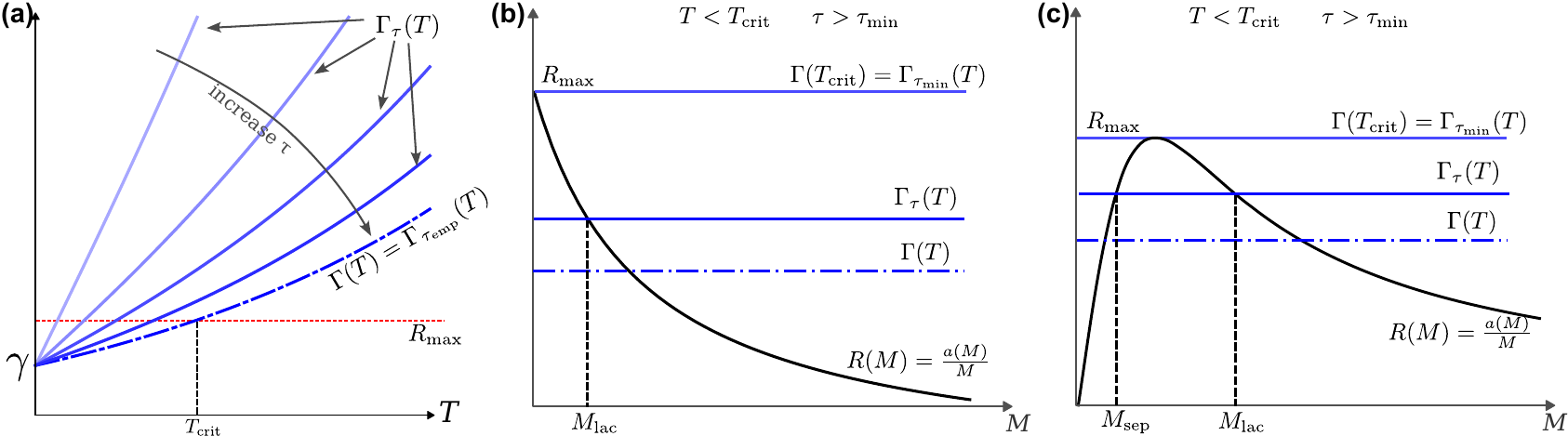}
  \caption{(a) Behaviour of $\Gamma(T)$ and $\Gamma_\tau(T)$ as functions of $T$. (b),(c) Determination of $M_\mathrm{lac}$ and $M_\mathrm{sep}$ for a decreasing-efficiency response (b) and a unimodal-efficiency response (c).}
  \label{fig:amtStim_setDur_curves}
\end{figure}
%--------------------------------

\paragraph{Steady-state condition and minimum removal duration.}

\begin{figure}[t]
  \centering
  \includegraphics[width=0.8\textwidth]{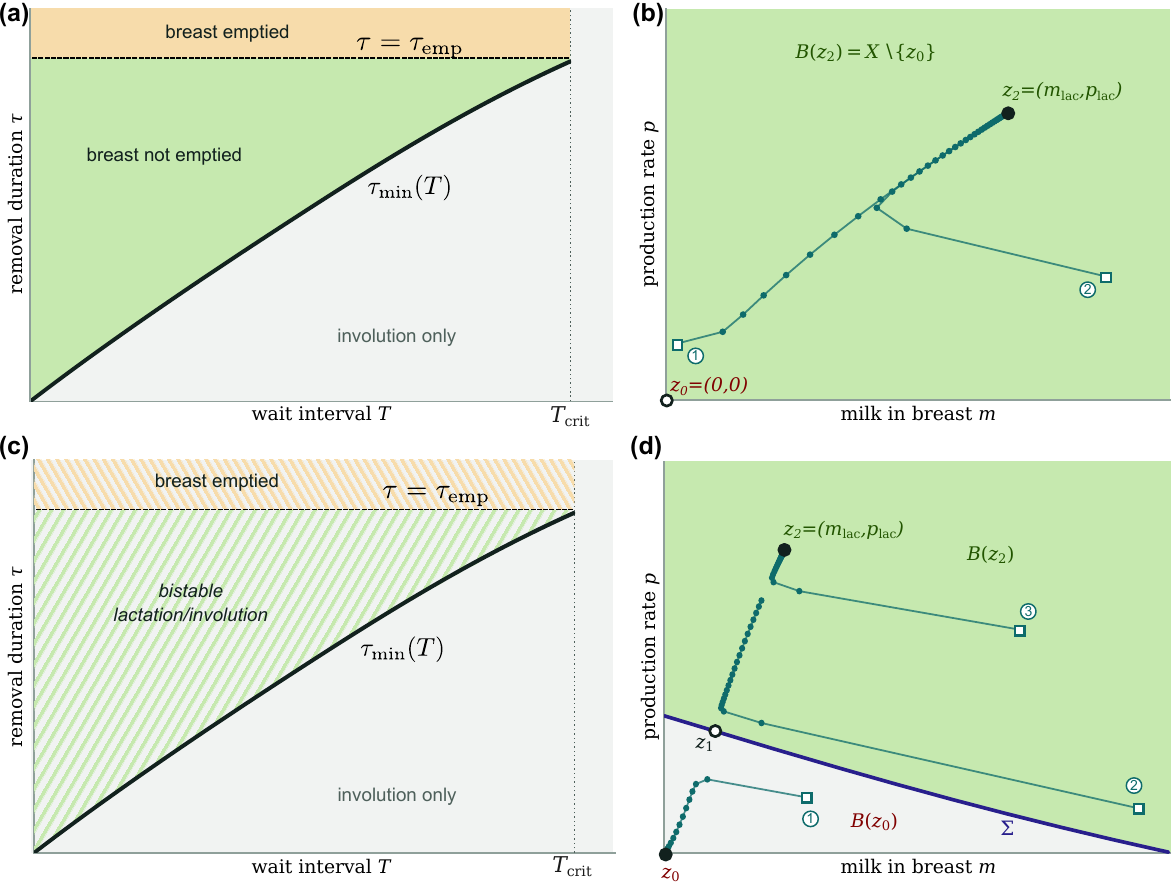}
  \caption{(a),(c) Partition of schedule space $(T,\tau)$ for a set-duration protocol using the same coloring conventions as in Fig.~\ref{fig:amtStim_setTar_dec_schedule} and Fig.~\ref{fig:amtStim_setTar_uni_schedule}. Shown is a Michaelis-Menten response \eqref{eqn:mich_ment} as example for a decreasing efficiency response in (a) and a Hill-response \eqref{eqn:hill} as example for a unimodal efficiency response in (c). The line $\tau_\mathrm{min}$ separates the schedule space and the line $\tau=\tau_\mathrm{emp}$ marks complete emptying of the breast. (b),(d) Phase plane for the decreasing efficiency response in (a) (b) and the unimodal response of (c) (d) for the case $\tau_\mathrm{min}(T)<\tau<\tau_\mathrm{emp}$. The color conventions are the same as in Fig.~\ref{fig:amtStim_setTar_dec_phase} and Fig.~\ref{fig:amtStim_setTar_uni_phase}. Shown are fixed points, the basins of attraction for the lactation state $z_2$ in green and for the involution state $z_0$ in grey, as well as example orbits (1)-(3).}
  \label{fig:amtStim_setDur}
\end{figure}

Writing the fixed-point condition \eqref{eqn:dursteady} in terms of the amount removed at
steady state $M_*=p_*Gr$ gives
\begin{equation}
\label{eqn:amtStim_setDur_fp}
    R(M_*)=\Gamma_\tau(T,\tau),
\end{equation}
where
\begin{equation}
\label{eq:Gammatau}
    \Gamma_\tau(T,\tau)
    :=
    \frac{\Gamma(T)}{r(T,\tau)}.
\end{equation}
Thus $\Gamma_\tau$ is the stimulation efficiency required to balance
production loss for a set-duration schedule.
As \(\tau\) increases from \(0\) to \(\tau_\mathrm{emp}\), \(r(\tau)\) increases from \(0\) to \(1\). Hence, for each fixed $T>0$, $\Gamma_\tau(T)$ decreases from $\infty$ to $\Gamma(T)$ as \(\tau\) increases from \(0\) to \(\tau_\mathrm{emp}\). Thus longer removal sessions reduce the stimulation efficiency required for maintenance.\newline
%For later use in the explicit Michaelis-Menten and Hill formulas below, we also define
%\begin{equation}
%\label{eq:kappatau}
%    \kappa_\tau
%    :=
%    \frac{R_\mathrm{max}}{\Gamma_\tau}
%    =\kappa r.
%\end{equation}
The following can be shown using basic monotonicity properties of $\Gamma_\tau$ and $\Gamma$, see Fig.~\ref{fig:amtStim_setDur_curves}: There exists at least one solution $M_*>0$  to \eqref{eqn:amtStim_setDur_fp} if and only if the following three, nested conditions hold (ignoring equality cases)
$$
\gamma<R_\mathrm{max},\quad T<T_\mathrm{crit},\quad \tau> \tau_\mathrm{min}(T).
$$
A positive wait-interval threshold $T_\mathrm{crit}>0$ only exists for $\gamma<R_\mathrm{max}$ and is defined (as in Sec.~\ref{sec:amount_stim_target}) as solution to
$$
\Gamma(T_\mathrm{crit})=R_\mathrm{max}.
$$
The minimal removal duration $0<\tau_\mathrm{min}(T)<\tau_\mathrm{emp}$ only exists for $T<T_\mathrm{crit}$ and is defined as solution to 
\[
    \Gamma_{\tau_{\mathrm{min}}}(T)=R_{\mathrm{max}}.
\]
This equation can be solved for $\tau_\mathrm{min}$, yielding
\begin{equation}
\label{eq:taumin}
    \tau_{\mathrm{min}}(T)
    =
    \tau_\mathrm{emp}\frac{\Gamma(T)(1-d(T))}
         {R_{\mathrm{max}}-d(T)\Gamma(T)}.
\end{equation}
We see that as $T$ increases from $0$ to $T_\mathrm{crit}$, $\tau_{\mathrm{min}}(T)$ increases from $0$ to $\tau_\mathrm{emp}$. Note that $\tau_{\mathrm{min}}(T)$  depends on the stimulation-response only through $R_\mathrm{max}$, compare curves in Fig.~\ref{fig:amtStim_setDur}(a),(c).\newline

\begin{comment}

\textcolor{red}{
Away from the threshold case, a positive solution $M_*>0$ to \eqref{eqn:amtStim_setDur_fp} exists only
$$
\Gamma_\tau<R_\mathrm{max}.
$$
A removal duration satisfying this inequality exists if and only if
\[
    \Gamma(T)<R_{\mathrm{max}} \iff T<T_{\mathrm{crit}}.
\]
Thus the critical wait threshold is exactly the same as for the
set-target protocol. If $T>T_{\mathrm{crit}}$, no removal duration can
sustain a positive steady state. For $0<T<T_{\mathrm{crit}}$ there is a unique minimum removal duration
$\tau_{\mathrm{min}}(T)$ determined by
\[
    \Gamma_{\tau_{\mathrm{min}}}(T,\tau_{\mathrm{min}})=R_{\mathrm{max}}.
\]
For the linear removal law
$h(\tau)=1-\nu\tau$, one has
\[
    r(T,\tau)
    =
    \frac{\nu\tau}{1-d+d\nu\tau},
\]
and hence
\begin{equation}
\label{eq:taumin}
    \tau_{\mathrm{min}}(T)
    =\frac{1-d(T)}{\nu(\kappa-d(T))}=
    \frac{\Gamma(T)(1-d(T))}
         {\nu\bigl(R_{\max}-d(T)\Gamma(T)\bigr)}.
\end{equation}
Therefore, for $0<T<T_{\rm crit}$,
\[
    \Gamma_\tau<R_{\max}
    \quad\Longleftrightarrow\quad
    \tau>\tau_{\mathrm{min}}(T).
\]
At the endpoint $\tau=\tau_\mathrm{emp}$, one has $h=0$ and $r=1$, so every
session completely empties the breast and
$\Gamma_\tau=\Gamma$. The fixed-point condition then reduces to $R(M)=\Gamma$, the same condition obtained for an emptying steady state under
set-target removal.}\newline

\end{comment}

% ---------------------------------------------------------------- FIGURE ---
%--------------------------------

The analysis in App.~\ref{app:amount_stim_duration} in Lem.~\ref{lem:durconv} and Lem.~\ref{lem:durjac} shows two general facts used below: A positive steady state on a decreasing branch of \(R\), i.e. for $R'(M_*)<0$, is locally asymptotically stable, whereas one on an increasing branch, i.e. for $R'(M_*)>0$, is a saddle. Moreover, every orbit converges to a fixed point.

% ================================================================ DECREASING EFFICIENCY
\paragraph{Decreasing stimulation efficiency.}
Assume that $R$ decreases strictly from $R(0)=R_\mathrm{max}$ to zero. For $0<T<T_\mathrm{crit}$, $R(M)=\Gamma_\tau$ has a unique positive solution $M_\mathrm{lac}$ exactly when $\tau_\mathrm{min}(T)<\tau<\tau_\mathrm{emp}$, see Fig.~\ref{fig:amtStim_setDur_curves}(b), and the corresponding production rate is
\[
  p_\mathrm{lac}=\frac{M_\mathrm{lac}}{G r}.
\]

\paragraph{Example: concave Michaelis--Menten response.}
For $a(M)=a_\mathrm{max}M/(M+\mu_M)$ (see \eqref{eqn:mich_ment}) this root is explicit:
\begin{equation}\label{eq:durrootconc}
  M_\mathrm{lac}=\mu_M(\kappa_\tau-1),\quad \kappa_\tau=\frac{R_\mathrm{max}}{\Gamma_\tau}.
\end{equation}
See Fig.~\ref{fig:amtStim_setDur}(a),(b) for a summary of the results of the following theorem:

\begin{theorem}\label{thm:durconc}
Let $R(M)=a(M)/M$ be a decreasing-efficiency response as above and write $z_0=(0,0)$.
\begin{enumerate}
  \item[(i)] If $T>T_\mathrm{crit}$, or if $0<T<T_\mathrm{crit}$ and
$\tau<\tau_\mathrm{min}(T)$, then $z_0$ is the unique fixed point of $F$ in $X$ and it is
    globally asymptotically stable.
  \item[(ii)] If $0<T<T_\mathrm{crit}$ and $\tau_\mathrm{min}(T)<\tau<\tau_\mathrm{emp}$, then $F$ has exactly two fixed points, $z_0$ and the interior
    state $z_2=(m_\mathrm{lac},p_\mathrm{lac})$, where $M_\mathrm{lac}$ is the unique solution of $R(M)=\Gamma_\tau$,
    $$
    p_\mathrm{lac}=\frac{M_\mathrm{lac}}{G r},\quad m_\mathrm{lac}=\frac{p_\mathrm{lac}Gh}{1-dh}.
    $$
   The state $z_0$ is unstable and the basin of attraction of $z_2$, $B(z_2)=X\setminus\{z_0\}$.
\end{enumerate}
\end{theorem}

For a proof see App.~\ref{app:amount_stim_duration}.
% ================================================================ UNIMODAL EFFICIENCY
\paragraph{Unimodal stimulation efficiency.}
Assume that $R(0)=0$, increases to a unique maximum and then decreases to zero. For $0<T<T_\mathrm{crit}$ and $\tau_\mathrm{min}(T)<\tau<\tau_\mathrm{emp}$, the equation $R(M)=\Gamma_\tau$ has two positive roots
$M_\mathrm{sep}<M_\mathrm{lac}$, see Fig.~\ref{fig:amtStim_setDur_curves}(c).% Both corresponding states are interior by \eqref{eqn:dursteady}.

\paragraph{Example: sigmoidal Hill response.}
For $a(M)=a_\mathrm{max}M^2/(M^2+\mu_M^2)$ as in \eqref{eqn:hill}, the two steady-state removed amounts are
\begin{equation}
  \label{eqn:durrootsig}
  M_\mathrm{sep}=\mu_M\left(\kappa_\tau-\sqrt{\kappa_\tau^2-1}\right),
  \quad
  M_\mathrm{lac}=\mu_M\left(\kappa_\tau+\sqrt{\kappa_\tau^2-1}\right),\quad \kappa_\tau=\frac{R_\mathrm{max}}{\Gamma_\tau}
\end{equation}
See Fig.~\ref{fig:amtStim_setDur}(c),(d) for a summary of the results of the following theorem and corollary:
%--------------------

\begin{theorem}\label{thm:dursig}
Let $R(M)=a(M)/M$ be a unimodal-efficiency response as above and assume $0<T<T_\mathrm{crit}$ and $\tau_\mathrm{min}(T)<\tau<\tau_\mathrm{emp}$. Let
\[
  M_\mathrm{sep}<M_\mathrm{lac}
\]
be the two solutions of $R(M)=\Gamma_\tau$ and define
\[
  p_\mathrm{sep}=\frac{M_\mathrm{sep}}{G r},\quad m_\mathrm{sep}
  =\frac{p_\mathrm{sep}Gh}{1-dh},
  \qquad
  p_\mathrm{lac}=\frac{M_\mathrm{lac}}{G r},\quad m_\mathrm{lac}
  =\frac{p_\mathrm{lac}Gh}{1-dh}.
\]
Set $z_0=(0,0)$, $z_1=(m_\mathrm{sep},p_\mathrm{sep})$ and $z_2=(m_\mathrm{lac},p_\mathrm{lac})$.

\begin{enumerate}
\item[(i)] \textbf{Fixed points and stability.}
The map $F$ has exactly the three fixed points $z_0,z_1,z_2$.
The states $z_0$ and $z_2$ are locally asymptotically stable, whereas
$z_1$ is a hyperbolic saddle.
\item[(ii)] \textbf{Global dynamics.}
Every orbit in $X$ converges to one of the three fixed points. Hence
\[
  X
  =
  B(z_0)\sqcup \Sigma\sqcup B(z_2),
  \qquad
  \Sigma:=W^s(z_1),
\]
where $B(z)$ and $W^s(z)$ denotes the basin of attraction and stable manifold of a fixed point $z$ respectively.
Moreover, $B(z_0)$ is an open lower set and $B(z_2)$ is an open
upper set with respect to the componentwise order on $X$.

\item[(iii)] \textbf{Separatrix.}
The stable set $\Sigma$ is a one-dimensional forward-invariant curve. It can be written as the graph of a continuous, strictly
decreasing function
\[
  p=\sigma(m),
  \qquad m\in I_\Sigma,
\]
where $I_\Sigma\subset[0,\infty)$ is an interval containing
$m_\mathrm{sep}$.
\end{enumerate}
\end{theorem}

For a proof, see App.\ref{app:amount_stim_duration}.

\begin{corollary}\label{cor:dursig}
Let $R$ be a unimodal-efficiency response. If $T>T_\mathrm{crit}$, or if $0<T<T_\mathrm{crit}$ and
$\tau<\tau_\mathrm{min}(T)$,
then $z_0=(0,0)$ is the unique fixed point and is globally
asymptotically stable.
\end{corollary}

% ================================================================ INTERPRETATION
\paragraph{Interpretation.}

The critical wait threshold \(T_{\rm crit}\) is unchanged by the removal protocol (compare Sec.~\ref{sec:amount_stim_target}): if \(T>T_{\rm crit}\), no removal duration can sustain lactation. For \(0<T<T_{\rm crit}\), set-duration removal introduces a second threshold, the minimum session duration \(\tau_{\mathrm{min}}(T)\). Sessions shorter than this threshold remove too little milk to generate sufficient stimulation, whereas longer sessions admit positive lactation states. Since \(\tau_{\mathrm{min}}(T)\) increases from \(0\) to \(\tau_\mathrm{emp}\) as \(T\) increases from \(0\) to \(T_{\rm crit}\), longer waits must be compensated by longer removal sessions. As under set-target removal, decreasing stimulation efficiency produces a unique positive attractor whenever maintenance is possible, whereas unimodal efficiency creates bistability between involution and sustained lactation.

\subsection{Duration-dependent stimulation, set-duration protocol}\label{sec:dur_stim_duration}

Next we examine the case in which stimulation depends on the length of a session rather than on the volume removed, so that each session contributes a fixed increment $a(\tau)$ to the production rate. Note that since we modelled the milk remaining in the breast as
$$
\text{available milk}\times h(\tau),
$$
$1-h(\tau)$ is the fraction of available milk removed and can therefore be interpreted as a measure of the degree of breast emptying. Since $h$ is a function of $\tau$, \emph{duration-dependent stimulation} can also be interpreted as \emph{degree-of-emptying-dependent stimulation} as long as $\tau<\tau_\text{emp}$. For larger values $\tau>\tau_\text{emp}$, i.e. if the session duration and therefore stimulation continues past complete emptying, this equivalence no longer holds.

\paragraph{The map becomes affine.}

Retaining the fixed removal duration for the milk equation, the model is
\begin{equation}\label{eq:Ftaustim}
  F(m,p)=\Big(x\,h,\;\;p\,g+a(\tau)\Big),
  \qquad x=x(m,p)=m\,d+p\,G .
\end{equation}
Only the second component has changed from \eqref{eq:Fdur}. Since it no longer refers to the state, the resulting map is \emph{affine},
\begin{equation}\label{eq:taustimaffine}
  F(z)=Lz+b,
  \qquad
  L=\begin{pmatrix} dh & Gh\\[1mm] 0 & g\end{pmatrix},
  \qquad
  b=\begin{pmatrix}0\\ a(\tau)\end{pmatrix},
\end{equation}
with the zero in the lower-left corner showing that there is no feedback from the milk present to production. This causes the dynamics to be particularly easy to understand.

\begin{theorem}\label{thm:taustim}
Assume $a(\tau)>0$. Then the map \eqref{eq:Ftaustim} has a unique fixed point
$z_\mathrm{lac}=(m_\mathrm{lac},p_\mathrm{lac})$ in $X$, given by
\begin{equation}\label{eq:taustimfp}
  p_\mathrm{lac}=\frac{a(\tau)}{1-g},
  \qquad
  m_\mathrm{lac}=\frac{p_\mathrm{lac}\,G\,h}{1-dh},
  \qquad
  M_\mathrm{lac}=Gr\,p_\mathrm{lac},
\end{equation}
where $r$ is defined in \eqref{eqn:r} and $z_\mathrm{lac}$ is globally asymptotically stable on $X$ for every \(T>0\) and removal duration \(\tau\) with $a(\tau)>0$. The involution state is not a steady state.
\end{theorem}

\paragraph{Interpretation.}

When the production stimulation depends on the removal duration (and not the amount removed), and a set-duration protocol is used, the milk production becomes an autonomous linear recursion with a globally attracting positive equilibrium, and the involution state $(m,p)=(0,0)$ ceases to exist. Thus, under the exponential decay assumptions of the baseline model, every finite wait interval and every initial condition lead to convergence to a lactation state. In particular, unlike the amount-dependent stimulation models, there is no threshold in removal frequency below which involution occurs. The mechanism is the loss of state-dependent feedback: When stimulation depends on the removed amount, a supply that has fallen supplies less milk, hence receives less stimulation, and may fall further; that self-reinforcement is what allows involution to be an attractor.

% ================================================================
\subsection{Duration-dependent stimulation, set-target protocol}
\label{sec:dur_stim_target}

We finally consider duration-dependent stimulation under a set-target removal protocol with schedule $(T,M_\mathrm{tar})$. As we will see, this combination can lead to a long-term behaviour that is qualitatively different from the ones observed in the preceding three sections.

\medskip
As previously, we let $M_{\mathrm{tar}}>0$ denote the prescribed target and
write $x=dm+Gp$ for the milk available immediately before a removal session. We use the linear removal law $h(\tau)=\max\{1-\tau/\tau_\mathrm{emp},0\}$.

\medskip
\noindent If $x>M_{\mathrm{tar}}$, the target is reached before emptying and
the required session duration, see also Sec.~\ref{sec:model}, is
\[
    \tau(x)=\tau_\mathrm{emp}\frac{M_{\mathrm{tar}}}{x}<\tau_\mathrm{emp}.
\]
If $x\leq M_{\mathrm{tar}}$, the target cannot be reached and we assume that
the session continues until complete emptying, so that $\tau=\tau_\mathrm{emp}$.

\medskip
The overall resulting map is therefore
\begin{equation}
\label{eq:durstim_target_map}
F(m,p)=
\begin{cases}
\displaystyle
\left(
x-M_{\mathrm{tar}},\;
gp+a\!\left(\tau_\mathrm{emp}\frac{M_{\mathrm{tar}}}{x}\right)
\right),
& x>M_{\mathrm{tar}},\\[3mm]
\displaystyle
\left(
0,\;
gp+a(\tau_\mathrm{emp})
\right),
& x\leq M_{\mathrm{tar}}.
\end{cases}
\end{equation}
Unlike the corresponding duration-dependent stimulation under a set-duration protocol, the map is non-linear since the duration needed to reach the prescribed target depends on the current state through $x$. As in Sec.~\ref{sec:amount_stim_target}, we can define the emptying region $E=\{z\in X:\;x(z)\le M_\mathrm{tar}\}$ and the target-met region as $E^c=X\setminus E$.

\paragraph{Fixed points in the emptying region $E$.} On $E$ the map $F$ becomes
\begin{equation*}
  F(m,p)=\big(0,\;p\,g+a(\tau_\mathrm{emp})).
\end{equation*}
The \emph{emptying fixed point} $z_\mathrm{emp}=(0,p_{\mathrm{emp}})$ with associated removed amount $M_{\mathrm{emp}}$, is uniquely given by
\begin{equation}
\label{eq:durstim_target_empty}
    m_{\mathrm{emp}}=0,
    \qquad
    p_{\mathrm{emp}}
    =
    \frac{a(\tau_\mathrm{emp})}{1-g},\qquad M_{\mathrm{emp}}(T)=\frac{a(\tau_\mathrm{emp})}{\Gamma(T)}.
    %\qquad
    %M_{\mathrm{emp}}(T)<M_{\mathrm{tar}}.
\end{equation}
It exists in the emptying region precisely when $x=p_{\mathrm{emp}}G=M_{\mathrm{emp}}(T)\leq M_\mathrm{tar}$. Note that since $a(\tau_\mathrm{emp})>0$, the involution state $(m,p)=(0,0)$ is not a steady state.

\paragraph{Fixed points in the target-met region $E^c$.} On $E^c$ the map $F$ becomes
\begin{equation*}
  F(m,p)=\big(x-M_\mathrm{tar},\;p\,g+a\!\left(\tau_\mathrm{emp}\frac{M_{\mathrm{tar}}}{x}\right)\big).
\end{equation*}
It is useful to characterize a target-met steady state by its associated
session duration $\tau\in(0,\tau_\mathrm{emp})$. Using $M_\mathrm{tar}=x(1-h)$, we can rewrite the fixed point equations for $m$ and $p$,
$$
m=x-M_\mathrm{tar},\qquad p=gp+a\left(\tau_\mathrm{emp}\frac{M_{\mathrm{tar}}}{ x}\right)
$$
in terms of $\tau$ and $p$, giving
\begin{equation}
\label{eq:durstim_target_fp_system}
    (1-g)p=a(\tau),
    \qquad
    M_{\mathrm{tar}}=G\,r(T,\tau)\,p,
\end{equation}
where $r(T,\tau)=\frac{1-h}{1-dh}$ as in \eqref{eqn:r}. Eliminating $p$, we can write a single equation for $\tau$
\begin{equation}
\label{eq:durstim_target_fp_tau}
    M_{\mathrm{tar}}
    =
    \frac{r(T,\tau)a(\tau)}{\Gamma(T)}.
\end{equation}
Note that this equation is related to \eqref{eqn:amtStim_setDur_fp}. The resulting solution $\tau$ depends on the scheduling parameters $(T,M_\mathrm{tar})$. Since both $a(\tau)$ and $r(T,\tau)$ are strictly increasing in $\tau$, the
right-hand side of \eqref{eq:durstim_target_fp_tau} is strictly increasing
from $0$ to $M_{\mathrm{emp}}(T)=\frac{a(\tau_\mathrm{emp})}{\Gamma(T)}$ as $\tau$ increases from $0$ to $\tau_\mathrm{emp}$, see also Fig.~\ref{fig:protocol_comparison}(d). Hence for 
\begin{equation}
    \label{eqn:durStim_setTarget_ex}
M_\mathrm{tar}< M_{\mathrm{emp}}(T)   
\end{equation}
a unique target-met steady state exist and, once $\tau$ is determined, is given by
\begin{equation}
\label{eqn:durstim_target_fp}
    p_\mathrm{tar}
    =
    \frac{a(\tau)}{1-g(T)},
    \qquad
    x_\mathrm{tar}
    =
    \frac{a(\tau)}
         {\Gamma(T)(1-d(T)h(\tau))},
    \qquad
    m_\mathrm{tar}
    =
    h(\tau)\,x_\mathrm{tar}.
\end{equation}

Notice that \eqref{eq:durstim_target_fp_tau} is precisely the steady
removed amount obtained under set-duration removal with duration $\tau$. We discuss how steady states under different protocols are connected in detail in Sec.~\ref{sec:protocol_correspondence}.\newline

To formulate the following theorem, which characterizes fixed point stability, we define the \emph{elasticity of the stimulation response}
$E_a$ and the corresponding \emph{critical elasticity}
$E_{\mathrm{crit}}$ by
\begin{equation}
\label{eq:durstim_elasticity_Ecrit}
    E_a(\tau)
    :=
    \frac{\tau a'(\tau)}{a(\tau)},
    \qquad
    E_{\mathrm{crit}}(T,\tau)
    :=\frac{1+d(T)}
         {1-d(T)h(\tau)}\frac{1+g(T)}
         {1-g(T)}.
\end{equation}

\begin{theorem}[Set-target removal with duration-dependent stimulation]
\label{thm:durstim_target}
Assume that $a(0)=0$ and that $a$ is continuously differentiable and
strictly increasing.

\begin{enumerate}
\item[(i)]
If
\[
    0<M_{\mathrm{tar}}<M_{\mathrm{emp}}(T)
    =
    \frac{a(\tau_\mathrm{emp})}{\Gamma(T)},
\]
there exists a unique target-met steady state. It is parameterized by
the unique $\tau\in(0,\tau_\mathrm{emp})$ satisfying
\eqref{eq:durstim_target_fp_tau} and given by \eqref{eqn:durstim_target_fp}.

\item[(ii)]
The target-met steady state is locally asymptotically stable if
\begin{equation}
\label{eq:durstim_target_stab}
    E_a(\tau)<E_{\mathrm{crit}}(T,\tau),    
\end{equation}
and unstable if $E_a(\tau)>E_{\mathrm{crit}}(T,\tau)$.
At $E_a(\tau)=E_{\mathrm{crit}}(T,\tau)$ one eigenvalue of the Jacobian is $-1$, while the second equals $-dg$. The fixed point is then nonhyperbolic, and its
stability cannot be determined from the
linearization alone.
Moreover, $\lambda=-1$ is the only possible local stability-loss
mechanism of the target-met steady state.

\item[(iii)]
If $M_{\mathrm{tar}}>M_{\mathrm{emp}}(T)$, the unique steady state is the emptied
state \eqref{eq:durstim_target_empty}, which is locally asymptotically
stable.
\end{enumerate}
\end{theorem}

The proof is given in App.~\ref{app:dur_stim_target}.

\paragraph{Stimulation efficiency and the possibility of instability.}

It is easy to check that \(E_{\mathrm{crit}}>1\) and, for fixed \(\tau\), decreases as a function of $T$ with limits $\infty$ for $T\rightarrow 0^+$, and \(1\) for $T\rightarrow \infty$. Therefore, if $E_a(\tau)>1$ for some value of $\tau$, we can fix $\tau$ and define $T_\mathrm{thresh}>0$ by $E_a(\tau)=E_\mathrm{crit}(T_\mathrm{thresh},\tau)$. Then for all $T>T_\mathrm{thresh}$ it holds that $E_a(\tau)>E_\mathrm{crit}(T,\tau)$, i.e. \eqref{eq:durstim_target_stab} is violated and hence the target-met state is unstable for such pairs $(T,\tau)$. One can then use \eqref{eq:durstim_target_fp_tau} to obtain the corresponding schedule variables $(T,M_\mathrm{tar})$.

\medskip
Next, we connect $E_a(\tau)$ with the shape of the stimulation efficiency: In analogy to the amount-dependent models, we can define the
\emph{stimulation efficiency} (which is now the production stimulation per removal duration)
\begin{equation}
\label{eq:Rtau}
    R_d(\tau)
    :=
    \frac{a(\tau)}{\tau}.
\end{equation}
Unlike in the amount-dependent models, $R_d$ does not directly
determine steady-state existence. However, since it connects to the elasticity of the stimulation response via
$$
E_a(\tau)=1+\frac{\tau R'_d(\tau)}{R_d(\tau)}
$$
it provides a simple criterion for whether unstable target-met steady states can occur. This allows to formulate the following corollary.

\begin{corollary}
\label{cor:durstim_target_instability}
An unstable target-met steady state exists for some schedule $(T,M_{\mathrm{tar}})$
if and only if
\[
    R_d'(\tau)>0
\]
for some $\tau\in(0,\tau_\mathrm{emp})$. In particular, if $R_d$ non-increasing on $(0,\tau_\mathrm{emp})$, every
target-met steady state is locally asymptotically stable.
\end{corollary}

This shows that only responses where the duration efficiency $R_d$ increasing on some interval, can create this type of instability.

\paragraph{Example: Michaelis-Menten response.}
For the Michaelis--Menten response of \eqref{eqn:mich_ment} taking durations as an argument
\[
    a(\tau)
    =
    a_{\mathrm{max}}\frac{\tau}{\tau+\mu_\tau},
\]
we have
\[
    R_d(\tau)
    =
    \frac{a_{\mathrm{max}}}{\tau+\mu_\tau},
\]
which is strictly decreasing. Hence every target-met steady state is
locally asymptotically stable.

\paragraph{Example: Hill response.}
For the Hill response of \eqref{eqn:hill}
\begin{equation}
\label{eq:durstim_hill}
    a(\tau)
    =
    a_{\mathrm{max}}
    \frac{\tau^2}{\tau^2+\mu_\tau^2},
\end{equation}
we have
\[
    R_d(\tau)
    =
    a_{\mathrm{max}}
    \frac{\tau}{\tau^2+\mu_\tau^2},
\]
which increases for $\tau<\mu_\tau$ and decreases for
$\tau>\mu_\tau$. 

Consequently, instability is possible only on the increasing-efficiency
branch $\tau<\mu_\tau$. Furthermore, since the increasing-efficiency branch always intersects the admissible interval \((0,\tau_\mathrm{emp})\), Corollary~\ref{cor:durstim_target_instability} implies that unstable target-met steady states exist for some schedules for every Hill response of this form.\newline

\begin{figure}[t]
    \centering
    \includegraphics[width=\textwidth]{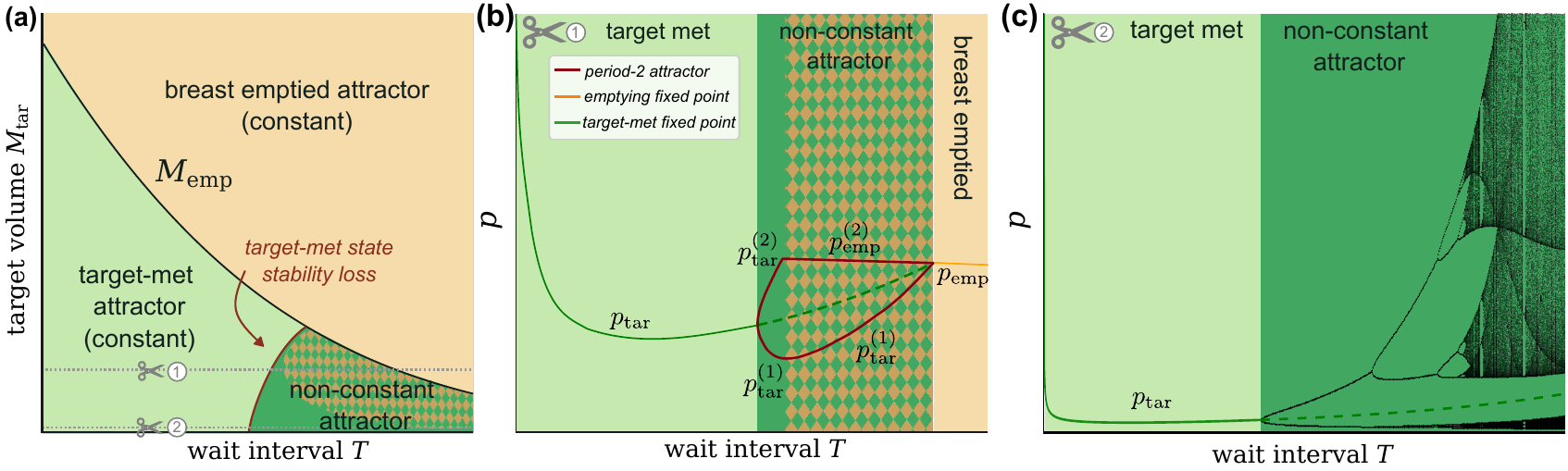}
    \caption{
        Duration-dependent stimulation under a set-target protocol with a unimodal-efficiency response, Hill response \eqref{eqn:hill}. (a) Long-term behaviour in schedule space $(T,M_\mathrm{tar})$. The curve $M_\mathrm{emp}(T)$ separates schedules for which a target-met fixed point exists from those for which it does not. For $M_\mathrm{tar}>M_\mathrm{emp}(T)$, the emptied fixed point is stable (orange). For $M_\mathrm{tar}<M_\mathrm{emp}(T)$, the dark red line separates the region where the target-met fixed point is stable (pale green) and the region where it is not. In the region, where the target-met fixed point exists, but is unstable, dark green denotes schedules for which a non-constant target-met attractor is observed numerically. The chequered region denotes schedules for which the observed non-constant attractor contains at least one emptying state. The horizontal dashed-grey lines indicate the slices shown in (b) and (c). (b),(c) Numerical bifurcation diagrams. Slices through schedule space at fixed $M_\mathrm{tar}$. Shown are the target-met fixed point, the emptied fixed point, and the numerically observed asymptotic values of $p$ as functions of $T$. Solid and dashed portions of the green curve indicate stable and unstable parts of the target-met fixed-point. In (b), the two members of the observed period-2 orbit are marked by superscripts $(1)$ and $(2)$. Parameters used (not calibrated to data): $\delta=\gamma=0.15$/h, $a_\mathrm{max}=50$ml/h, $\mu_\tau=0.5$h, $\tau_\mathrm{emp}=1/3$h with slice position $M_\mathrm{tar}=12.5$ml (b) and $M_\mathrm{tar}=0.01$ml (c). For (c), 720 regularly spaced $T$-values between 0 and 30 were used with starting values close to the unstable target-met fixed point. In each case 4000 iterations were performed and the last 2000 iterates were kept for plotting. The transition to non-constant dynamics
can be explored interactively in Online Resource 1.
    }
    \label{fig:durstim_tar_schedule}
\end{figure}

For the Hill response the stability boundary \(E_a=E_{\mathrm{crit}}\) can be evaluated explicitly; the resulting expression is given in App.~\ref{app:dur_stim_target} in \eqref{eqn:durstim_hill_flip}. Mapping this boundary through \eqref{eq:durstim_target_fp_tau} gives the stability boundary in \((T,M_{\mathrm{tar}})\)-space shown as dark red line in Fig.~\ref{fig:durstim_tar_schedule}(a). Since we have determined the existence of a region where no stable fixed point exists, we next discuss non-constant long-term behaviours.

\begin{figure}[t]
    \centering
    \includegraphics[width=\textwidth]{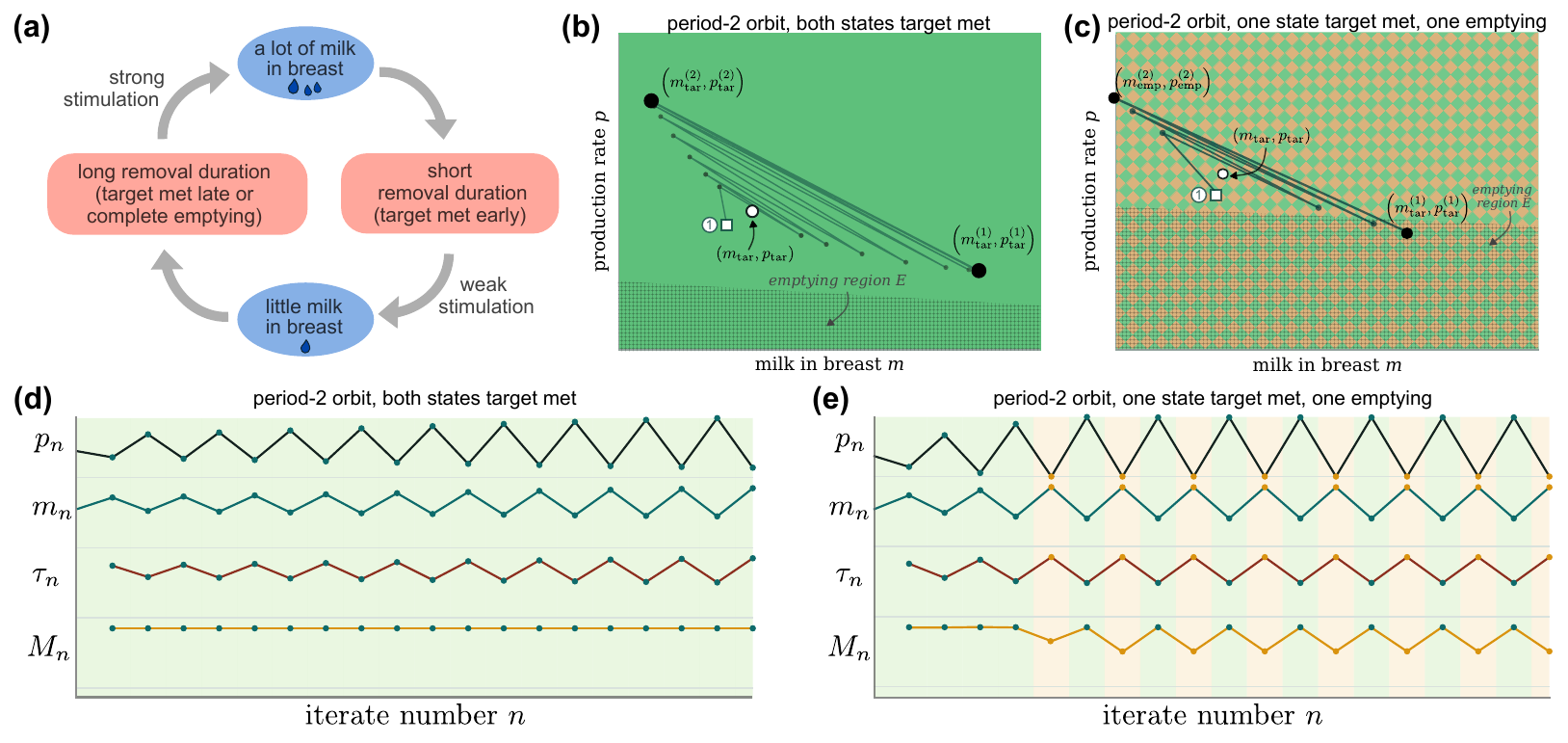}
    \caption{Duration-dependent stimulation under a set-target
protocol with a unimodal-efficiency response, illustrated with the Hill
response in \eqref{eqn:hill}. (a) Schematic explanation of the feedback mechanism that can generate stable period-2 dynamics.
    (b),(c) Phase planes. The grey lattice pattern indicates the emptying region $E$. Filled circles denote the numerically observed stable period-2 orbit, open circle denotes the unstable target-met fixed point. In (b), both members of the attracting period-2 orbit are target-met. In (c), one member is target-met and the other is an emptying state. Representative trajectories illustrate convergence towards the period-2 attractor. (d),(e) Qualitative behaviour of $m_n$, $p_n$, $\tau_n$ and milk removed $M_n$ as a function of iteration number $n$ of the orbits shown in (b) (panel (d)) and in (c) (panel (e)). Background shading and dot color indicate target-met or emptying behaviour.}
    \label{fig:durstim_tar_phase}
\end{figure}

\paragraph{Non-constant long-term behaviour.}

The occurrence of an eigenvalue \(\lambda=-1\) at the stability boundary suggests a possible flip (period-doubling) bifurcation. We do not pursue a formal bifurcation classification here, but use numerical simulations to explore the long-term dynamics for the Hill example above. The found behaviour is indeed consistent with a supercritical flip, with a stable period-$2$ orbit appearing beyond the fixed-point stability boundary: Fig.~\ref{fig:durstim_tar_schedule}(b) and Fig.~\ref{fig:durstim_tar_phase}(b)-(e) show that the period-2 attractor can either consist of two target-met states, Fig.~\ref{fig:durstim_tar_phase}(b),(d), or contain one target-met and one emptying state, Fig.~\ref{fig:durstim_tar_phase}(c),(e).

\medskip
Further numerical exploration revealed a striking result: For small targets $M_\mathrm{tar}$, the system appears to undergo a series of period-doubling events as $T$ increases. We also observed apparent chaotic behaviour, see Fig.~\ref{fig:durstim_tar_schedule}(c). In App.~\ref{app:dur_stim_target} we derive the system's small target limit. The resulting approximation is the Maynard Smith-Slatkin population model \cite{Smith1973}, which shows structural similarities with the logistic map, which is known to have the Feigenbaum-cascade as bifurcation diagram.

\begin{remark}[Timescales and non-constant dynamics]
The instability described above cannot occur at arbitrarily short wait
intervals: We assume $E_\mathrm{max}>1$ (otherwise the instability cannot occur). If the stimulation elasticity is bounded by
$E_a(\tau)\leq E_{\mathrm{max}}$ and since

\[
E_{\mathrm{crit}}(T,\tau)
=
\frac{(1+d)(1+g)}
     {(1-dh)(1-g)}
>
\frac{1+g}{1-g}.
\]
instability requires
\[
    T>
    \frac{1}{\gamma}
    \log\left(\frac{E_{\mathrm{max}}+1}{E_{\mathrm{max}}-1}\right).
\]
For the Hill response considered here, $E_{\mathrm{max}}=2$, giving the
necessary condition
\[
    T>\frac{\log 3}{\gamma}.
\]
Thus instability requires waits that are sufficiently
long relative to the decay timescale of the production rate. Whether this regime is physiologically relevant is a separate question.
The period-doubling cascade observed numerically for very small targets
should therefore primarily be interpreted as demonstrating a possible
dynamical consequence of the proposed feedback mechanism, rather than
as a prediction that such behaviour occurs in human lactation.
\end{remark}

\paragraph{Interpretation.}
The non-steady dynamics arise from a feedback that is absent under
set-duration removal. If more milk is available before a set-target
session, the prescribed target is reached more quickly. This shortens
the session and, under duration-dependent stimulation, reduces the
stimulation of subsequent production. Conversely, a lower available
amount produces a longer and more strongly stimulating session. If stimulation increases sufficiently strongly with session duration, this negative feedback can overshoot, causing the production and milk amounts to alternate between successive sessions, see Fig.~\ref{fig:durstim_tar_phase}(a).\newline

Thus the two duration-dependent protocols generate the same family of
positive steady states but can have qualitatively different dynamics.
Under set-duration removal the map is affine and its positive steady
state is always globally stable under the baseline exponential decay
assumptions. Under set-target removal the state-dependent session
duration makes the map non-linear, and increasing duration efficiency can
destabilize the steady state and generate persistent non-steady
behaviour. See also the discussion in Sec.~\ref{sec:protocol_correspondence}.

\subsection{Steady-state correspondence and differences between removal protocols}
\label{sec:protocol_correspondence}

In the preceding sections, we analysed the long-term behaviour under two removal protocols: setting the target session output and setting the removal duration. Although these protocols define different dynamical maps, their positive steady states can be compared through the realised session output \(M_*\) and removal duration \(\tau_*\). A target-met fixed point under set-target removal with \(M_{\mathrm{tar}}=M_*\) can be associated with the set-duration schedule that prescribes its realised duration \(\tau=\tau_*\). Conversely, a set-duration fixed point with \(\tau\le \tau_\mathrm{emp}\) corresponds to a target-met set-target fixed point with \(M_{\mathrm{tar}}=M_*\). This correspondence concerns the location of fixed points; their stability and resulting long-term dynamics need not be preserved between protocols.

\medskip
The correspondence follows from a general balance relation.

\begin{lemma}[Steady-state balance]
\label{lem:Mday_balance}
Consider any positive steady state of the model, and let \(M_*>0\), \(\tau_*\), and \(a_*\) denote the amount of milk removed, the removal duration, and the resulting stimulation per session, respectively. Then
\begin{equation}
  \label{eqn:steady_state_balance}
\frac{a_*}{M_*}
=
\frac{\Gamma(T)}{r(T,\tau_*)}.
\end{equation}
In particular, at complete emptying \(r(T,\tau_*)=1\), so that
\[
\frac{a_*}{M_*}=\Gamma(T).
\]
\end{lemma}

For a proof see App.~\ref{app:protocol_comparison}.

\medskip
The lemma allows the fixed points of the two removal protocols to be represented in the common
\((T,M_*)\)-plane. Next, we will discuss what this means for the two stimulation mechanisms.

\paragraph{Amount-dependent stimulation.}
For amount-dependent stimulation, where \(a_*=a(M_*)\), the balance relation \eqref{eqn:steady_state_balance}, solved for $\tau_*$, gives
\begin{equation}
\label{eqn:protocol_correspondence_amount}
\tau_*(T,M_*)
=
\tau_\mathrm{emp}
\frac{1-d(T)}
{\frac{R(M_*)}{\Gamma(T)}-d(T)},
\qquad
R(M_*)\geq\Gamma(T).
\end{equation}
Equality \(R(M_*)=\Gamma(T)\) corresponds to complete emptying,
\(\tau_*=\tau_\mathrm{emp}\).\newline

For a decreasing efficiency response, Eq.~\eqref{eqn:protocol_correspondence_amount} shows that the positive pre-emptying steady states of the two protocols are in one-to-one correspondence and have the same qualitative stability. Thus, although the protocols use different schedule variables, they describe the same family of attracting positive states.

\begin{figure}[t]
  \centering
  \includegraphics[width=\textwidth]{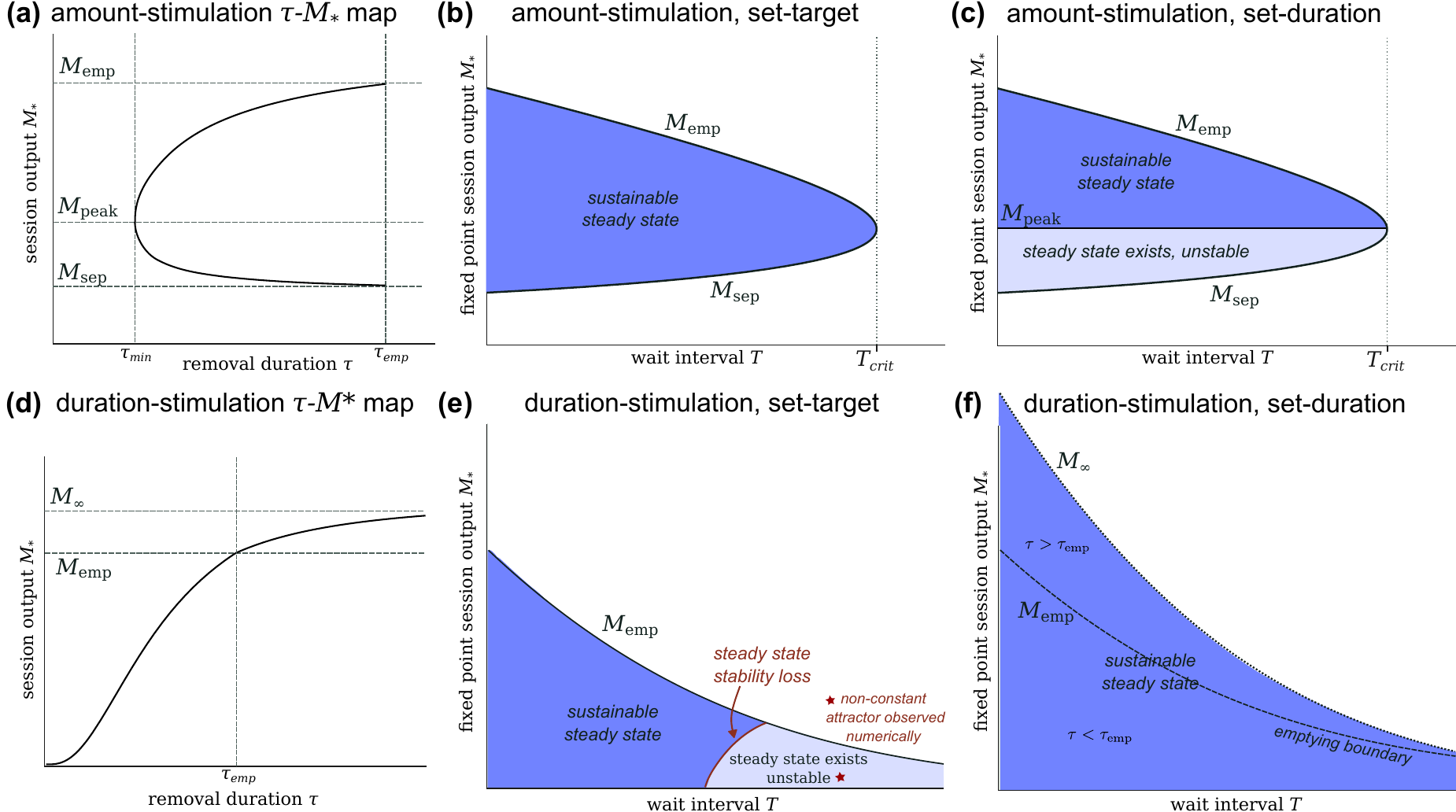}
  \caption{Effect of protocol for unimodal stimulation efficiency. (a),(d) Shown are the maps between $\tau$ and $M_*$ given by \eqref{eqn:protocol_correspondence_amount} and \eqref{eqn:protocol_correspondence_duration}. (b)-(c) and (e)-(f): The existence and stability of positive fixed points in $(T,M_*)$-space for different stimulation mechanisms and removal protocols. Dark blue denotes an existing, stable state, light blue an existing, unstable state. In (f) $M_\infty$ is defined as $M_\infty=a_{\mathrm{max}}/\Gamma(T)$.}
  \label{fig:protocol_comparison}
\end{figure}

\medskip
For amount-dependent stimulation with a unimodal efficiency response however, there are fundamental differences: Fig.~\ref{fig:protocol_comparison}(a) shows the relation between removal duration and fixed-point session output at fixed \(T\). We see that the steady-state correspondence is not one-to-one. For a range of removal durations, one fixed point lies below \(M_{\mathrm{peak}}\) and one above it. Fig.~\ref{fig:protocol_comparison}(b), (c) compare the existence and stability of the corresponding fixed points under set-target and set-duration removal: Under set-target removal, the target-met positive fixed points on both sides of \(M_{\mathrm{peak}}\) are stable (Fig.~\ref{fig:protocol_comparison}(b)). Under set-duration removal, however, the lower branch \(M_*<M_{\mathrm{peak}}\) is unstable, whereas the upper branch \(M_*>M_{\mathrm{peak}}\) is stable (Fig.~\ref{fig:protocol_comparison}(c)). The two protocols therefore permit the same fixed-point outputs over this range, but do not assign the same stability to them.

\paragraph{Duration-dependent stimulation.}

For duration-dependent stimulation, where \(a_*=a(\tau_*)\), we obtain from \eqref{eqn:steady_state_balance} for $M_*$
\begin{equation}
\label{eqn:protocol_correspondence_duration}
M_*(T,\tau_*)
=
\frac{r(T,\tau_*)a(\tau_*)}{\Gamma(T)}.
\end{equation}
At complete emptying,
\[
\tau_*=\tau_{\mathrm{emp}}
\qquad\Longleftrightarrow\qquad
M_*=M_{\mathrm{emp}}(T)
=\frac{a(\tau_{\mathrm{emp}})}{\Gamma(T)}.
\]

Here, a different distinction occurs: Fig.~\ref{fig:protocol_comparison}(d) shows how the right-hand side of \eqref{eqn:protocol_correspondence_duration} increases monotonically with $\tau_*$. However, set-duration removal permits prescribed durations \(\tau>\tau_\mathrm{emp}\). The breast is then already emptied, but increasing \(\tau\) can continue to increase stimulation, so that the fixed-point output extends beyond \(M_{\mathrm{emp}}(T)\) and approaches \(M_\infty(T):=a_{\mathrm{max}}/\Gamma(T)\). Set-target removal has no corresponding post-emptying duration regime because its realised duration cannot exceed \(\tau_{\mathrm{emp}}\). This is a consequence of the modelling assumption that for under a set-target protocol the removal is stopped, once the breast is emptied. For decreasing efficiency these states (whenever they exist) remain stable under both protocols. For unimodal efficiency, however, the set-target fixed point can lose stability (Fig.~\ref{fig:protocol_comparison}(e)); numerical simulations reveal non-constant long-term dynamics in this regime. The corresponding set-duration fixed point remains stable (Fig.~\ref{fig:protocol_comparison}(f)).

\paragraph{Interpretation.} It is an interesting and perhaps surprising observation that given a steady state, the choice of removal protocol affects the state's stability, and hence whether it would be observable as long-term behaviour. To understand this, one has to view the protocols through the lens of perturbations. We will use the amount-dependent stimulation as an example: For the set-target protocol, as long as the dynamics stay in the target-met region, the amount pumped and therefore the stimulation stay constant at $M_\mathrm{tar}$ and $a(M_\mathrm{tar})$. Therefore there is no feedback from the amount of milk in the breast $m$ to the production rate $p$ (hence a zero in the lower-left corner of the Jacobian in App.~\ref{app:amount_stim_target}). Both $p$ and $m$ have a stabilizing effect on themselves (hence the entries $d<1$ and $g<1$ on the diagonal of the Jacobian in App.~\ref{app:amount_stim_target}), leading to overall stability. For the set-duration protocol the situation is different: Changes in $m$ and $p$ \emph{do} affect the amount pumped during a set duration $\tau$ and feed back onto the production rate (hence the terms involving $a'(M)$ in the lower row of the Jacobian in App.~\ref{app:amount_stim_duration}). Then, if $R'(M)>0$, this causes the instability.

\section{Daily Output}
\label{sec:daily_output}

Having characterised the possible long-term states, we next ask how these translate into milk obtained over time. For this we move away from per session behaviour and rather ask the question of how the schedule $(T,M_\mathrm{tar})$ or $(T,\tau)$ influences the \emph{daily milk output}. Guiding questions are: Which schedules lead to the same daily milk output? How can daily milk output be maximized under certain constraints?

\medskip
If $M_n$ and $\tau_n$ are the session output and session duration of the $n$-th removal session, the long-term average daily milk output is defined as
$$
M_\mathrm{DAY}=24\lim_{N\rightarrow \infty}\frac{\sum_{n=1}^N M_n}{\sum_{n=1}^N (T+\tau_n)},
$$
whenever this limit exists.\newline

If the system converges to a steady state with session duration $\tau_*$ and session output $M_*$, the formula simplifies to 
\begin{align}
    \label{eqn:Mday_steady}
    M_\mathrm{DAY}=\frac{24}{T+\tau_*}M_*.
\end{align}
For cases of bistability, we assume the convergence was to the non-trivial state.\newline
Similarly, for a general period-$k$ orbit with session durations $(\tau_*^{(1)},\dots \tau_*^{(k)})$ and session outputs $(M_*^{(1)},\dots M_*^{(k)})$ we obtain
\begin{align}
    \label{eqn:Mday_periodic}
M_\mathrm{DAY}=\frac{24}{T+ \frac{1}{k}\sum_{j=1}^k \tau_*^{(j)}}\frac{1}{k}\sum_{j=1}^k M^{(j)}_*.
\end{align}
\paragraph{Interpretation.} For both \eqref{eqn:Mday_steady} and \eqref{eqn:Mday_periodic} the first factor represents the average number of sessions per day and the second factor the average session output. In general both factors can depend on the scheduling parameters. This can create a trade-off between the amount removed per session and the number of daily sessions. For example, increasing the target amount or removal duration may increase the milk obtained in an individual session, but it may also increase the duration of each removal cycle and therefore reduces the number of sessions possible per day. This trade-off is the focus of the analysis below.

%----------

\paragraph{On the treatment of the two removal protocols.} As discussed in Sec.~\ref{sec:protocol_correspondence}, before complete emptying the two removal protocols, set-target and set-duration, describe the same positive steady states, although their stability can differ. To make the following analysis more readable, we formulate the main results for the set-duration protocol and then state explicitly how they translate to set-target removal. The main exception is Sec.~\ref{ssec:daily_non_steady}, where non-constant long-term dynamics under set-target removal are considered directly.

\subsection{Universal properties of steady-state daily output}

Unless stated otherwise, we assume a set-duration protocol and consider daily output at a stable steady state. In models with bistability, we assume convergence to the positive lactation state. The daily output is given by
\begin{equation}
    \label{eqn:dailyOutput_setDur}
M_\mathrm{DAY}(T,\tau)=\frac{24 M_*(T,\tau)}{T+\tau},    
\end{equation}
where the session output $M_*(T,\tau)$ is given by \eqref{eqn:amtStim_setDur_fp} for amount-dependent stimulation and by \eqref{eq:taustimfp} for duration-dependent stimulation.\newline

We start by inspecting how the \emph{session} output $M_*(T,\tau)$ depends on the schedule parameters.

\begin{lemma}[Monotonicity of steady-state session output]
    \label{lem:sessionOutput_monotonicity}
Consider a stable steady-state under a set-duration protocol. Let \(M_*(T,\tau)>0\) be the associated session output. Then the following hold.
\begin{enumerate}
    \item[(i)] The session output decreases with $T$,
    \[
    \frac{\partial M_*}{\partial T}< 0.
    \]
    \item[(ii)] The session output increases with $\tau$,
    \[
    \frac{\partial M_*}{\partial \tau}\geq 0.
    \]
    More precisely, \(\partial_\tau M_*>0\) before complete emptying. Beyond complete emptying, \(M_*\) is
    constant for amount-dependent stimulation and continues to increase
    for duration-dependent stimulation.
\end{enumerate}
\end{lemma}

For a proof see App.~\ref{app:daily_output}.

\medskip
A simple consequence of this Lemma is that
\[
\frac{\partial M_{\mathrm{DAY}}}{\partial T}<0.
\]

\paragraph{Interpretation.}
The two monotonicity results in Lem.~\ref{lem:sessionOutput_monotonicity} show the different effects of the schedule parameters $T$ and $\tau$. Increasing the wait interval \(T\) lengthens the removal cycle while decreasing the session output,
so it always has a negative effect on daily output. In contrast, increasing \(\tau\) before complete emptying increases the amount removed per session, but also lengthens the removal cycle. The resulting effects oppose each other, leading to a trade-off. Next we inspect this trade-off in more detail, i.e. we investigate how the function $\tau\mapsto M_\mathrm{DAY}(T,\tau)$ behaves. A particularly interesting question is whether a maximum is reached before, at, or after emptying the breast.

\medskip
By differentiating $M_\mathrm{DAY}(T,\tau)$ with respect to $\tau$, it follows directly that
\begin{equation}
    \label{eqn:daily_Q_crit}
Q>1 \iff \text{$M_\mathrm{DAY}$ increases with $\tau$}    
\end{equation}
where we have defined,
\[
Q(T,\tau)
=
(T+\tau)
\frac{\partial_\tau M_*(T,\tau)}
     {M_*(T,\tau)}.
\]
\noindent $Q$ can be interpreted as the ratio
\[
Q :=
\frac{\text{relative increase in session output}}
     {\text{relative increase in cycle duration}}.
\]
The guiding principle for the next sections is therefore
\[
Q>1:\ M_{\mathrm{DAY}}\text{ increases},\qquad
Q=1:\ \text{critical point},\qquad
Q<1:\ M_{\mathrm{DAY}}\text{ decreases}.
\]

\begin{figure}[t]
  \centering
  \includegraphics[width=\textwidth]{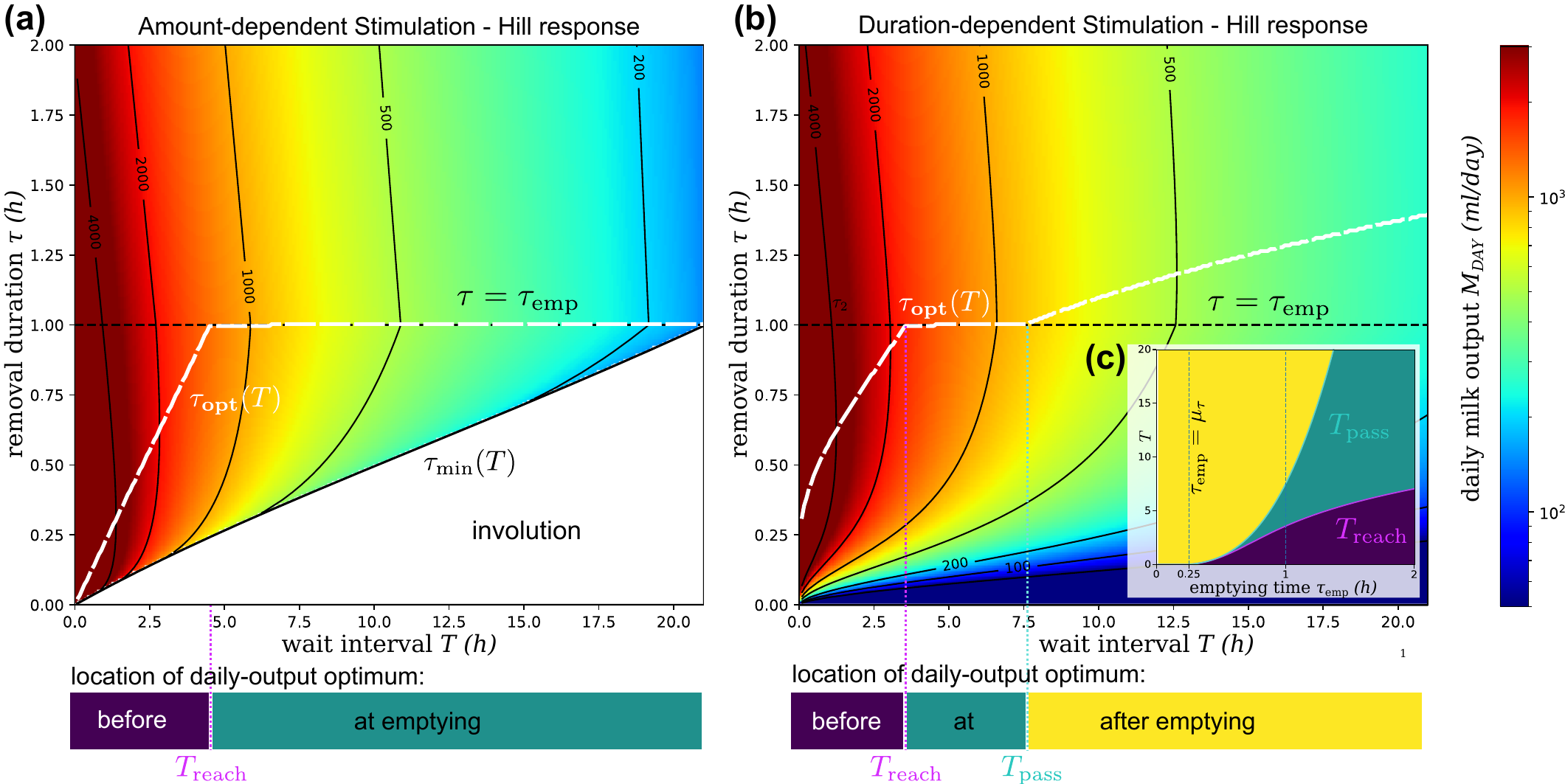}
  \caption{Daily milk output under set-duration removal. (a),(b): Steady-state $M_{\mathrm{DAY}}$ as a function of wait interval ($T$) and removal duration ($\tau$) for the Hill response with amount-dependent (a) and duration-dependent stimulation (b). Colours and contours show $M_{\mathrm{DAY}}$; the white dashed curve marks the maximizing duration $\tau_{\mathrm{opt}}(T)$, and the black dashed line complete emptying, $\tau=\tau_{\mathrm{emp}}$. In (a), the solid black curve marks the lower boundary $\tau_{\mathrm{min}}(T)$ of positive steady states. Bars below (a),(b) indicate whether the maximum occurs before, at, or after complete emptying, with transitions at $T_{\mathrm{reach}}$ and $T_{\mathrm{pass}}$. (c) Regime diagram for duration-dependent stimulation in the $(T,\tau_{\mathrm{emp}})$-plane; the curves $T_{\mathrm{reach}}$ and $T_{\mathrm{pass}}$ separate the three regimes, the vertical line marks $\tau_{\mathrm{emp}}=\mu_\tau$. Parameters: $\delta=0.03$/h, $\gamma=0.1$/h, $a_\mathrm{max}=37.5$ml/h, $\mu_M=125$ml, $\mu_\tau=0.25$h, $\tau_\mathrm{emp}=1$h (in (a), (b))}
  \label{fig:dailyOutput_dur}
\end{figure}

\paragraph{Level sets and constrained schedules.}
The level sets of \(M_{\mathrm{DAY}}(T,\tau)\) in Fig.~\ref{fig:dailyOutput_dur}, show which schedules $(T,\tau)$ produce the same long-term daily output. The figure also shows \(\tau_{\mathrm{opt}}(T)\), the removal duration maximizing \(M_{\mathrm{DAY}}(T,\tau)\) for fixed \(T\). Characterizing  $\tau_\mathrm{opt}(T)$ is the main focus of the next sections. What can we learn from the level sets and why is $\tau_\mathrm{opt}(T)$ such as a useful quantity?

\medskip
The level sets and \(\tau_{\mathrm{opt}}(T)\) can be used to answer several natural scheduling questions:
\begin{itemize}
    \item For a given daily output target, which schedule $(T,\tau)$ allows for the maximum wait interval $T$ between sessions? \emph{Answer: Intersect $T\mapsto\tau_\mathrm{opt}(T)$ with the corresponding level set. The intersection point gives the wanted schedule.}   
    \item For a given wait interval $T$, what is the maximal achievable daily output?\newline
    \emph{Answer: $M_\mathrm{DAY}(T,\tau_\mathrm{opt}(T))$.} 
    \item For a given wait interval $T$ and a given daily output target, what is the minimal removal duration possible? \emph{Answer: Intersect the level set with the vertical line $T=\text{const}$. If there are several possible removal durations, pick the smallest one. If the daily output target is bigger than  $M_\mathrm{DAY}(T,\tau_\mathrm{opt}(T))$, there is no such schedule.} 
\end{itemize}
These are just some examples of how the level sets and the function $\tau_\mathrm{opt}(T)$ can be used to answer practical questions about scheduling. In particular, they show that the optimal removal duration $\tau_\mathrm{opt}(T)$ is a useful quantity for comparing schedules. Next, we characterize $\tau_\mathrm{opt}(T)$, starting with amount-dependent stimulation.

\subsection{Amount-dependent stimulation}

In this section we consider amount-dependent stimulation $a=a(M)$. To determine the location of the maximum of $\tau\mapsto M_\mathrm{DAY}(T,\tau)$ for fixed $T$, it is useful to characterize its growth or decay behaviour near $\tau=\tau_\mathrm{min}$, and near $\tau=\tau_\mathrm{emp}$. The following Lemma shows what happens near $\tau_\mathrm{min}$.

\begin{lemma}[Daily output near $\tau_\mathrm{min}$]
\label{lem:dailyOutput_amtStim_tauMin}
Fix $0<T<T_\mathrm{crit}$. Along a stable positive steady state under set-duration removal, \(\tau \mapsto M_\mathrm{DAY}(T,\tau)\) is strictly increasing for \(\tau>\tau_{\mathrm{min}}(T)\) sufficiently close to \(\tau_{\mathrm{min}}(T)\).
\end{lemma}
For a proof see App.~\ref{app:daily_output}.\newline

Thus the daily output $M_\mathrm{DAY}(T,\tau)$ initially increases as \(\tau\) moves away from \(\tau_{\mathrm{min}}\). To determine whether it reaches a maximum before complete emptying, we inspect its growth or decay as \(\tau\) approaches \(\tau_{\mathrm{emp}}\). This is characterized by the value of $Q$ just before emptying. We obtain
\begin{equation}
    \label{eqn:QAminus}
    Q_A^\mathrm{pre}(T)
    :=\lim_{\tau\rightarrow\tau_\mathrm{emp}^-}Q(T,\tau)=
    -\frac{(1+T/\tau_\mathrm{emp})(1-d(T))\Gamma(T)}{M_{\mathrm{emp}}(T)R'(M_{\mathrm{emp}}(T))}.
\end{equation}
Now \eqref{eqn:daily_Q_crit} tells us that
$$
Q_A^\mathrm{pre}(T)>1 \iff \partial_\tau M_\mathrm{DAY}(T,\tau)>0\quad \text{for $\tau<\tau_\mathrm{emp}$ sufficiently close to $\tau_\mathrm{emp}$}.
$$
We are now ready to state the main result of this section, which answers the question of when a pre-emptying maximum of $\tau\mapsto M_\mathrm{DAY}(T,\tau)$ exists and how it is characterized. Note that the proposition contains a concavity assumption on $\tau\mapsto M_*(T,\tau)$, which ensures that the sign of \(\partial_\tau M_\mathrm{DAY}\) can change at most once. Fig.~\ref{fig:dailyOutput_dur}(a) demonstrates the main statements of Prop.~\ref{prop:dailyOutput_amtStim} and Cor.~\ref{cor:dailyOutput_amtStim_T}.

\begin{proposition}[Maximal daily output under amount-dependent stimulation]
\label{prop:dailyOutput_amtStim}
Assume that \(R\) is a decreasing, or unimodal-efficiency response, and that \(R_{\mathrm{max}}>\gamma\). Fix \(0<T<T_{\mathrm{crit}}\) and consider positive stable steady states under a set-duration protocol. Further assume that before emptying 
\[
    \partial_\tau^2 M_*(T,\tau)<0.
\]
Then \(\tau\mapsto M_{\mathrm{DAY}}(T,\tau)\) has a unique maximum $\tau_\mathrm{opt}$. Its position relative to
complete emptying is determined as follows:
\[
\begin{array}{lll}
Q_A^\mathrm{pre}(T)<1
&\Longrightarrow&
\text{the maximum occurs before complete emptying, $\tau_\mathrm{opt}<\tau_\mathrm{emp}$},
\\[2mm]
Q_A^\mathrm{pre}(T)\geq 1
&\Longrightarrow&
\text{the maximum occurs at complete emptying, $\tau_\mathrm{opt}=\tau_\mathrm{emp}$.}
\end{array}
\]
Whenever the maximum occurs before complete emptying, \(\tau_{\mathrm{opt}}\) is given as solution to
\[
Q(T,\tau_\mathrm{opt})=1.
\]
\end{proposition}

The proof is given in App.~\ref{app:daily_output}.

%------------------

\begin{corollary}[Dependence of the maximum on the wait interval]
\label{cor:dailyOutput_amtStim_T}
Under the assumptions of Prop.~\ref{prop:dailyOutput_amtStim},
\[
    Q_A^\mathrm{pre}(T)<1
\]
for all sufficiently small \(T>0\), whereas
\[
    Q_A^\mathrm{pre}(T)>1
\]
for \(T<T_{\mathrm{crit}}\) sufficiently close to \(T_{\mathrm{crit}}\).
Consequently, the daily-output maximum lies before complete emptying
for sufficiently short wait intervals and is attained at complete
emptying for wait intervals sufficiently close to \(T_{\mathrm{crit}}\). In particular, there exists at least one
\(T_{\mathrm{reach}}\in(0,T_{\mathrm{crit}})\) satisfying
\[
    Q_A^\mathrm{pre}(T_{\mathrm{reach}})=1,
\]
at which the maximum reaches the complete-emptying boundary.
\end{corollary}

\paragraph{Corresponding result for set-target removal.}
Before complete emptying, the set-target and set-duration protocols describe the same steady states, although their stability can differ; see Sec.~\ref{sec:protocol_correspondence} and Fig.~\ref{fig:protocol_comparison}(a)-(c).
\medskip

Consider first stable set-target steady states whose corresponding set-duration steady states are also stable. These are precisely the states satisfying $R'(M_*)<0$. For such states, the realized removal duration
\(\tau_*(T,M_{\mathrm{tar}})\) increases with the session target; see Fig.~\ref{fig:protocol_comparison}(a). Hence \(M_{\mathrm{tar}}=M_*(T,\tau)\) provides an order-preserving reparametrization of the set-duration steady states, and the results above for \(M_{\mathrm{DAY}}\) transfer directly to the set-target protocol.

\medskip
Stable set-target steady states whose corresponding set-duration states are unstable, i.e. states satisfying $R'(M_*)>0$ do not generate an additional daily-output maximum. The reason is that for these states, the realized removal duration \(\tau_*(T,M_{\mathrm{tar}})\) \emph{decreases} as \(M_{\mathrm{tar}}\) increases; see Fig.~\ref{fig:protocol_comparison}(a). This also follows directly by differentiating the relation \eqref{eqn:protocol_correspondence_amount}. Thus increasing \(M_{\mathrm{tar}}\) simultaneously increases the amount removed per session and shortens the removal session, so \(M_{\mathrm{DAY}}\) increases with \(M_{\mathrm{tar}}\).

\medskip
Consequently, whenever the set-duration maximum occurs before complete emptying, the corresponding set-target maximum is attained at
\[
M_{{\mathrm{tar},opt}}
=
M_*(T,\tau_{\mathrm{opt}}).
\]
If the set-duration maximum occurs at complete emptying, then under set-target removal the daily output first reaches its maximum at
\[
M_{\mathrm{tar}}=M_{\mathrm{emp}}(T)
=
M_*(T,\tau_{\mathrm{emp}})
\]
and remains constant for all \(M_{\mathrm{tar}}\geq M_{\mathrm{emp}}(T)\).

\paragraph{Examples.}
For both the Michaelis--Menten and Hill responses, the function
\(\tau\mapsto M_*(T,\tau)\) is strictly concave. This can be seen by differentiating the steady state session output $M_\mathrm{lac}$, given in \eqref{eq:durrootconc} and \eqref{eqn:durrootsig}, with respect to $\tau$, and using that before emptying $\partial_\tau r>0$ and $\partial^2_\tau r<0$. Hence Prop.~\ref{prop:dailyOutput_amtStim}
applies.\newline

\noindent\emph{Michaelis--Menten response:} We have $R(M)=a_{\mathrm{max}}/(M+\mu_M)$ and $R_{\mathrm{max}}=a_{\mathrm{max}}/\mu_M$. A direct calculation using \(R(M_{\mathrm{emp}})=\Gamma(T)\) gives
\[
M_{\mathrm{emp}}(T)R'(M_{\mathrm{emp}}(T))
=
-\Gamma(T)
\left(
1-\frac{\Gamma(T)}{R_{\mathrm{max}}}
\right).
\]
\noindent\emph{Hill response:} We have $R(M)=a_\mathrm{max} M/(M^2+\mu_M^2)$ and $R_\mathrm{max}=a_\mathrm{max}/(2\mu_M)$. We obtain
\[
M_{\mathrm{emp}}(T)R'(M_{\mathrm{emp}}(T))
=
-\Gamma(T)
\sqrt{
1-
\left(\frac{\Gamma(T)}{R_{\mathrm{max}}}\right)^2
}.
\]
Substituting these results into \eqref{eqn:QAminus} and using Cor.~\ref{cor:dailyOutput_amtStim_T} shows that \(T_{\mathrm{reach}}\) is characterized by
\begin{equation}
\label{eqn:Mday_MMHill}
(1+T_{\mathrm{reach}}/\tau_{\mathrm{emp}})
(1-d(T_{\mathrm{reach}}))
=
\begin{cases}
1-\dfrac{\Gamma(T_{\mathrm{reach}})}{R_{\mathrm{max}}},
& \text{Michaelis--Menten},\\[3mm]
\sqrt{
1-\left(
\dfrac{\Gamma(T_{\mathrm{reach}})}{R_{\mathrm{max}}}
\right)^2
},
& \text{Hill}.
\end{cases}
\end{equation}
Examining the monotonicity of the right and the left hand sides as functions of $T$, we see that there always exists a unique 
\(T_{\mathrm{reach}}\in(0,T_{\mathrm{crit}})\). Thus, for \(0<T<T_{\mathrm{reach}}\) the daily-output optimum occurs before complete emptying, whereas for
\(T_{\mathrm{reach}}<T<T_{\mathrm{crit}}\) it is attained at complete emptying,
compare Fig.~\ref{fig:dailyOutput_dur}(a).

\subsection{Duration-dependent stimulation}

Suppose now that stimulation depends on the removal duration,
\(a=a(\tau)\). Under the set-duration protocol, the steady-state balance
lemma, Lem.~\ref{lem:Mday_balance}, gives
\begin{equation}
\label{eqn:balance_durStim}
M_*(T,\tau)
=
\frac{r(T,\tau)a(\tau)}{\Gamma(T)}.
\end{equation}

We know from Lem.~\ref{lem:sessionOutput_monotonicity} that session output $M_*(T,\tau)$ increases with removal duration $\tau$. At the same time, increasing the removal duration also decreases the number of sessions per day.
\medskip
%since both $\tau\mapsto r(T,\tau)$ and $a(\tau)$ are increasing \(M_*(T,\tau)\) is strictly increasing with \(\tau\), i.e. session output increases with removal duration.

As before, the quantity \(Q(T,\tau)\) defined above determines whether
\(M_{\mathrm{DAY}}(T,\tau)\) is increasing or decreasing, see \eqref{eqn:daily_Q_crit}. We characterize the behaviour of \(Q\) with respect to \(\tau\) in the following Lemma. Note that the condition on the stimulation elasticity \(E_a(\tau)\) plays a role analogous to the concavity assumption in Prop.~\ref{prop:dailyOutput_amtStim}; it ensures that the sign of \(\partial_\tau M_\mathrm{DAY}\) can change at most once.

\begin{lemma}[Behaviour of \(Q\) under duration-dependent stimulation]
\label{lem:Q_durStim}
Fix \(T>0\) and let \(M_*(T,\tau)>0\) be the session output as a stable positive steady state. Define the stimulation elasticity as
$$
E_a(\tau)
:=
\frac{\tau a'(\tau)}{a(\tau)},
$$
and suppose that \(E_a(\tau)>0\) is strictly decreasing, with $\lim_{\tau\to\infty}E_a(\tau)=0$.\newline
Then \(Q(T,\tau)\) is strictly decreasing with \(\tau\) both before
and after complete emptying. Moreover,
\[
\lim_{\tau\to0^+}Q(T,\tau)=\infty,
\qquad
\lim_{\tau\to\infty}Q(T,\tau)=0.
\]
At complete emptying, the one-sided limits are
\begin{align}
Q_D^\mathrm{pre}(T)
&:=
\lim_{\tau\to\tau_{\mathrm{emp}}^-}Q(T,\tau)
=
(1+T/\tau_\mathrm{emp})
\left[
E_a(\tau_{\mathrm{emp}})+1-d(T)
\right],
\label{eqn:QDminus}
\\
Q_D^\mathrm{post}(T)
&:=
\lim_{\tau\to\tau_{\mathrm{emp}}^+}Q(T,\tau)
=
(1+T/\tau_\mathrm{emp})E_a(\tau_{\mathrm{emp}}).
\label{eqn:QDplus}
\end{align}
In particular,
\[
Q_D^\mathrm{post}(T)<Q_D^\mathrm{pre}(T).
\]
\end{lemma}

For a proof see App.~\ref{app:daily_output}.

\paragraph{Interpretation.}
The two quantities \(Q_D^\mathrm{pre}\) and \(Q_D^\mathrm{post}\) describe the growth or decay of daily output immediately before and after complete emptying. Their difference
arises because increasing the removal duration before emptying increases both
the amount removed and the stimulation, whereas after emptying it can only
increase stimulation.\newline

The following proposition now characterizes the maximum of $\tau\mapsto M_\mathrm{DAY}(T,\tau)$. See also Fig.~\ref{fig:dailyOutput_dur}(b) for a demonstration of its main results.

\begin{proposition}[Maximal daily output under duration-dependent stimulation]
\label{prop:dailyOutput_durStim}
Fix \(T>0\) and suppose the assumptions of Lem.~\ref{lem:Q_durStim} hold. Then \(\tau\mapsto M_{\mathrm{DAY}}(T,\tau)\) has a unique maximum. Its position relative to
complete emptying is determined as follows:
\[
\begin{array}{lll}
Q_D^\mathrm{pre}(T)<1
&\Longrightarrow&
\text{the maximum occurs before complete emptying, $\tau_\mathrm{opt}<\tau_\mathrm{emp}$},
\\[2mm]
Q_D^\mathrm{post}(T)\leq1\leq Q_D^\mathrm{pre}(T)
&\Longrightarrow&
\text{the maximum occurs at complete emptying, $\tau_\mathrm{opt}=\tau_\mathrm{emp}$},
\\[2mm]
Q_D^\mathrm{post}(T)>1
&\Longrightarrow&
\text{the maximum occurs after complete emptying, $\tau_\mathrm{opt}>\tau_\mathrm{emp}$}.
\end{array}
\]
Whenever the maximum occurs away from the emptying boundary, it is attained
at the unique \(\tau_{\mathrm{opt}}\) satisfying
\[
Q(T,\tau_{\mathrm{opt}})=1.
\]
\end{proposition}

The proof is given in App.~\ref{app:daily_output}.

\begin{corollary}[Dependence of the maximum on the wait interval]
\label{cor:dailyOutput_durStim_T}
Under the assumptions of Prop.~\ref{prop:dailyOutput_durStim}, both
\(Q_D^\mathrm{pre}(T)\) and \(Q_D^\mathrm{post}(T)\) are strictly increasing with \(T\), and $Q_D^\mathrm{post}(T)<Q_D^\mathrm{pre}(T)$ for all $T>0$.
Moreover,
\[
\lim_{T\to0^+}Q_D^\mathrm{pre}(T)
=
\lim_{T\to0^+}Q_D^\mathrm{post}(T)
=
E_a(\tau_{\mathrm{emp}}),
\]
whereas
\[
\lim_{T\to\infty}Q_D^\mathrm{pre}(T)
=
\lim_{T\to\infty}Q_D^\mathrm{post}(T)
=
\infty.
\]
\textbf{Case: $E_a(\tau_{\mathrm{emp}})<1$.} There exist unique wait intervals
\[
T_{\mathrm{reach}}<T_{\mathrm{pass}}
\]
satisfying
\[
Q_D^\mathrm{pre}(T_{\mathrm{reach}})=1,
\qquad
Q_D^\mathrm{post}(T_{\mathrm{pass}})=1.
\]
Consequently, the location of the maximum is
\[
\begin{cases}
\text{before complete emptying},
& T<T_{\mathrm{reach}},
\\
\text{at complete emptying},
& T_{\mathrm{reach}}\leq T\leq T_{\mathrm{pass}},
\\
\text{after complete emptying},
& T>T_{\mathrm{pass}}.
\end{cases}
\]
\textbf{Case: $E_a(\tau_{\mathrm{emp}})\geq 1$.} The maximum lies after complete emptying for every \(T>0\).
\end{corollary}

\paragraph{Corresponding result for set-target removal.}
All of the above refers to steady states, however we know from Sec.~\ref{sec:dur_stim_target} that duration-dependent stimulation under a set-target protocol can lead to non-constant long-term dynamics. We will discuss daily output for non-constant long-term dynamics in the next section, and restrict our discussion here to the case of stable fixed points.

\medskip
Before complete emptying, set-target and set-duration removal describe the
same positive steady states, see Sec.~\ref{sec:protocol_correspondence}. For duration-dependent stimulation,
\(M_*(T,\tau)\) is strictly increasing with \(\tau\), so the corresponding
session target $M_{\mathrm{tar}}=M_*(T,\tau)$ provides an equivalent parametrization, see Fig.~\ref{fig:protocol_comparison}(d). Hence a pre-emptying fixed-point daily maximum at \(\tau_{\mathrm{opt}}\) under set-duration removal corresponds to a unique set-target maximum at
\[
M_{{\mathrm{tar},opt}}=M_*(T,\tau_{\mathrm{opt}}).
\]
The protocols differ once complete emptying is reached. Under set-duration
removal, the prescribed duration can exceed \(\tau_{\mathrm{emp}}\) (compare Fig.~\ref{fig:protocol_comparison}(d)), and continued
stimulation can therefore shift the maximum beyond complete emptying. Under
the set-target formulation, the session terminates at complete emptying.
Consequently, if \(E_a(\tau_{\mathrm{emp}})<1\), the set-target fixed-point daily maximum lies before
complete emptying for \(T<T_{\mathrm{reach}}\), whereas for
\(T\geq T_{\mathrm{reach}}\) the fixed-point daily output first reaches its maximum at
\(M_{\mathrm{tar}}=M_{\mathrm{emp}}(T)\) and remains constant for
\(M_{\mathrm{tar}}\geq M_{\mathrm{emp}}(T)\). If
\(E_a(\tau_{\mathrm{emp}})\geq1\), this latter case applies for every \(T>0\).

\paragraph{Examples.} 

For the \emph{Michaelis--Menten} duration response,
\[
a(\tau)
=
a_{\mathrm{max}}\frac{\tau}{\tau+\mu_\tau},
\qquad
E_a(\tau)
=
\frac{\mu_\tau}{\tau+\mu_\tau},
\]
so \(E_a\) fulfils the assumptions of Prop.~\ref{prop:dailyOutput_durStim}, and $E_a(\tau_\mathrm{emp})<1$. Hence, for set-duration removal, the maximum
moves from before, to at, and finally to after complete emptying as
\(T\) increases. For set-target removal, sufficiently short wait intervals have
a pre-emptying maximum with respect to the session target, whereas for sufficiently long wait intervals the daily output is maximized at complete emptying.\newline 

For the \emph{Hill} duration response,
\[
a(\tau)
=
a_{\mathrm{max}}\frac{\tau^2}{\tau^2+\mu_\tau^2},
\qquad
E_a(\tau)
=
\frac{2\mu_\tau^2}{\tau^2+\mu_\tau^2},
\]
which again fulfils the assumptions of Prop.~\ref{prop:dailyOutput_durStim}, with
\[
E_a(\tau_\mathrm{emp})<1 \iff \mu_\tau<\tau_\mathrm{emp}.
\]
If \(\mu_\tau < \tau_\mathrm{emp}\) the same sequence of optimization regimes occurs as for the Michaelis--Menten response. If \(\mu_\tau\geq \tau_\mathrm{emp}\), complete
emptying is optimal for all set-target schedules, whereas under
set-duration removal the optimum always lies after complete emptying. See Fig.~\ref{fig:dailyOutput_dur}(b),(c) for a visual summary of these results.

\begin{figure}[p]
  \centering
  \includegraphics[width=0.8\textwidth]{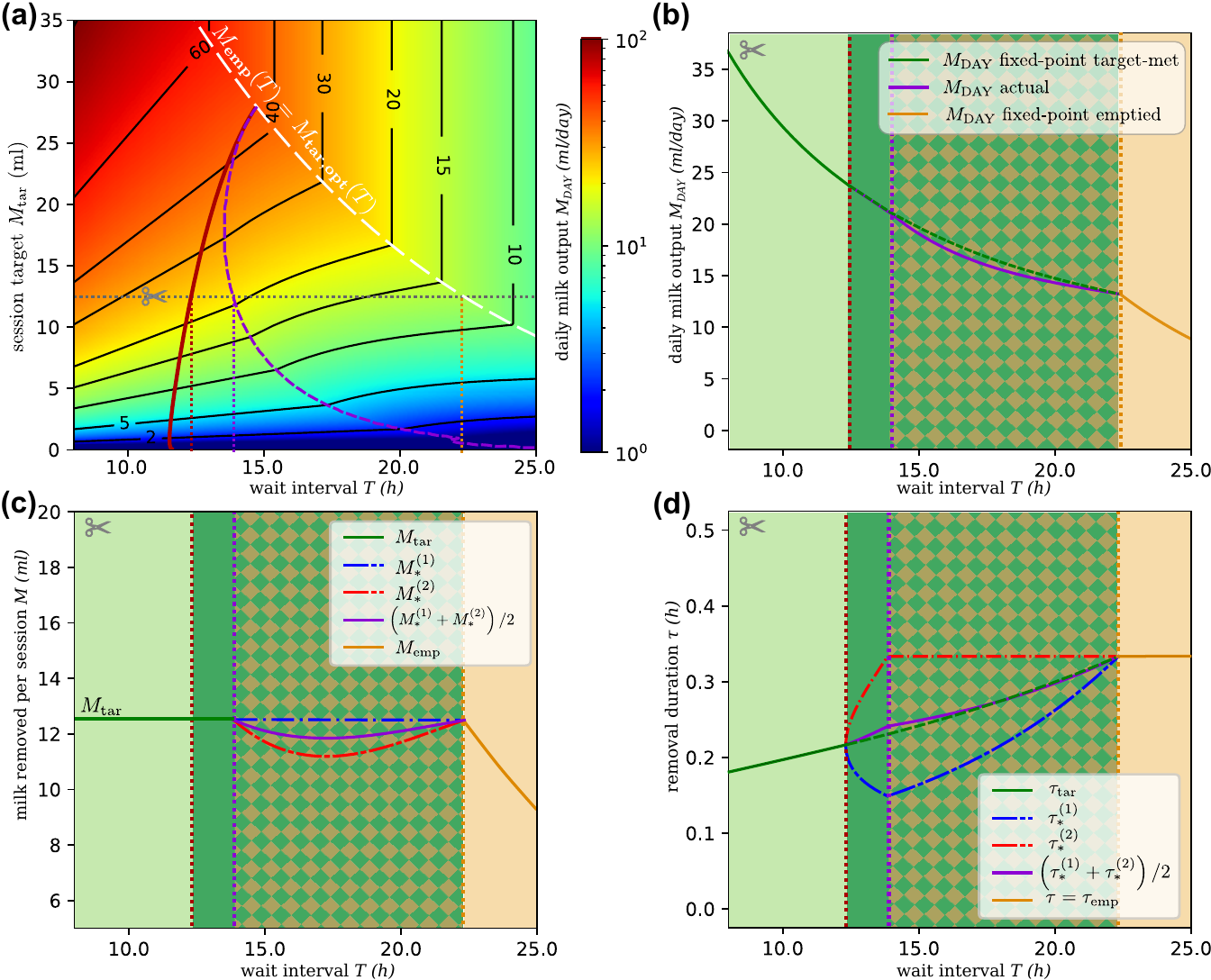}
  \caption{Daily milk output on non-constant attractors under duration-dependent stimulation and set-target removal. (a) Numerically determined long-term daily milk output \(M_{\mathrm{DAY}}\) in \((T,M_{\mathrm{tar}})\)-schedule space for the Hill response. Colours and black contours indicate \(M_{\mathrm{DAY}}\). The black curve \(M_{\mathrm{emp}}(T)\) separates target-met and emptied fixed-point branches. The red curve marks the instability boundary of the target-met fixed point, the dashed-purple curve the (numerically determined) boundary at which the attracting orbit first contains an emptying state. The horizontal dotted line indicates the slice \(M_{\mathrm{tar}}=12.5\)ml used in (b)-(d). (b) Daily milk output along this slice, comparing the numerically realized attractor with the unstable target-met fixed-point and the emptying fixed-point. (c) Milk removed per session along the same slice. \(M^{(1)}\) and \(M^{(2)}\) denote the two period-2 values and the additional curve their mean. (d) Corresponding removal durations \(\tau^{(1)}\) and \(\tau^{(2)}\), their mean, the target-met fixed-point duration, and the complete-emptying duration \(\tau_\mathrm{emp}\). Vertical dotted lines in (b)-(d) mark, from left to right, the period-doubling point, the first occurrence of an emptying state on the attracting orbit, and the transition to the emptied fixed point. Parameters used: $\delta=0.15$/h, $\gamma=0.15$/h, $a_\mathrm{max}=50$ml/h, $\mu_\tau=0.5$h, $\tau_\mathrm{emp}=1/3$h.}
  \label{fig:dailyOutput_nonConst}
\end{figure}

\subsection{Daily output on non-constant attractors.}
\label{ssec:daily_non_steady}

As shown in Sec.~\ref{sec:dur_stim_target}, under duration-dependent stimulation with set-target removal, the target-met fixed point may lose stability, giving rise to non-constant attractors. However, so far all results above are based on fixed points.

\paragraph{Pre-emptying maxima are always stable.}

We start by discussing how the position of the fixed-point daily output maximum with respect to $M_\mathrm{tar}$ relates to the instability boundary of the target-met fixed point. The following Lemma gives an answer.

\begin{lemma}[Stability of a pre-emptying daily-output maximum]
\label{lem:stable_preemptying_max}
Fix \(T>0\) and consider duration-dependent stimulation under a
set-target protocol. Suppose that the fixed-point daily output
\[
M_{\mathrm{tar}}\mapsto M_{\mathrm{DAY}}(T,M_{\mathrm{tar}})
\]
has a pre-emptying maximum, i.e.
\[
0<M_{{\mathrm{tar},opt}}(T)<M_{\mathrm{emp}}(T).
\]
Then the corresponding target-met fixed point is locally asymptotically
stable.
\end{lemma}

For a proof see App.~\ref{app:daily_output}.

\paragraph{Interpretation.} The above Lemma shows that a pre-emptying maximum of the daily output never lies in the region, where the target-met fixed point is unstable.

\paragraph{Numerical exploration of daily output under non-constant attractors.}

We next examine numerically the daily output under non-constant attractors. We chose a Hill response using the parameters given in Fig.~\ref{fig:dailyOutput_nonConst}. For the chosen parameters, \(E_a(\tau_{\mathrm{emp}})>1\). Hence the previous results imply that the fixed-point daily-output maximum lies at complete emptying for all \(T>0\), while part of the target-met fixed-point region is unstable. Beyond the fixed-point stability boundary, daily output is evaluated numerically from the asymptotic attractor. For this example, accounting for the non-constant attractors does not change the maximizing set-target schedule: daily output is still first maximized at complete emptying.

\medskip
In the example in Fig.~\ref{fig:dailyOutput_nonConst} we consider the behaviour of the daily output along a slice through schedule space with constant $M_\mathrm{tar}$. Along this slice the only observed non-constant long-term behaviour is a 2-periodic attractor. In this case \eqref{eqn:Mday_periodic} gives
\begin{align*}
M_\mathrm{DAY}=\frac{24}{T+ \frac{1}{2}\left(\tau_*^{(1)}+\tau_*^{(2)}\right)}\frac{1}{2}\left(M^{(1)}_*+M^{(2)}_*\right).
\end{align*}

Fig.~\ref{fig:dailyOutput_nonConst} shows the resulting output and the numerically determined maximizing schedule. We observe that the daily output under the non-steady dynamics is qualitatively similar to the daily output corresponding to the unstable target-met fixed point, Fig.~\ref{fig:dailyOutput_nonConst}(a),(b). Inspecting how the individual and mean session outputs, and individual and mean session durations are affected, Fig.~\ref{fig:dailyOutput_nonConst}(c) and (d) respectively, we see that the observed (small) difference in daily output is mainly driven by periodic attractors that involve emptied-breast states: Here the mean session output is lowered, and the mean session duration is increased, compared to the unstable target-met fixed point. This leads to the observed small reduction in daily output.

\subsection{Interpretation}

In this section, we shifted the focus from milk removed per session to
the long-term daily milk output \(M_{\mathrm{DAY}}\), and asked how this
quantity depends on the removal schedule. For set-target removal, the
schedule is controlled by \((T,M_{\mathrm{tar}})\), whereas for set-duration
removal it is controlled by \((T,\tau)\).

\medskip
The effect of the wait interval \(T\) is consistent across all model variants: Holding the second schedule parameter fixed, increasing \(T\) decreases steady-state daily milk output. Increasing the amount or duration of removal can creates a trade-off. Longer removal sessions, or larger removal targets, can increase the amount of milk obtained per session, but also tends to lengthen the removal cycle and thereby reduce the number of sessions per day. As a result,
\(M_{\mathrm{DAY}}\) can be non-monotone in the second schedule
parameter.\newline

Fig.~\ref{fig:dailyOutput_dur} illustrates how the location of this optimum depends on the
stimulation mechanism. For amount-dependent stimulation, sufficiently
short wait intervals favour removal that stops before complete
emptying. As \(T\) increases, the optimum reaches complete emptying and
remains there. Duration-dependent stimulation can show an additional
regime: after the optimum has reached complete emptying, sufficiently
long wait intervals can favour continuing the removal session beyond
the time required to empty the breast. In this regime no additional
milk is removed during the extended part of the session, but the
additional removal duration continues to stimulate subsequent
production. For Hill-type duration responses, the existence of a
pre-emptying optimum also depends on the emptying time $\tau_\mathrm{emp}$; for sufficiently short emptying times, this regime disappears.\newline

The preceding results describe daily output for stable steady-states.
Duration-dependent stimulation under set-target removal can, however,
lose fixed-point stability and generate non-constant attractors.
Fig.~\ref{fig:dailyOutput_nonConst} shows that the onset of period-doubling need not produce a substantial change in daily milk output. In the numerical example examined, while both points of the period-2 attractor remain target-met, daily output remains close to the value obtained from the formal fixed-point continuation. A clearer reduction occurs once one point of the attractor reaches complete emptying. However, these changes did not affect the maximum of the daily output with respect to the second scheduling variable.\newline

%While both points of the period-2 attractor remain target-met, the amount removed in each session remains \(M_{\mathrm{tar}}\), and the mean removal duration stays close to that of the corresponding fixed point. Daily output therefore remains close to the value obtained from the formal fixed-point continuation. A clearer reduction occurs once one point of the attractor reaches complete emptying: the mean amount removed per session then drops below \(M_{\mathrm{tar}}\), while the mean removal duration increases slightly. Thus, the appearance of non-constant dynamics does not necessarily lead to a noticeable change indaily output.\newline

Taken together, the results show that maximizing output per session and
maximizing output per day are distinct objectives. Inspecting the level-sets of $M_\mathrm{DAY}$ in Fig.~\ref{fig:dailyOutput_dur}(a),(b) and Fig.~\ref{fig:dailyOutput_nonConst}(a) we see that for a given
\(M_{\mathrm{DAY}}\), there can be several schedules with different
wait intervals, removal targets, and removal durations that produce
similar long-term output. One can therefore choose among these schedules based on additional criteria, such as the time spent removing milk or the frequency of removal sessions.

\section{Comparison and summary of model predictions}
\label{sec:compare}

\begin{figure}[t]
  \centering
  \includegraphics[width=\textwidth]{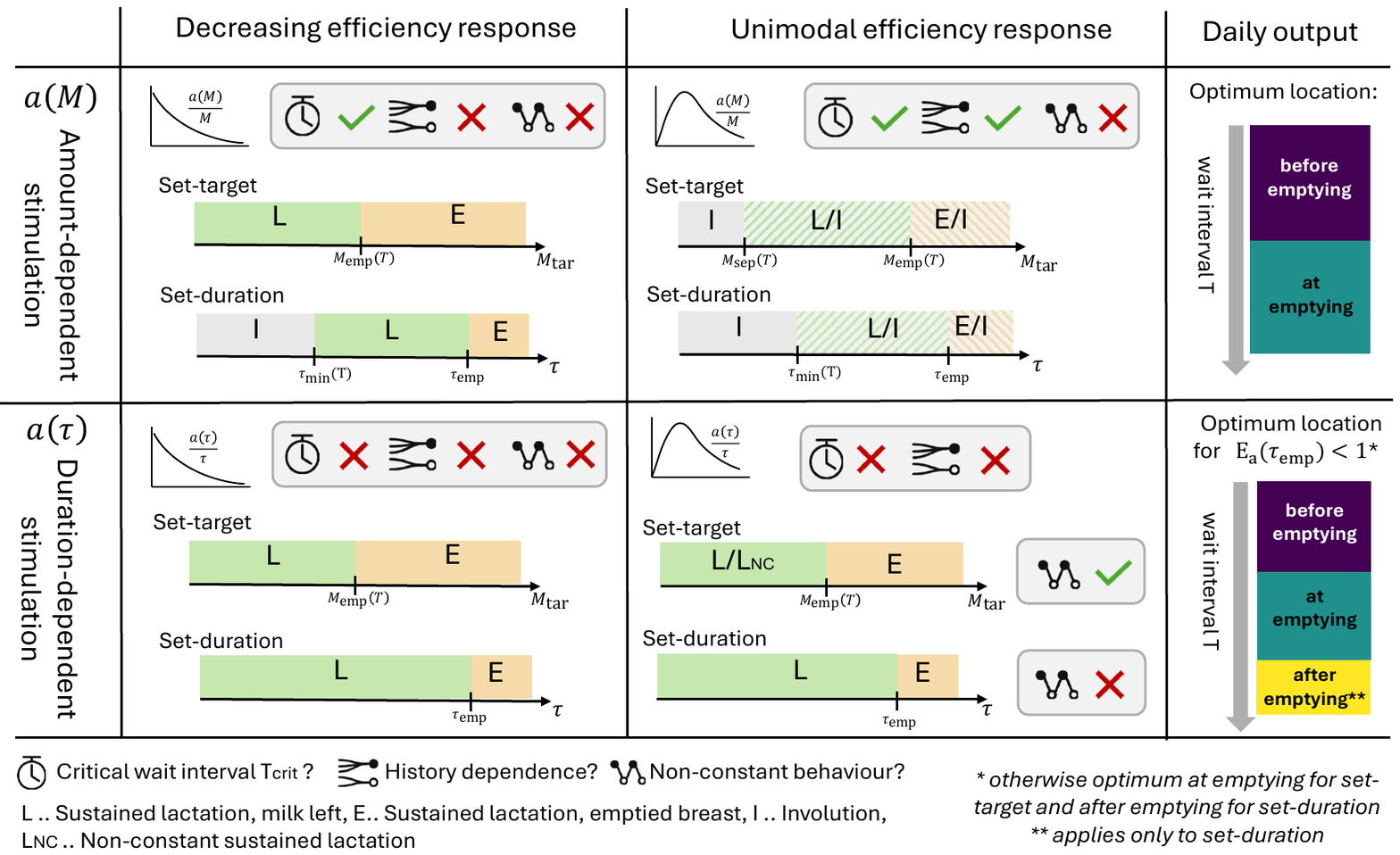}
  \caption{Comparison of long-term model behaviour and daily milk output across the dimensions: Stimulation-dependence, removal protocol and stimulation efficiency response.}
  \label{fig:model_comparison}
\end{figure}

After analysing the individual model variants, we now compare which
qualitative predictions are robust and which depend on the assumed
stimulation mechanism, stimulation-efficiency response, or removal
protocol. Fig.~\ref{fig:model_comparison} provides an overview of the
resulting long-term behaviour and daily milk output. In particular, we distinguish whether a
model predicts a critical wait interval, dependence of the long-term
outcome on the initial state, and non-constant long-term dynamics.

\paragraph{Duration-dependent stimulation lacks direct state-dependent
feedback.}
A principal distinction is whether stimulation depends on the amount of
milk removed or on removal duration. With amount-dependent stimulation,
a reduction in milk production leads to less milk being available for
removal, which in turn reduces stimulation of subsequent production.
This feedback allows low-production states to reinforce themselves and
ultimately permits involution. For duration-dependent stimulation this direct coupling between the amount removed and subsequent stimulation is absent. Consequently, for the regular positive-removal schedules considered here, the duration-dependent models do not admit the involution state as a long-term outcome.

\paragraph{Amount-dependent stimulation yields a critical wait
interval.}
For amount-dependent stimulation, sustained lactation requires the
stimulation generated per unit milk removed to compensate for the loss
of production between removal sessions. This gives a robust result: the same critical
wait interval \(T_{\mathrm{crit}}\) for both removal protocols and for both
efficiency classes. If the maximal stimulation efficiency is smaller than the rate of production loss, no positive wait interval can sustain lactation. Otherwise, positive sustained states are possible only for wait intervals with $0<T<T_{\mathrm{crit}}$. The critical wait interval shortens if production loss or milk loss between sessions is large, and lengthens if the maximal stimulation efficiency is large.

\paragraph{History dependence requires a unimodal efficiency response.}
The distinction between decreasing and unimodal efficiency for amount-dependent stimulation is particularly important at small removals. For a decreasing efficiency
response, stimulation is most efficient for small amounts and whenever the wait interval is below \(T_{\mathrm{crit}}\) every nonzero initial condition converges to sustained
lactation. For a unimodal efficiency response, efficiency is small at small
removals, increases before reaching its maximum and only then decays. In the
amount-dependent models this has two effects: First, a minimal amount of milk has to be removed per session to allow for a sustained lactation state to exist. Second, even if the sustained lactation state exists, the long-term outcome depends on the initial milk
amount and production rate: If these are too low, the system will converge to the involution state. In contrast, duration-dependent stimulation does not produce bistability between involution and sustained lactation, since the involution state is absent under continued stimulation.

\paragraph{Set-target and set-duration protocols are qualitatively
similar for decreasing efficiency, but can differ substantially for
unimodal efficiency.}

In this work we compared two removal protocols, setting the session target or setting the removal duration. For decreasing efficiency responses, changing between the two protocols has relatively little effect on the sustained long-term states predicted by the model. One distinction arises for duration-dependent stimulation, where prescribed removal durations can extend beyond complete emptying and thereby support larger steady-state session outputs. The differences become more pronounced for unimodal efficiency responses. For amount-dependent stimulation, some steady states that are stable under set-target removal are unstable under set-duration removal. For duration-dependent stimulation, corresponding steady states can likewise differ in stability, and set-target removal can additionally generate non-constant long-term dynamics. Thus, prescribing an amount or a duration can change how deviations from a sustained state feed back into subsequent removal sessions, even when the corresponding steady-state session output and removal duration are the same.

\paragraph{Non-constant long-term behaviour occurs only for
duration-dependent stimulation with unimodal efficiency under a
set-target protocol.}
Within the model variants analysed here, all amount-dependent models
converge to steady states. Duration-dependent stimulation under
set-duration removal likewise has a stable steady state. Non-constant
long-term dynamics arise only when duration-dependent stimulation with
a unimodal efficiency response is combined with set-target removal. In this case the realised removal duration depends on the amount of milk available at the start of the session, and this state-dependent duration feeds back into subsequent stimulation. Numerical simulations then revealed attracting non-constant solutions, including period-doubled states.

\paragraph{Daily output reflects a trade-off between session output and session frequency.}
Across all model variants, increasing the wait interval decreases daily milk output, since it
lengthens the removal cycle without increasing the amount obtained per session. In contrast,
increasing the removal target or removal duration can increase session output, but simultaneously
reduces the number of sessions that can be performed per day. Consequently, for
sufficiently short wait intervals the daily-output maximum can occur before complete emptying, see the right column of Fig.~\ref{fig:model_comparison}: the gain from extracting more milk per session is then outweighed by the additional time required for removal. As the wait interval increases, the optimum shifts towards complete emptying. For duration-dependent stimulation, continued removal after emptying can still increase subsequent production, so that under set-duration removal the optimum can shift further, from before to at and eventually after
complete emptying. Thus, the stimulation mechanism determines not only whether lactation can be sustained, but
also how the competing benefits of more intensive removal and more frequent removal affect
long-term milk output.

\section{Discussion}
\label{sec:discussion}

\subsection{Comparing model predictions to reality}

\paragraph{The big picture.}
We developed and analysed a hybrid model of the feedback between milk removal and subsequent production. For repeated constant schedules, we characterised the long-term behaviour and identified which qualitative predictions depend on the assumed stimulation mechanism and removal protocol (see Sec.~\ref{sec:compare} and Fig.~\ref{fig:model_comparison} and \emph{Interactive Lactation Explorer} - Online Resource 1). The predictions are broadly consistent with the physiology of established lactation, where milk removal regulates production and prolonged milk stasis can lead to involution \cite{Truchet2017,Daly1996,Kent2012}. However, the underlying physiology involves hormonal, mechanical and local regulatory mechanisms. Our stimulation function therefore represents the combined effect of a removal session rather than any specific physiological pathway. The models distinguish between amount- and duration-dependent stimulation, decreasing and unimodal stimulation efficiency, and two alternative removal protocols. These distinctions lead to several potentially testable predictions.

\paragraph{Amount-dependent versus duration-dependent stimulation.}
Amount-dependent stimulation introduces a direct state-dependent feedback: declining production reduces the milk available for removal and hence subsequent stimulation, potentially reinforcing the decline. This allows for involution and leads to a critical wait interval $T_{\mathrm{crit}}$. In contrast, duration-dependent stimulation lacks this coupling in our formulation and does not permit involution under continued stimulation. Observations of self-reinforcing production loss despite continued removal, or of a critical wait interval beyond which more intensive removal cannot sustain lactation, would therefore favour the amount-dependent models.

\medskip
However, this distinction should be interpreted cautiously. First, removal amount and duration are intrinsically correlated, making the two mechanisms difficult to distinguish experimentally. Second, our removal law assumes a state-independent emptying time $\tau_{\mathrm{emp}}$. If emptying duration depended on the available milk, duration-dependent stimulation could itself become state dependent and potentially alter the low-production dynamics.

\paragraph{Decreasing and unimodal stimulation efficiency.}
The two efficiency classes differ primarily in their response to small inputs: decreasing efficiency is maximal at small removal amounts or durations, whereas unimodal efficiency reaches its maximum at an intermediate scale. Distinguishing these alternatives requires measurements spanning small and intermediate inputs, ideally through longitudinal data relating removal to subsequent production.

\medskip
For amount-dependent stimulation, unimodal efficiency additionally permits bistability: whether lactation is sustained can depend on the initial milk amount and production rate. This is consistent with the idea that lactation must first be sufficiently established before a given schedule can maintain it. However, our model does not describe postpartum establishment or its endocrine regulation. Bistability therefore demonstrates that an establishment-like threshold can arise from the feedback structure, but does not establish unimodal stimulation efficiency as the underlying physiological mechanism. History dependence could also arise from additional physiological variables absent from the model.

\medskip
Further predictions may help distinguish the models. For amount-dependent stimulation, the maximal sustainable session output decreases with increasing wait interval. Non-constant long-term dynamics, on the other hand, occur only for duration-dependent stimulation with unimodal efficiency under a set-target protocol. Observing persistent oscillations under approximately repeated schedules would therefore be of interest, although distinguishing them from biological variability may be difficult.

%\paragraph{Data needed to distinguish the models.}
%Several types of data would be particularly informative. First, measurements of removed milk as a function of removal duration and the amount initially available could test the assumed removal law, including whether a meaningful emptying time $\tau_{\mathrm{emp}}$ exists and whether it depends on breast state. Second, longitudinal feeding or pumping data could be used to fit and compare the alternative stimulation models directly. Third, targeted experiments that partially decouple removal amount and removal duration would be especially valuable for distinguishing $a(M)$ from $a(\tau)$. Finally, experiments spanning small to intermediate removal amounts or durations would be important for distinguishing decreasing from unimodal stimulation efficiency. Together, such data would allow the qualitative model classes analysed here to be connected to quantitative model selection and parameter estimation.

\paragraph{Parameter estimates.}
Available estimates suggest a characteristic decline timescale of milk production
of about $48\,\mathrm{h}$ \cite{Truchet2017}, which we use as an
order-of-magnitude proxy for $\gamma^{-1}\approx48\,\mathrm{h}$.
Further, \citet{Meier2016} reported that efficient breast pumps remove
approximately $85\%$ of the available milk within $15$ min. Under our
linear removal law, this corresponds to
$\nu\approx3.4\,\mathrm{h}^{-1}$ and hence to a model emptying time of
$\tau_{\mathrm{emp}}=1/\nu\approx18$ min. \citet{Kent1999} measured a production rate of about $18$ ml/h per breast during exclusive breastfeeding. 
During established lactation, exclusively breastfed infants typically consume approximately 700--800 mL of milk per day, distributed over roughly 7--8 feeding occasions, although substantial variation exists \cite{Kent2006,Rios2023}. 

\medskip
To estimate the stimulation parameters, we assume that the milk-loss timescale is of similar order to the production-loss timescale, $\delta\sim\gamma$, and use the
steady state of a session that empties the breast, for which
$M^*=G(T)\,p^*$ and $a(M^*)=p^*\bigl(1-e^{-\gamma T}\bigr)$. For $p^*\approx18\,\mathrm{ml/h}$
and $T\approx3\,\mathrm{h}$, this gives $M^*\approx51\,\mathrm{ml}$, consistent with
the intake data above, and $a(M^*)\approx1.1\,\mathrm{ml/h}$. Assuming
$\mu_M\approx100\,\mathrm{ml}$, this yields $a_{\max}\approx3$--$5\,\mathrm{ml/h}$ and
$R_{\max}\approx0.032\,\mathrm{h}^{-1}$ for the Michaelis--Menten response and
$R_{\max}\approx0.027\,\mathrm{h}^{-1}$ for the Hill response. Both exceed
$\gamma\approx0.021\,\mathrm{h}^{-1}$, so a sustained lactation state exists, with
critical wait intervals $T_\text{crit}$ of roughly one to two days.
Eq.~\eqref{eqn:Mday_MMHill} then gives $T_\text{reach}\approx2$--$3\,\mathrm{h}$:
for shorter wait intervals, terminating removal before complete emptying
maximizes daily milk output. Consistently with our assumption of an emptying
steady state at $T\approx3\,\mathrm{h}$, this lies just below typical feeding
intervals. Thus, this non-trivial model prediction may occur in a
physiologically and practically relevant regime.

\medskip
In contrast, for the Hill duration-response model the necessary condition for loss of fixed-point
stability, $T>\log(3)/\gamma$, gives $T\gtrsim53\,\mathrm{h}$. Thus, with
these order-of-magnitude estimates, the non-constant dynamics found in the
duration-dependent set-target model are unlikely to occur under typical
feeding or pumping schedules. A quantitative assessment will require
parameter estimation from longitudinal data, in particular for the
stimulation-response parameters that are not currently constrained.

\subsection{Milk output and schedule design}

\paragraph{Maximising daily milk output.} Sustaining lactation is only one possible objective of a removal schedule; in practice, the amount of milk obtained over time is also important. Increasing the session target or removal duration may increase the amount of milk obtained per session, but at the cost of lengthening the full wait--removal cycle and thereby reducing session frequency. The resulting trade-off can make daily output non-monotone with respect to removal intensity. In the examples analysed here, schedules that maximise daily output involve either complete emptying of the breast or, for sufficiently short wait intervals, terminating individual sessions before the breast is emptied. This mirrors a familiar distinction in impulsive harvesting, where maximising the yield of a single harvest and maximising sustained yield per unit time are likewise different objectives \cite{Zhang2003, Lawson2023}; in contrast to harvesting, however, removal here also stimulates future production. These results depend on model assumptions and should not be interpreted as direct breastfeeding or pumping recommendations. They do, however, illustrate why maximising milk obtained per session, maximising milk obtained per day, and minimising the practical burden of removal are different objectives.

\paragraph{Schedule choice depends on objective.}
More generally, a given daily milk output can be achieved by different removal
schedules. This flexibility may be as relevant in applications as maximizing
output itself. Depending on the practical objective, one might choose a
schedule that achieves an adequate output while minimizing the number of
sessions or total pumping time, allowing longer uninterrupted intervals,
limiting prolonged breast fullness, or satisfying constraints imposed by work, sleep, or
other activities. The present analysis provides a mechanistic basis for such
schedule-design problems, but quantitatively meaningful optimization will
require physiological parameterization and extension to non-constant schedules.

\subsection{Model assumptions and limitations}

\paragraph{Model structure.} The simplicity that makes the model analytically tractable also imposes important limitations. We represent each breast independently, and by only two state variables, total milk volume and an effective production rate, without resolving individual lobes, alveoli, or spatial variation. We separate each cycle into a comparatively long continuous wait phase and a short removal phase represented as a discrete update. Production and milk loss during removal are neglected. This approximation is appropriate only when removal sessions occur on a substantially shorter timescale than the intervals between them. 

\paragraph{Stimulation.} The biological mechanisms connecting milk removal to future production are represented phenomenologically through a single stimulation function. In particular, the present formulation does not explicitly include the proposed negative feedback of breast fullness on ongoing milk production \cite{Peaker1996}. Further we let stimulation depend either on the amount removed or on the session duration, but not on both. For duration-dependent stimulation we additionally assume that stimulation can continue after complete emptying; whether post-emptying nipple or pump stimulation has a comparable effect on subsequent milk synthesis is not established.

\paragraph{Removal law.} Milk removal itself is described by a simple linear removal law. The removal law further assumes a state-independent removal-rate parameter, so that complete emptying occurs after the fixed duration $\tau_{\mathrm{emp}}=1/\nu$. In reality, removal rate and the time required to empty the breast may depend on the amount of milk available, milk-ejection dynamics, infant or pump characteristics, and other state-dependent factors. Moreover, actual breastfeeding and pumping sessions can contain multiple milk-ejection events, variable flow rates and interruptions. 

\paragraph{Variable schedules.} The present paper analyses repeated identical schedules. Real breastfeeding and pumping behaviour is highly variable: intervals between sessions differ throughout the day, night-time schedules differ from daytime schedules, individual sessions differ in duration and effectiveness, and both schedules and parameters can change systematically during establishment, maintenance, and weaning. Constant schedules provide a useful first step because they reveal fixed points, thresholds, bistability, and the dependence of long-term output on model assumptions in a form that can be analysed explicitly. They are best viewed as building blocks for studying realistic sequences of non-identical removal events.

\subsection{Outlook}

\paragraph{Model selection and parameter estimation.}
To determine the most likely stimulation mechanism, a natural next step is to fit the alternative model variants to longitudinal pumping data and use model-selection and predictive criteria to assess which stimulation mechanisms are supported by the data. This would also allow estimation of the physiological timescales, stimulation-response parameters, and removal characteristics required for quantitative predictions. In particular, distinguishing amount- from duration-dependent stimulation will require datasets with sufficient independent variation in removed amount and removal duration.

\paragraph{Analysis of variable schedules.} Analytically, the next step is to extend the analysis from constant to variable schedules. The hybrid formulation already allows the wait interval, removal duration, and removal target to vary from one session to the next, so realistic daily or multi-day schedules can be represented without changing the underlying model structure. This will make it possible to study questions that cannot be addressed by constant schedules, including whether occasional long gaps can be compensated by more intensive removal at other times, how rapidly the system responds to changes in schedule, and how the timing and ordering of sessions influence long-term production.

\paragraph{Practical impact.} Combined with parameter estimation from longitudinal human data, the same framework can ultimately be used to formulate practical optimisation problems. Relevant objectives include increasing production from a low-output state, maintaining a prescribed daily output with minimal pumping burden, adapting schedules to external constraints, and deliberately reducing production during weaning. The present theory does not yet provide individual recommendations for these questions. Rather, it identifies the dynamical mechanisms and thresholds that must be understood before such recommendations can be made and provides an analytically tractable foundation for subsequent data-driven and practice-oriented work. Such applications will require both model calibration and validation on independent longitudinal data.

\section{Statements and Declarations}

\subsection*{Competing interests}
The authors declare no competing interests.

\subsection*{Supplementary Information}
\label{sec:SI}
\textbf{Online Resource 1.} \emph{Interactive Lactation Explorer}: A stand-alone html file with which the analytical and numerical results can be tested, visualised and explored, see also \emph{Data and Code availability.}

\subsection*{Data and Code availability}
No experimental data were generated or analysed during the current study. A Jupyter notebook for numerical simulation of the model, as well as the \emph{Interactive Lactation Explorer} are available on Github \url{https://github.com/angelikamanhart/Lactation_model_analytical_paper}. For a direct link to the \emph{Interactive Lactation Explorer} follow \url{https://angelikamanhart.github.io/Lactation_model_analytical_paper/interactive_lactation_explorer.html}.

\subsection*{Author contribution}
AM designed the project and developed the models. AM and TM analysed the models. AM wrote and edited the paper.

\subsection*{Acknowledgements}
The authors thank Dominik St\"urzer for helpful discussions and feedback.

\subsection*{LLM disclosure}
During the preparation of this manuscript, the authors used ChatGPT (OpenAI) and Claude (Anthropic) for language editing and to assist in discussing, checking, and presenting mathematical arguments. In particular ChatGPT was used for the creation of the interactive \emph{Lactation Explorer}. All resulting text and mathematical content were reviewed and verified by the authors, who take full responsibility for the manuscript.

\subsection*{Funding}
No funding was received for conducting this study.

% appendix
\section{Appendix}
\label{sec:app}

Throughout this appendix, $\leq$ denotes the componentwise order on
$X=[0,\infty)^2$.

%-----------------------------

\subsection{Amount-dependent stimulation, set-target protocol: Auxiliary results and proofs}
\label{app:amount_stim_target}

\paragraph{Auxiliary results.}

%The state space $X=[0,\infty)^2$ is equipped with the %componentwise partial order: $(m,p)\le(\tilde m,\tilde p)$
%iff $m\le\tilde m$ and $p\le\tilde p$. \newline

The following facts are independent of the particular shape of $R$, and apply to both the decreasing efficiency response class and the unimodal efficiency response class. Throughout $F: X\rightarrow X$ is given by \eqref{eq:Fmap},
$$
F(m,p)=\Big(\max\{x-M_\mathrm{tar},\,0\},\;\;p\,g+a\big(\min\{M_\mathrm{tar},\,x\}\big)\Big),
$$
where $x(m,p) = m d + p G$.
%in both Sections~\ref{sec:fixedtarget_conc} (decreasing efficiency class) and~\ref{sec:fixedtarget_sig} (unimodal efficiency class).

\begin{lemma}[Monotonicity]\label{lem:monotone}
$F$ is order-preserving: $z\le\tilde z$ implies $F(z)\le F(\tilde z)$.
\end{lemma}

\begin{proof}
Both components of \eqref{eq:Fmap} are nondecreasing in $m$ and in $p$, since $x(m,p)=md+pG$ is,
$a$ is increasing, and $\min$ and $\max$ preserve monotonicity.
\end{proof}

\begin{lemma}[Boundedness]\label{lem:bounded}
Every orbit of $F$ is bounded. More precisely, for every $z\in X$,
\[
  \limsup_{n\to\infty}p_n\le\frac{a_\mathrm{max}}{1-g},
  \qquad
  \limsup_{n\to\infty}m_n\le
  \frac{1}{1-d}\max\left\{0,\;\frac{a_\mathrm{max}}{\Gamma}-M_\mathrm{tar}\right\}.
\]
\end{lemma}

\begin{proof}
Since $a\le a_\mathrm{max}$, the second component of \eqref{eq:Fmap} gives
$p_{n+1}\le p_ng+a_\mathrm{max}$, so $p_n\le g^np_0+a_\mathrm{max}(1-g^n)/(1-g)$, which yields the
first bound and shows $(p_n)$ is bounded. Fix $\varepsilon>0$ and $n_0$ with
$p_n\le a_\mathrm{max}/(1-g)+\varepsilon$ for $n\ge n_0$. Then, for $n\geq n_0$, the first component satisfies
$m_{n+1}\le\max\{m_nd+\frac{a_\mathrm{max}}{\Gamma}+\varepsilon G-M_\mathrm{tar},0\}$, and iterating this recursion from $n_0$ gives the second bound as $\varepsilon\downarrow0$.
\end{proof}

\begin{lemma}[Sandwich]\label{lem:sandwich}
Assume $R_\mathrm{tar}>\Gamma$ and set $z^\ast=(0,M_\mathrm{tar}/G)$. Then
$F(z^\ast)\ge z^\ast$, and every $z\ge z^\ast$ satisfies $F^n(z)\to z_\mathrm{tar}$.
\end{lemma}

\begin{proof}
We have $F(z^\ast)=\big(0,\,gM_\mathrm{tar}/G+a_\mathrm{tar}\big)$, and $R_\mathrm{tar}>\Gamma$ directly implies $F(z^\ast)\ge z^\ast$. By Lem.~\ref{lem:monotone} the sequence
$\big(F^n(z^\ast)\big)_n$ is nondecreasing, and it is bounded by Lemma~\ref{lem:bounded}, so it
converges to a fixed point $w\ge z^\ast$. Any fixed point with $m=0$ lies in $E$ and therefore has
$p\le M_\mathrm{tar}/G$; combined with $w\ge z^\ast$ this would force $p=M_\mathrm{tar}/G$ and 
hence $R_\mathrm{tar}=\Gamma$, which the hypothesis excludes. Since $w$ has $m>0$, it lies in the target-met region and hence $w=z_\mathrm{tar}$, the unique fixed point of $F$.

Let now $z\ge z^\ast$ and put $P=\max\{p,\,a_\mathrm{max}/(1-g)\}$,
$M=\max\{m,\,(PG-M_\mathrm{tar})/(1-d)\}$ and $z^{\ast\ast}=(M,P)$. Then $z^{\ast\ast}\ge z$ and
$F(z^{\ast\ast})\le z^{\ast\ast}$: the second component is at most $Pg+a_\mathrm{max}\le P$ by the choice of $P$, and the
first is at most $\max\{Md+PG-M_\mathrm{tar},0\}\le M$ by the choice of $M$. Hence
$\big(F^n(z^{\ast\ast})\big)_n$ is nonincreasing and converges to a fixed point $w'$. Since
$z^{\ast\ast}\ge z^\ast$, monotonicity gives $F^n(z^{\ast\ast})\ge F^n(z^\ast)\to z_\mathrm{tar}$, so
$w'\ge z_\mathrm{tar}$. Since $m_\mathrm{tar}>0$, this implies that the first component of $w'$ is positive and $w'$ therefore lies in the target-met region. Hence $w'=z_\mathrm{tar}$. Finally
$F^n(z^\ast)\le F^n(z)\le F^n(z^{\ast\ast})$ for all $n$, and both bounds converge to
$z_\mathrm{tar}$.
\end{proof}

\begin{lemma}[Linearisation]\label{lem:jacobian}
Let $z$ be a fixed point of $F$ with $p>0$.
\begin{enumerate}
  \item[(i)] If $z$ lies in the interior of $E$, then $z=(0,p)$ with $p=pg+a(pG)$,
    and $DF(z)$ has eigenvalues $0$ and $g+a'(pG)G$. Writing $M=pG$ for the volume removed at $z$,
    the state is asymptotically stable if $R'(M)<0$ and unstable if $R'(M)>0$.
  \item[(ii)] If $z$ lies in the interior of $X\setminus E$, then $DF(z)$ has eigenvalues $d$ and
    $g$, both in $(0,1)$, and $z$ is asymptotically stable.
\end{enumerate}
\end{lemma}

\begin{proof}
On the interior of $E$ the map is $(m,p)\mapsto\big(0,pg+a(x)\big)$, with Jacobian
\[
  \begin{pmatrix} 0 & 0\\ a'(x)\,d & g+a'(x)\,G\end{pmatrix},
\]
whose eigenvalues are $0$ and $g+a'(x)G$. A fixed point there has $m=0$ and $x=pG$ with $p$ fulfilling
$p=pg+a(pG)$. Setting $M=pG$ and using $R(M)=\Gamma$, we can rewrite $g+a'(x)\,G-1=G(a'(M)-\Gamma)=G\,M\,R'(M)$ showing that 
the condition $g+a'(pG)G<1$ is therefore precisely $R'(M)<0$. On the interior of $X\setminus E$ the map is the affine map \eqref{eq:affine}, whose linear part is upper triangular with diagonal entries $d$ and $g$.
\end{proof}

Further we define $E_0=\{0\}\times\big[0,\,M_\mathrm{tar}/G\big]$ so that $E_0$ is the part of the $p$-axis lying in $E$.

\begin{lemma}[Entry into the emptying region]
\label{lem:entry}

\begin{enumerate}
\item[(i)] If $R_\mathrm{tar}<\Gamma$, then every orbit enters $E$
in finite time. Moreover, if $R_\mathrm{tar}\leq\Gamma$, then
\[
F(E)\subseteq E_0.
\]

\item[(ii)] If $R_\mathrm{tar}>\Gamma$, then the target-met fixed point
$z_\mathrm{tar}=(m_\mathrm{tar},p_\mathrm{tar})$ lies in the interior
of $E^c$, and every orbit that never enters $E$ converges to
$z_\mathrm{tar}$.
\end{enumerate}
\end{lemma}

\begin{proof}
Eq.~\eqref{eqn:target_met} shows that $m_\mathrm{tar}<0$ if $R_\mathrm{tar}<\Gamma$ and
$m_\mathrm{tar}>0$ if $R_\mathrm{tar}>\Gamma$.

Suppose an orbit never enters $E$. It is then generated entirely by
the now affine target-met map $F$. Since the linear part of $F$ has
eigenvalues $d,g\in(0,1)$,
\[
F^n(z)\to z_\mathrm{tar}.
\]
If $R_\mathrm{tar}<\Gamma$, this is impossible for an orbit contained
in $X$, since $m_\mathrm{tar}<0$. Hence every orbit enters $E$ in
finite time.

If $R_\mathrm{tar}\leq\Gamma$ and $(m,p)\in E$, then
$p\leq M_\mathrm{tar}/G$ and $x(m,p)\leq M_\mathrm{tar}$.
Therefore
\[
pg+a(x)
\leq
g\frac{M_\mathrm{tar}}G+a(M_\mathrm{tar})
\leq
\frac{M_\mathrm{tar}}G,
\]
so $F(E)\subseteq E_0$.

Finally, if $R_\mathrm{tar}>\Gamma$, then $m_\mathrm{tar}>0$.
Since at a target-met fixed point
\[
m_\mathrm{tar}
=
x(z_\mathrm{tar})-M_\mathrm{tar},
\]
we have $x(z_\mathrm{tar})>M_\mathrm{tar}$, so
$z_\mathrm{tar}\in\operatorname{int}(E^c)$. Any orbit which never
enters $E$ is an orbit of $F$ and therefore converges to
$z_\mathrm{tar}$.
\end{proof}

%\paragraph{Lemma: Dynamics on the $p$-axis.}

%Only the scalar dynamics on the $p$-axis is specific to the %response class; the remaining
%ingredients are Lemmas~\ref{lem:monotone}--\ref{lem:sandwich}.

\begin{lemma}[Dynamics on the $p$-axis, decreasing-efficiency case]\label{lem:axisconc}
Let $q\in[0,M_\mathrm{tar}/G]$. While the orbit of $(0,q)$ remains in $E_0=\{0\}\times [0,M_\mathrm{tar}/G]$, its second component evolves under the
increasing scalar map $\Phi(q)=qg+a(qG)$, with
\[
  \Phi(q)-q=qG\big(R(qG)-\Gamma\big).
\]
Then:
\begin{enumerate}
  \item[(a)] if $R_\mathrm{max}<\Gamma$, the orbit remains in $E_0$, decreases strictly for $q>0$ and converges
    to $(0,0)$;
  \item[(b)] if $R_\mathrm{max}>\Gamma$ and $M_\mathrm{tar}\ge M_\mathrm{emp}$, the orbit remains in $E_0$ and
    converges to $(0,p_\mathrm{emp})$ for every $q>0$;
  \item[(c)] if $R_\mathrm{max}>\Gamma$ and $M_\mathrm{tar}<M_\mathrm{emp}$, the orbit leaves $E_0$ after finitely
    many steps for every $q>0$.
\end{enumerate}
\end{lemma}

\begin{proof}
For $(0,q)\in E_0$, $F(0,q)=(0,\Phi(q))$, and $\Phi$ is increasing because $a$ is increasing. The
identity above shows that the sign of $\Phi(q)-q$ is exactly the sign of $R(qG)-\Gamma$.

(a) If $R_\mathrm{max}<\Gamma$, then $R(qG)<\Gamma$ for every $q>0$, so $\Phi(q)<q$ for every $q>0$. Thus $E_0$
is forward invariant and the orbit decreases to a fixed point, necessarily $(0,0)$.

(b) If $M_\mathrm{tar}\ge M_\mathrm{emp}$, then $E_0$ contains $(0,p_\mathrm{emp})$ and is forward
invariant by Lemma~\ref{lem:entry}(i), since
$R_\mathrm{tar}\le\Gamma$. By the strict decrease of $R$, $\Phi(q)>q$ below
$p_\mathrm{emp}$ and $\Phi(q)<q$ above it. Hence every positive orbit in $E_0$ converges monotonically
to $(0,p_\mathrm{emp})$.

(c) If $M_\mathrm{tar}<M_\mathrm{emp}$, then $R(qG)>\Gamma$ throughout the positive part of $E_0$, so
$\Phi(q)>q$. If a positive orbit remained in $E_0$ forever, it would increase to a fixed point in
$E_0$, but no such positive fixed point exists. Hence it leaves $E_0$ after finitely many steps.
\end{proof}

\paragraph{Proof of Thm.~\ref{thm:amtStim_tar_dec}.}
Remember that for $T>0$ it holds that $R_\mathrm{max}>\Gamma \iff T<T_\mathrm{crit}$.
\begin{proof}
\textbf{Fixed points.} A fixed point in $E$ has $m=0$ and, for $p>0$, satisfies
$R(pG)=\Gamma$. By the assumed shape of $R$, there is no positive solution for $R_\mathrm{max}<\Gamma$ and
exactly one, $p=p_\mathrm{emp}$, for $R_\mathrm{max}>\Gamma$; it is admissible in $E$ exactly when
$M_\mathrm{tar}\ge M_\mathrm{emp}$. A fixed point outside $E$ must be
$z_\mathrm{tar}=(m_\mathrm{tar},p_\mathrm{tar})$ of \eqref{eqn:target_met}; it is admissible exactly when
$R_\mathrm{tar}>\Gamma$, i.e. when
$M_\mathrm{tar}<M_\mathrm{emp}$. This gives the stated fixed points.\newline

\textbf{Stability.} Since $a'(0)=R(0)=R_\mathrm{max}$, the nonzero eigenvalue at $z_0$ is
\[
  g+Ga'(0)=g+(1-g)\frac{R_\mathrm{max}}{\Gamma},
\]
so $z_0$ is stable for $R_\mathrm{max}<\Gamma$ and unstable for $R_\mathrm{max}>\Gamma$. Since $R$ is strictly decreasing,
$R'(M_\mathrm{emp})<0$, so by Lemma~\ref{lem:jacobian}(i) the positive $p$-axis state is stable
whenever it lies in the interior of $E$. The interior target-met state is stable
by Lemma~\ref{lem:jacobian}(ii).

\textbf{Global convergence.} For $R_\mathrm{max}<\Gamma$, one has
$R_\mathrm{tar}<\Gamma$, so Lemma~\ref{lem:entry}(i) sends every orbit into
$E_0$, after which Lemma~\ref{lem:axisconc}(a) gives convergence to $z_0$.\newline

Now let $R_\mathrm{max}>\Gamma$ and $z\ne z_0$. If $M_\mathrm{tar}>M_\mathrm{emp}$ then $R_\mathrm{tar}<\Gamma$ and Lem.~\ref{lem:entry}(i) sends every orbit into $E_0$, after which Lem.~\ref{lem:axisconc}(b) gives convergence to $z_2=(0,p_\mathrm{emp})$ (see Fig.~\ref{fig:amtStim_setTar_dec_phase}(b), orbits (1) and (2)).\newline

Finally, for $R_\mathrm{max}>\Gamma$, $z\ne z_0$ and $M_\mathrm{tar}<M_\mathrm{emp}$ there are two routes: Either the orbit stays in $E^c$, in which case Lem.~\ref{lem:entry}(ii) gives convergence to $z_2=z_\mathrm{tar}$ directly (see Fig.~\ref{fig:amtStim_setTar_dec_phase}(a), orbits (2)). If the orbit enters $E$, its next iterate is $(0,q)$ with $q>0$ ($q$ cannot be $0$ since $F^{-1}(\{z_0\})=\{z_0\}$). If $q\geq M_\mathrm{tar}/G$, it dominates $z^*$ and Lem.~\ref{lem:sandwich} applies directly. If $0<q<M_\mathrm{tar}/G$, Lem.~\ref{lem:axisconc}(c) implies that the orbit leaves $E_0$ after finitely many iterations, thereby producing a point $(0,\tilde q)$ with $\tilde q>M_\mathrm{tar}/G$. Lem.~\ref{lem:sandwich} then gives convergence to $z_\mathrm{tar}$ (see Fig.~\ref{fig:amtStim_setTar_dec_phase}(a), orbits (1)).\newline
Hence in both cases $B(z_2)=X\setminus\{z_0\}$.
\end{proof}

\paragraph{Dynamics on the $p$-axis: unimodal-efficiency case.}

As in the decreasing-efficiency case, only the scalar dynamics on the $p$-axis is specific to the response class; the
remaining ingredients are Lemmas~\ref{lem:monotone}--\ref{lem:sandwich}. We note for later use
that $M_\mathrm{tar}>M_\mathrm{sep}$ gives $a_\mathrm{tar}>a(M_\mathrm{sep})=p_\mathrm{sep}(1-g)$, hence
\begin{equation}\label{eq:pbar}
  p_\mathrm{tar}>p_\mathrm{sep}.
\end{equation}

\begin{lemma}[Dynamics on the $p$-axis, unimodal-efficiency]\label{lem:axis}
Let $q\in[0,M_\mathrm{tar}/G]$ and assume $0<T<T_\mathrm{crit}$ and $M_\mathrm{tar}>M_\mathrm{sep}$. While the orbit of $(0,q)$ remains in $E_0$, its first component
vanishes and its second evolves under the increasing scalar map $\Phi(q)=qg+a(qG)$, which satisfies
$$\Phi(q)-q=qG\left(R(qG)-\Gamma\right).$$
Consequently:
\begin{enumerate}
  \item[(a)] if $q<p_\mathrm{sep}$, the orbit remains in $E_0$, decreases strictly when $q>0$ and converges to
    $z_0$;
  \item[(b)] if $q>p_\mathrm{sep}$ and $M_\mathrm{tar}\ge M_\mathrm{emp}$, the orbit remains in $E_0$
    and converges to $z_2=(0,p_\mathrm{emp})$;
  \item[(c)] if $q>p_\mathrm{sep}$ and $M_\mathrm{tar}<M_\mathrm{emp}$, the orbit leaves $E_0$ after
    finitely many steps.
\end{enumerate}
\end{lemma}

\begin{proof}
For $(0,q)\in E_0$, $F(0,q)=(0,\Phi(q))$, and $\Phi$ is increasing because $a$ is increasing. The
identity above shows that the sign of $\Phi(q)-q$ is exactly the sign of $R(qG)-\Gamma$.
The assumed unimodal shape of $R$ makes this difference negative on $(0,p_\mathrm{sep})$,
positive on $(p_\mathrm{sep},p_\mathrm{emp})$ and negative on $(p_\mathrm{emp},\infty)$.
\begin{itemize}
  \item[(a)] If $0<q<p_\mathrm{sep}$ then $\Phi(q)<q$ and, since $\Phi$ is increasing with fixed
point $p_\mathrm{sep}$, also $\Phi(q)<p_\mathrm{sep}$. As $p_\mathrm{sep}<M_\mathrm{tar}/G$ the orbit
stays in $E_0$; being decreasing and bounded below it converges to a fixed point of $\Phi$ in
$[0,p_\mathrm{sep})$, necessarily $0$.
  \item[(b)] Here $M_\mathrm{tar}/G\ge p_\mathrm{emp}$, so $E_0\supseteq\{0\}\times[0,p_\mathrm{emp}]$,
and $E_0$ is forward invariant by Lemma~\ref{lem:entry}(i), whose hypothesis
$R_\mathrm{tar}\le\Gamma $ holds by \eqref{eq:window} since
$M_\mathrm{tar}\ge M_\mathrm{emp}$. The orbit is monotone by the sign of
$R(qG)-\Gamma$ (increasing if $q<p_\mathrm{emp}$, decreasing if $q>p_\mathrm{emp}$) and
bounded, hence converges to a fixed point of $\Phi$ in $(p_\mathrm{sep},M_\mathrm{tar}/G]$, which
must be $p_\mathrm{emp}$.
  \item[(c)] Now $M_\mathrm{tar}/G<p_\mathrm{emp}$, so $\Phi$ has no fixed point in
$(p_\mathrm{sep},M_\mathrm{tar}/G]$ and $\Phi(q)>q$ there. The orbit therefore increases strictly
while it remains in $E_0$; were it to remain in $E_0$ for all time it would converge to a fixed point
of $\Phi$ in that interval, a contradiction. Hence some iterate exceeds $M_\mathrm{tar}/G$.
\end{itemize}
\end{proof}

\paragraph{Additional statement about the shape of the separatrix.}
It holds that,
\[
  \Sigma=\bigcup_{k=1}^{N}\Sigma_k,
\]
where the $\Sigma_k$ are strictly decreasing arcs, consecutive arcs
share an endpoint, the first arc lies in $E$, and subsequent arcs are
obtained by successive inverse images under the affine target-met map.
The number of non-degenerate arcs is
\begin{equation}\label{eq:Ncount}
  N=\left\lceil
    \frac{\log\!\big(1-(1-g)\,p_\mathrm{sep}/a_\mathrm{tar}\big)}
         {\log g}
  \right\rceil.
\end{equation}
In particular, $N=1$ if and only if
\[
  M_\mathrm{tar}\ge M_c,
  \qquad
  M_c:=a^{-1}(p_\mathrm{sep}),
\]
where $a^{-1}$ denotes the inverse of the strictly increasing
stimulation function on its range, with
$a^{-1}(y)=+\infty$ when $y\ge a_\mathrm{max}$.

\paragraph{Proof of Thm.~\ref{thm:amtStim_tar_uni} and Additional Statement about Separatrix.}

\begin{proof}
\textbf{(i)} Let $z=(m,p)$ be a fixed point.\newline
If $z\in E$ then the first component of
\eqref{eq:Fmap} forces $m=0$, so $x=pG\le M_\mathrm{tar}$ and $p=pg+a(pG)$, i.e. $R(pG)=\Gamma$ for $p>0$; by the unimodal-efficiency assumption the solutions are $p\in\{0,p_\mathrm{sep},p_\mathrm{emp}\}$, admissible precisely when
$pG\le M_\mathrm{tar}$. This retains $p=0$ always, $p=p_\mathrm{sep}$ by the hypothesis
$M_\mathrm{sep}<M_\mathrm{tar}$, and $p=p_\mathrm{emp}$ exactly when $M_\mathrm{tar}\ge
M_\mathrm{emp}$.\newline
If instead $z\notin E$ then $F$ is affine near $z$, and \eqref{eqn:target_met} gives $z=z_\mathrm{tar}$; this is
admissible iff $m_\mathrm{tar}>0$, i.e.\ iff $R_\mathrm{tar}>\Gamma$, which by
\eqref{eq:window} means $M_\mathrm{tar}<M_\mathrm{emp}$. The two cases are complementary, so there
are exactly three fixed points, as claimed.\newline

For stability we apply Lemma~\ref{lem:jacobian}. At $z_0$ we have $a'(0)=0$, so the non-zero
eigenvalue is $g<1$. Since $R$ is unimodal with $R(M_\mathrm{sep})=R(M_\mathrm{emp})=\Gamma$ and
$M_\mathrm{sep}<M_\mathrm{emp}$, we have $R'(M_\mathrm{sep})>0$ and $R'(M_\mathrm{emp})<0$; by
Lemma~\ref{lem:jacobian}(i), $z_1=(0,p_\mathrm{sep})$ is therefore unstable, and $(0,p_\mathrm{emp})$
is stable whenever it lies in the interior of $E$ (i.e.\ for $M_\mathrm{tar}>M_\mathrm{emp}$);
and $z_2=(m_\mathrm{tar},p_\mathrm{tar})$ is stable by Lemma~\ref{lem:jacobian}(ii).

\smallskip
\textbf{(ii)} Let $z\in X$ and apply Lemma~\ref{lem:entry}.\newline

\emph{Case $R_\mathrm{tar}<\Gamma$.} In this case Lem.~\ref{lem:entry}(i) shows that every orbit will enter $E$ in finite time. Then, at the next iteration the orbit enters $E_0$ at some point $(0,q)$ with $q\leq M_\mathrm{tar}/G$. Now Lem.~\ref{lem:axis}(a) and (b) assert convergence to $z_0$ if $q<p_\mathrm{sep}$ (see Fig.~\ref{fig:amtStim_setTar_uni_phase}(b), orbit (1)), or convergence to $z_2=(0,p_\mathrm{emp})$ for $q>p_\mathrm{sep}$ (see Fig.~\ref{fig:amtStim_setTar_uni_phase}(b), orbits (2) and (3)). For $q=p_\mathrm{sep}$ the orbit will be fixed at $z_1=(0,p_\mathrm{sep})$.\newline

\emph{Case $R_\mathrm{tar}>\Gamma$.} We follow analogous arguments as in the proof of Thm.~\ref{thm:amtStim_tar_dec}: Suppose $z\neq z_0$ and the orbit never enters $E$. Then Lem.~\ref{lem:entry} shows convergence to $z_2=z_\mathrm{tar}\in E^c$ (see Fig.~\ref{fig:amtStim_setTar_uni_phase}(a), orbit (3)). 
If the orbit enters $E$, its next iterate is $(0,q)$ with $q>0$ (since $F^{-1}(\{z_0\})=\{z_0\}$). If $q\geq M_\mathrm{tar}/G$, it dominates $z^*$ and Lemma~\ref{lem:sandwich} applies directly. If $0<q<M_\mathrm{tar}/G$, Lemma~\ref{lem:axis}(a) and (c) imply that for $q<p_\mathrm{sep}$, the orbit converges to $z_0$ (see Fig.~\ref{fig:amtStim_setTar_uni_phase}(a), orbit (1)), and for $q>p_\mathrm{sep}$ the orbit will leave $E_0$ after finitely many iterations, thereby producing a point $(0,\tilde q)$ with $\tilde q>M_\mathrm{tar}/G$. Lemma~\ref{lem:sandwich} then gives convergence to $z_\mathrm{tar}$ (see Fig.~\ref{fig:amtStim_setTar_uni_phase}(a), orbit (2)). Finally, for $q=p_\mathrm{sep}$ the orbit will be fixed at $z_1=(0,p_\mathrm{sep})$.\newline

Hence every orbit converges to $z_0$, $z_1$ or $z_2$. Let
$$
\Sigma:=\{z\in X : F^n(z)\rightarrow z_1\},
$$
then $X=B(z_0)\ \sqcup\ \Sigma\ \sqcup\ B(z_2)$.\newline

The preceding case distinction also shows that every orbit in $\Sigma$ reaches $z_1$ in finitely many iterations. By definition $\Sigma=W^s(z_1)$. The sets $B(z_0)$ and $B(z_2)$ are open because $z_0$ and $z_2$ are asymptotically stable and $F$ is continuous. $B(z_0)$ is a lower set: if
$z\le w\in B(z_0)$ then $0\le F^n(z)\le F^n(w)\to z_0$, using Lemma~\ref{lem:monotone}. $B(z_2)$ is an upper set: if
$z\ge w\in B(z_2)$ then $F^n(z)\ge F^n(w)\to z_2$, and since $F^n(z)$ converges to one of the
three fixed points, its limit is $\ge z_2$ and hence equals $z_2$.

\smallskip
\textbf{(iii)} 
By the proof of part~(ii), $z\in\Sigma$ if and only if the orbit
of $z$ enters $E$ for the first time at some step $k-1\ge0$ with
entry value $p_\mathrm{sep}$. We define the \emph{entry map} by
\begin{equation}
  \Psi(m,p)=p\,g+a\big(x(m,p)\big).
\end{equation}
Points with $k=1$ are those in $E$ with
$\Psi(z)=p_\mathrm{sep}$.
Hence
\[
  \Sigma_1=\{z\in E:\ \Psi(z)=p_\mathrm{sep}\}.
\]
Solving $\Psi(m,p)=pg+a(x)=p_\mathrm{sep}$ for $p$ in terms of $x$ and substituting into $x=md+pG$ gives
\begin{equation}\label{eq:arc}
  p(x)=\frac{p_\mathrm{sep}-a(x)}{g},
  \qquad
  m(x)=\frac{1}{d}\left(
    x-\frac{G}{g}\big(p_\mathrm{sep}-a(x)\big)
  \right),
\end{equation}
for $x\in \big[M_\mathrm{sep},\,\min\{M_\mathrm{tar},a^{-1}(p_\mathrm{sep})\}\big]$.\newline
At $x=M_\mathrm{sep}$ one gets $(p,m)=(p_\mathrm{sep},0)$ so $\Sigma_1$ starts at $z_1$. Differentiating,
\[
  p'(x)=-\frac{a'(x)}{g}<0,
  \qquad
  m'(x)=\frac{1}{d}
  \left(1+\frac{G}{g}a'(x)\right)>0,
\]
so $\Sigma_1$ is a strictly decreasing arc. The admissible range
of $x$ is bounded above by $M_\mathrm{tar}$, to remain in $E$, and by
$a^{-1}(p_\mathrm{sep})$, to keep $p\ge0$.\newline

Points whose first entry occurs after further affine steps are obtained
by pulling the preceding arc back through $F$. For the geometric arc
construction we include the boundary $x=M_\mathrm{tar}$ so that
consecutive arcs share endpoints; boundary points themselves are
already contained in the earlier arc. Inverting \eqref{eq:affine} gives
\begin{equation}\label{eq:pullback}
  p=\frac{p'-a_\mathrm{tar}}{g},
  \qquad
  m=\frac{1}{d}\Big(m'+M_\mathrm{tar}-pG\Big).
\end{equation}
The constraint $p\ge0$ is $p'\ge a_\mathrm{tar}$. For such a preimage,
\[
  x=md+pG=m'+M_\mathrm{tar}.
\]
Hence it lies strictly outside $E$ whenever $m'>0$, and on
$\partial E$ when $m'=0$. The latter case is precisely the
shared-endpoint convention just described.

The map \eqref{eq:pullback} is increasing in $p'$ and, along an arc
traversed with $p'$ decreasing and $m'$ increasing, produces $p$
decreasing and $m$ increasing. Hence every $\Sigma_k$ is a strictly
decreasing arc.

Let
\[
  \pi_k:=\max\{p:(m,p)\in\Sigma_k\}.
\]
Since $\Sigma_{k+1}$ is obtained by pulling back $\Sigma_k$ through
$F$, its maximal-$p$ endpoint is the inverse image of the maximal-$p$
endpoint of $\Sigma_k$. Thus
\[
  \pi_1=p_\mathrm{sep},
  \qquad
  \pi_{k+1}=\frac{\pi_k-a_\mathrm{tar}}{g},
\]
and therefore
\[
  \pi_k
  =
  p_\mathrm{tar}
  -(p_\mathrm{tar}-p_\mathrm{sep})g^{-(k-1)}.
\]
The corresponding inverse-image point is the common endpoint of
$\Sigma_k$ and $\Sigma_{k+1}$. The pullback can be continued as a
non-degenerate arc precisely while $\pi_k>a_\mathrm{tar}$.\newline

Using the expression for $\pi_k$, this condition is equivalent to
\[
  g^k>
  1-\frac{(1-g)p_\mathrm{sep}}{a_\mathrm{tar}}.
\]
By \eqref{eq:pbar}, the right-hand side lies in $(0,1)$, so this
inequality fails for all sufficiently large $k$ and the chain of arcs
is finite. Consequently,
\[
  N=
  \left\lceil
    \frac{\log\!\big(1-(1-g)p_\mathrm{sep}/a_\mathrm{tar}\big)}
         {\log g}
  \right\rceil,
\]
as stated in \eqref{eq:Ncount}. If the logarithmic quotient is an
integer, equality corresponds to a degenerate pullback consisting only
of the terminal endpoint already contained in $\Sigma_N$, so the
ceiling formula still counts the non-degenerate arcs correctly.

Finally, $N=1$ is equivalent to
\[
  g\le
  1-\frac{(1-g)p_\mathrm{sep}}{a_\mathrm{tar}},
\]
or equivalently $a_\mathrm{tar}\ge p_\mathrm{sep}$.
Since $a$ is strictly increasing, this is equivalent to
\[
  M_\mathrm{tar}\ge a^{-1}(p_\mathrm{sep})=M_c.
\]
Geometrically, this means that $\Sigma_1$ reaches the $m$-axis before
leaving $E$.

Each $\Sigma_k$ is a continuous, strictly decreasing arc, consecutive
arcs meet only at their common endpoints, and the final arc terminates
on the $m$-axis. Hence their union is the graph of a continuous, strictly decreasing
function
\[
    \sigma:[0,\bar m_\mathrm{sep}]\to[0,p_\mathrm{sep}],
\]
with $\sigma(0)=p_\mathrm{sep}$ and
$\sigma(\bar m_\mathrm{sep})=0$.
\end{proof}

\begin{remark}\label{rem:junction}
The junction $\Sigma_1\cap\Sigma_2$ lies on $\partial E$, at $x=M_\mathrm{tar}$, whereas every
later junction lies strictly outside $E$. Accordingly $\Sigma_1$ is the only arc whose extent is
constrained by the geometry of $E$; the remaining arcs are determined by \eqref{eq:pullback}
alone.
\end{remark}

The branch count $N$ has a natural reading in terms of how far back the history matters. When $N=1$, one session suffices to decide which side of the unstable production level the trajectory enters. When $N\ge 2$, the target caps the stimulation below $p_\mathrm{sep}$, and the outcome can depend on the accumulated effect of several successive sessions. The criterion $M_\mathrm{tar}\ge M_c$ is exactly the statement that one maximal session is sufficient.

\paragraph{Proof of Cor.~\ref{cor:regime1}}

\begin{proof}
In both cases $R(M)<\Gamma$ for $0<M\le M_\mathrm{tar}$: if $T>T_\mathrm{crit}$ because
$R\le R_\mathrm{max}<\Gamma$, and if $M_\mathrm{tar}<M_\mathrm{sep}$ by \eqref{eq:window}.
Hence there is no fixed point $(0,p)$ with $0<pG\le M_\mathrm{tar}$ and, since
$R_\mathrm{tar}<\Gamma$, no target-met fixed point, so $z_0$ is the unique fixed point;
it is locally asymptotically stable by Lemma~\ref{lem:jacobian}(i), since $a'(0)=0$.
By Lemma~\ref{lem:entry}(i) every orbit enters $E$ and thereafter remains in $E_0$, where
$\Phi(q)-q=qG\bigl(R(qG)-\Gamma\bigr)<0$ for $q>0$; hence every orbit decreases to $0$.
\end{proof}

\subsection{Amount-dependent stimulation, set-duration protocol: Auxiliary results and proofs}
\label{app:amount_stim_duration}

Throughout this subsection we assume $0<\tau<\tau_\mathrm{emp}$, and hence
$0<h<1$. The map $F:X\rightarrow X$ is given by \eqref{eq:Fdur},
\begin{equation*}
  F(m,p)=\Big(x\,h,\;\;p\,g+a\big(x(1-h)\big)\Big),\quad x=md+pG.
\end{equation*}

\paragraph{Auxiliary results.}

We start by noting that $F$ is globally injective, and its inverse on the image $F(X)$ is explicit:
\begin{equation}\label{eq:Finv}
  x=\frac{m'}{h},\qquad p=\frac{p'-a\big(x(1-h)\big)}{g},\qquad m=\frac{x-pG}{d}.
\end{equation}

\begin{lemma}\label{lem:durbasic}
The map \eqref{eq:Fdur} is order-preserving on $X$. Moreover, if $z\le\tilde z$ componentwise and $z\neq\tilde z$, then both components
of $F(z)$ are strictly smaller than the corresponding components of
$F(\tilde z)$.
The map is injective, with inverse \eqref{eq:Finv} defined on $F(X)$,
and every orbit is bounded, with
\[
  \limsup_{n\to\infty}p_n\le\frac{a_\mathrm{max}}{1-g},
  \qquad
  \limsup_{n\to\infty}m_n\le\frac{a_\mathrm{max}\,h}{\Gamma(1-dh)}.
\]
Moreover, for every $z\in X\setminus\{0\}$ the image $F(z)$ has both coordinates strictly
positive.
\end{lemma}

\begin{proof}
If $z\le\tilde z$ and $z\neq\tilde z$, then
$x(z)<x(\tilde z)$ since $d,G>0$. Hence the first components are
strictly ordered; the second components are also strictly ordered
because $a$ is strictly increasing. Inverting \eqref{eq:Fdur} in the order $x$, then $p$, then $m$ gives \eqref{eq:Finv}.
For the bounds, $p_{n+1}\le p_ng+a_\mathrm{max}$ gives the first, and
$m_{n+1}\le m_ndh+p_nGh$ then gives the second on iterating. Finally $z\ne0$ gives $x(z)>0$, hence
$m'=x h>0$ and $p'\ge a(x(1-h))>0$.
\end{proof}

\begin{lemma}[Global convergence]\label{lem:durconv}
Every orbit of the set-duration map $F$ converges to a fixed point.
\end{lemma}

\begin{proof}
By Lem.~\ref{lem:durbasic}, every orbit is bounded and $F$ is
injective. Moreover, $X=\mathbb{R}_+^2$ is convex,
$F$ is $C^1$, and $DF(z)$ is entrywise nonnegative for every
$z\in X$, with
\[
  \det DF(z)=dhg>0.
\]
Thus $F$ satisfies the cooperative analogue of the hypotheses
$(H+)$ in \cite{Smith1998}. By
\citet[Corollary~4.4]{Smith1998}, every orbit with compact closure
converges to a fixed point. Since $X$ is closed and every orbit is
bounded, its closure is compact and contained in $X$. Hence every
orbit converges to a fixed point.
\end{proof}

%---------

\begin{lemma}[Linearisation]\label{lem:durjac}
For $z=(m,p)\in X$, set
\[
  M=(1-h)x(z).
\]
Then
\[
DF(z)=\begin{pmatrix}
dh & Gh\\
a'(M)\,d(1-h) & g+a'(M)\,G(1-h)
\end{pmatrix},\qquad \det DF(z)=dhg.
\]
At a fixed point with production rate $p_*$ and removed amount
$M_*=Gr p_*$, the eigenvalues of $DF$ are real and positive, with product
$dhg\in(0,1)$. Moreover,
\begin{equation}\label{eq:durdetI}
\det(I-DF)
=
G(1-h)\bigl(\Gamma_\tau-a'(M_*)\bigr).
\end{equation}

\begin{enumerate}
\item[(i)] At the zero fixed point $z_0=(0,0)$, the state is locally
asymptotically stable if
\[
a'(0)<\Gamma_\tau,
\]
and unstable if $a'(0)>\Gamma_\tau$.

\item[(ii)] At a positive fixed point, where
$R(M_*)=\Gamma_\tau$,
\begin{equation}\label{eq:durdetR}
\det(I-DF)
=
-G(1-h)M_*R'(M_*).
\end{equation}
Hence the fixed point is locally asymptotically stable if
$R'(M_*)<0$, whereas if $R'(M_*)>0$ it is a hyperbolic saddle.

\end{enumerate}
In the equality cases $a'(0)=\Gamma_\tau$ or $R'(M_*)=0$, one
eigenvalue equals $1$.
\end{lemma}

\begin{proof}
The Jacobian and its determinant follow by direct computation. Its
characteristic discriminant is
\[
\bigl(dh-g-a'(M_*)G(1-h)\bigr)^2
+4Gh\,a'(M_*)d(1-h)\ge0,
\]
so the eigenvalues are real. Their product is $dhg>0$ and their sum
is positive, hence both eigenvalues are positive. Since $\det(I-DF)=1-\operatorname{tr}DF+\det DF$,
we obtain
\[
\begin{aligned}
\det(I-DF)
&=(1-dh)(1-g)-a'(M_*)G(1-h)\\
&=G(1-h)\bigl(\Gamma_\tau-a'(M_*)\bigr),
\end{aligned}
\]
where we used $\Gamma_\tau=(1-g)(Gr)$.

\medskip
Because the two eigenvalues are positive and have product
$dhg<1$, they both lie in $(0,1)$ if and only if
$\det(I-DF)>0$. This gives part~(i).

\medskip
At a positive fixed point, $R(M_*)=\Gamma_\tau$, and
\[
a'(M_*)=R(M_*)+M_*R'(M_*),
\]
so
\[
\det(I-DF)
=
-G(1-h)M_*R'(M_*).
\]
Thus $R'(M_*)<0$ implies that both eigenvalues lie in $(0,1)$.
If $R'(M_*)>0$, then $\det(I-DF)<0$, so one eigenvalue lies below
$1$ and the other above $1$; hence the fixed point is a hyperbolic
saddle.

If $\det(I-DF)=0$, one eigenvalue is $1$.
\end{proof}

\paragraph{Proof of Thm.~\ref{thm:durconc}.}

\begin{proof}

\textbf{Fixed points and stability.} A positive fixed point is equivalent to a solution of
$R(M)=\Gamma_\tau$. For decreasing $R$ there is no positive solution when
$R_\mathrm{max}<\Gamma_\tau$ and exactly one, $M_\mathrm{lac}$, when $R_\mathrm{max}>\Gamma_\tau$. Since
$a'(0)=R_\mathrm{max}$, Lemma~\ref{lem:durjac} shows that $z_0$ is stable for
$R_\mathrm{max}<\Gamma_\tau$ and unstable for $R_\mathrm{max}>\Gamma_\tau$. At the positive root,
$a'(M_\mathrm{lac})-\Gamma_\tau=M_\mathrm{lac}R'(M_\mathrm{lac})<0$, so Lemma~\ref{lem:durjac}
gives stability of $z_2$.

\textbf{(i)} Since $z_0$ is the unique fixed point,
Lemma~\ref{lem:durconv} implies that every orbit converges to $z_0$.

\textbf{(ii)} Since $a'(0)=R_\mathrm{max}>0$ and $h<1$, all four entries of $DF(0)$ are
positive, so $DF(0)$ has a strictly positive eigenvector $v$ for its spectral radius $\rho$, and
$\rho>1$ because $z_0$ is unstable. As $F$ is $C^1$ with $F(0)=0$,
\[
  F(\varepsilon v)=\varepsilon\rho v+o(\varepsilon)
  \qquad(\varepsilon\downarrow0),
\]
and since $v>0$ strictly and $\rho>1$, there is $\varepsilon_0>0$ with $F(\varepsilon v)\ge
\varepsilon v$ for all $\varepsilon\in(0,\varepsilon_0]$. For such $\varepsilon$ the orbit of
$\varepsilon v$ is nondecreasing and bounded, hence converges to a fixed point $\ge\varepsilon
v>0$, necessarily $z_2$.

Now let $z\in X\setminus\{0\}$. By Lem.~\ref{lem:durbasic},
$F(z)$ has both coordinates strictly positive, so
$F(z)\ge\varepsilon v$ for some
$\varepsilon\in(0,\varepsilon_0]$. Hence
\[
  F^{n+1}(z)\ge F^n(\varepsilon v)\longrightarrow z_2.
\]
By Lem.~\ref{lem:durconv}, the orbit of $z$ converges to a fixed
point. Since the only fixed points are $z_0$ and $z_2$, the above
inequality excludes convergence to $z_0$. Thus
$F^n(z)\to z_2$. Hence $B(z_2)=X\setminus\{z_0\}$.
\end{proof}

\paragraph{Proof of Thm.~\ref{thm:dursig}.}

\begin{proof}
  
Under the hypotheses \(0<T<T_{\mathrm{crit}}\) and \(\tau>\tau_{\mathrm{min}}(T)\), one has $R_\mathrm{max}>\Gamma_\tau$.

\textbf{(i)}
The positive fixed points are in one-to-one correspondence with the
positive solutions of $R(M)=\Gamma_\tau$. Since $R_\mathrm{max}>\Gamma_\tau$ and $R$ is unimodal, these are precisely $M_\mathrm{sep}$ and $M_\mathrm{lac}$. Together with $z_0$, this gives exactly the three fixed points stated above.

Since $R(0)=a'(0)=0$, the eigenvalues at $z_0$ are $dh$ and $g$, so
$z_0$ is locally asymptotically stable.
Since $R'(M_\mathrm{sep})>0$, Lem.~\ref{lem:durjac}(ii) shows that
$z_1$ is a hyperbolic saddle, whereas $R'(M_\mathrm{lac})<0$ implies
that $z_2$ is locally asymptotically stable.

\smallskip
\textbf{(ii)}
By Lem.~\ref{lem:durconv}, every orbit converges to a fixed point.
Since $z_0,z_1,z_2$ are the only fixed points,
\[
  X=B(z_0)\sqcup W^s(z_1)\sqcup B(z_2).
\]

The basins $B(z_0)$ and $B(z_2)$ are open because $z_0$ and $z_2$
are locally asymptotically stable. If $z\le w\in B(z_0)$, then
order preservation gives
\[
  0\le F^n(z)\le F^n(w)\longrightarrow z_0,
\]
hence $z\in B(z_0)$. Thus $B(z_0)$ is a lower set. Similarly, if $z\ge w\in B(z_2)$, then $F^n(z)$ converges to one of the three fixed points, while
\[
  F^n(z)\ge F^n(w)\longrightarrow z_2.
\]
Since no fixed point lies strictly above $z_2$, the limit must be
$z_2$. Hence $B(z_2)$ is an upper set.

\smallskip
\textbf{(iii)}
Let $0<\lambda_s<1<\lambda_u$ denote the eigenvalues of
$DF(z_1)$. Since $\lambda_s\lambda_u=dhg$,
\[
  \lambda_s=\frac{dhg}{\lambda_u}<dhg<dh.
\]
By the stable manifold theorem, $z_1$ has a one-dimensional local
stable manifold $W^s_{\mathrm{loc}}(z_1)$. An eigenvector associated with $\lambda_s$ may be chosen as
\[
  (v,w)=(Gh,\lambda_s-dh).
\]
Its slope is negative, since $\lambda_s<dh$. Hence, near $z_1$,
$W^s_{\mathrm{loc}}(z_1)$ is a strictly decreasing $C^1$ curve.

Since $F$ is injective and $\det DF=dhg>0$, it is a $C^1$
diffeomorphism onto its image. By the global stable-manifold
construction, $W^s(z_1)$ is obtained by backward continuation of $W^s_{\mathrm{loc}}(z_1)$ and is a connected one-dimensional curve.

Next we show that this curve cannot fold in either coordinate
direction. Suppose that $\Sigma$ contained two distinct points
\[
  z=(m_1,p_1),
  \qquad
  \widetilde z=(m_2,p_2)
\]
with $m_1\le m_2$ and $p_1\le p_2$. By Lem.~\ref{lem:durbasic}, both components of $F(z)$ are then strictly smaller than the corresponding components of
$F(\widetilde z)$. Since $F$ is injective, the two iterates remain
distinct, and repeated application of Lem.~\ref{lem:durbasic} gives
the same strict componentwise ordering for every positive iterate. On the other hand, both $z$ and $\widetilde z$ belong to
$W^s(z_1)$, so
\[
  F^n(z)\longrightarrow z_1,
  \qquad
  F^n(\widetilde z)\longrightarrow z_1.
\]
For all sufficiently large $n$, both iterates therefore lie on
$W^s_{\mathrm{loc}}(z_1)$. This is impossible, since the local stable
manifold is a strictly decreasing curve and hence cannot contain two
distinct points whose two coordinates are ordered in the same
direction.

Consequently, every vertical line intersects $\Sigma$ in at most one
point. Let
\[
  I_\Sigma
  :=
  \{m\ge0:\text{ there exists }p\ge0
      \text{ with }(m,p)\in\Sigma\}.
\]
Since $\Sigma$ is connected, $I_\Sigma$ is an interval, and there is
a function $\sigma:I_\Sigma\to[0,\infty)$ such that
\[
  \Sigma=\{(m,\sigma(m)):m\in I_\Sigma\}.
\]
If $m_1<m_2$ are in $I_\Sigma$, the argument above rules out
$\sigma(m_1)\le\sigma(m_2)$. Hence
\[
  \sigma(m_1)>\sigma(m_2),
\]
so $\sigma$ is strictly decreasing. Finally, connectedness of
$\Sigma$ excludes jump discontinuities of the monotone function
$\sigma$, and therefore $\sigma$ is continuous. Thus $\Sigma$ is the graph of a continuous, strictly decreasing
function, as claimed.
\end{proof}

\paragraph{Proof of Cor.~\ref{cor:dursig}.}

\begin{proof}
  Under either hypothesis of Cor.~\ref{cor:dursig}, one has
$\Gamma_\tau>R_\mathrm{max}$, equivalently $R_\mathrm{max}<\Gamma_\tau$.
Hence there is no positive solution of
$R(M)=\Gamma_\tau$. Hence $z_0$ is the unique fixed point, and Lem.~\ref{lem:durjac}(i) gives local asymptotic stability, while Lem.~\ref{lem:durconv} gives convergence of every orbit to $z_0$.
\end{proof}

\subsection{Duration-dependent stimulation, set-duration protocol: Auxiliary results and proofs}
\label{app:dur_stim_duration}

\paragraph{Proof of Thm.~\ref{thm:taustim}}

\begin{proof}
The linear part $L$ in \eqref{eq:taustimaffine} is upper triangular with diagonal entries $dh$ and
$g$, both in $(0,1)$, so its spectral radius is $\max\{dh,g\}<1$ and $I-L$ is invertible. Hence
$F$ has the unique fixed point $z_\mathrm{lac}=(I-L)^{-1}b$, and $z_n-z_\mathrm{lac}=L^n(z_0-
z_\mathrm{lac})\to0$ for every $z_0$, which is global asymptotic stability. Solving $(I-L)z=b$
componentwise gives $p_\mathrm{lac}(1-g)=a(\tau)$ and $m_\mathrm{lac}(1-dh)=p_\mathrm{lac}Gh$, that is
\eqref{eq:taustimfp}; the removed amount is $M_\mathrm{lac}=x(z_\mathrm{lac})(1-h)=G r
p_\mathrm{lac}$. Rewriting $L^n$ we can see the convergence for $dh\neq g$ more explicitly from
\begin{equation}\label{eq:taustimsol}
  p_n=g^n\big(p_0-p_\mathrm{lac}\big)+p_\mathrm{lac},
  \qquad
  m_n=(dh)^n\big(m_0-m_\mathrm{lac}\big)
      +Gh\,\frac{(dh)^n-g^n}{dh-g}
       \big(p_0-p_\mathrm{lac}\big)+m_\mathrm{lac}.
\end{equation}
The case $dh=g$ follows by taking the limit $dh\to g$.
In either case the convergence is exponential. Finally $F(0,0)=(0,a(\tau))\neq(0,0)$, so the origin is not fixed.
\end{proof}

%\input{app_dur_stim_set_dur_decay.tex}

%-------------------
\subsection{Duration-dependent stimulation, set-target protocol: Auxiliary results and proofs}
\label{app:dur_stim_target}

\paragraph{Proof of Thm.~\ref{thm:durstim_target}.}

\begin{proof}

\textbf{Existence of fixed points.} We have already established in Sec.~\ref{sec:dur_stim_target} that for $0<M_{\mathrm{tar}}<M_{\mathrm{emp}}(T)$ there exists a unique target-met fixed point as defined in \eqref{eqn:durstim_target_fp} with $\tau$ the unique solution to \eqref{eq:durstim_target_fp_tau}. We have also shown that if $M_{\mathrm{tar}}>M_{\mathrm{emp}}(T)$, there exists a unique emptying fixed point given by \eqref{eq:durstim_target_empty}. At $M_{\mathrm{tar}}=M_{\mathrm{emp}}(T)$ the two branches meet. Furthermore, $F(0,0)=(0,a(\tau_\mathrm{emp}))$, so the origin is not a fixed point.

\medskip
\textbf{Stability of the target-met fixed point.}
To determine local stability, we apply the Jury conditions \cite{Elaydi2005,Jury1974} for a
two-dimensional discrete map. For the Jacobian
\[
J=
\begin{pmatrix}
d & G\\
-cd & g-cG
\end{pmatrix},
\qquad
c:=\frac{\tau a'(\tau)}{x_\mathrm{tar}}>0,
\]
we have
\[
\det J=dg,
\qquad
\mathrm{Tr} J=d+g-cG.
\]
Local asymptotic stability requires
\[
1-\det J>0,\qquad
1-\mathrm{Tr} J+\det J>0,\qquad
1+\mathrm{Tr} J+\det J>0.
\]
The first two conditions are automatically satisfied, since
\[
1-\det J=1-dg>0, \qquad 1-\mathrm{Tr} J+\det J=(1-d)(1-g)+cG>0.
\]
Hence stability is determined entirely by the remaining condition
\[
1+\mathrm{Tr} J+\det J=(1+d)(1+g)-cG>0.
\]
The second Jury condition excludes a stability loss through \(\lambda=+1\), while \(\det J=dg<1\) excludes a complex-conjugate pair crossing the unit circle. Hence the only possible local stability loss is through \(\lambda=-1\).\newline

Using the definitions of $E_a$ and $E_\mathrm{crit}$ in \eqref{eq:durstim_elasticity_Ecrit}
and using \eqref{eqn:durstim_target_fp}, we can write
\[
    cG=(1-g)(1-dh)E_a(\tau),
\]
so that the remaining Jury condition is equivalent to
\[
    E_a(\tau)<E_{\mathrm{crit}}(T,\tau).
\]
At equality, one eigenvalue is $\lambda=-1$ and the other is
$\lambda=-dg\in(-1,0)$.

\medskip

\noindent\textbf{Stability of the emptying fixed points.}
For $M_{\mathrm{tar}}>M_{\mathrm{emp}}(T)$ the emptied fixed point lies
strictly inside the emptying region, where
\[
    DF=
    \begin{pmatrix}
        0&0\\
        0&g
    \end{pmatrix}.
\]
Its eigenvalues are $0$ and $g\in(0,1)$, so it is locally
asymptotically stable.
\end{proof}

\paragraph{Proof of Cor.~\ref{cor:durstim_target_instability}.}

\begin{proof}
Since $a(\tau)=\tau R_d(\tau)$, we have
\[
    E_a(\tau)
    =
    1+\frac{\tau R_d'(\tau)}{R_d(\tau)}.
\]
Because $E_{\mathrm{crit}}(T,\tau)>1$, any $\tau$ with
$R_d'(\tau)\le0$ gives
$E_a(\tau)\le1<E_{\mathrm{crit}}(T,\tau)$ and therefore a stable
target-met fixed point.

Conversely, suppose $R_d'(\tau_0)>0$ for some
$\tau_0\in(0,\tau_\mathrm{emp})$. Then $E_a(\tau_0)>1$. For fixed $\tau_0$,
$E_{\mathrm{crit}}(T,\tau_0)$ decreases continuously and strictly from
$\infty$ to $1$ as $T$ increases from $0$ to $\infty$. Hence there is a
unique $T_{\mathrm{thresh}}(\tau_0)$ such that
\[
    E_{\mathrm{crit}}(T_{\mathrm{thresh}},\tau_0)=E_a(\tau_0),
\]
and the corresponding target-met fixed point is unstable for
$T>T_{\mathrm{thresh}}(\tau_0)$. Choosing
\[
    M_{\mathrm{tar}}
    =
    \frac{r(T,\tau_0)a(\tau_0)}{\Gamma(T)}
\]
produces the required set-target schedule.
\end{proof}

\paragraph{Explicit instability boundary for the Hill response.}

For
\[
    a(\tau)
    =
    a_{\mathrm{max}}
    \frac{\tau^2}{\tau^2+\mu_\tau^2},
\]
one has
\[
    E_a(\tau)
    =
    \frac{2\mu_\tau^2}{\tau^2+\mu_\tau^2}.
\]
Substitution into
$E_a(\tau)=E_{\mathrm{crit}}(T,\tau)$, using
$h=1-\tau/\tau_\mathrm{emp}$, gives after rearrangement
\begin{equation}
\label{eqn:durstim_hill_flip}
    (1+d)(1+g)\tau^2
    -2\mu_\tau^2d(1-g)\tau/\tau_\mathrm{emp}
    +\mu_\tau^2(-1+3d+3g-dg)
    =0.
\end{equation}

\paragraph{Small target approximation}

We derived the small target limit as follows: 1. the exact target-met recurrence for $x_n$ is
$$
x_{n+1}=(d+g)x_n-d\,gx_{n-1}-d(1-g)M_\mathrm{tar}+Ga\left(\frac{M_\mathrm{tar}}{x_n}\tau_\mathrm{emp}\right).
$$
Rescaling $x_n=M_\mathrm{tar}^{2/3}z_n$ and letting $M_\mathrm{tar}\rightarrow 0$ gives
$$
z_{n+1}=(d+g)z_n-d\,gz_{n-1}+G a_\mathrm{max}\left(\frac{\tau_\mathrm{emp}}{\mu_\tau z_n}\right)^2.
$$
After $z_n=y_n\left[G a_\mathrm{max}\left(\frac{\tau_\mathrm{emp}}{\mu_\tau}\right)^2\right]^{1/3}$,
the leading order map is simply
$$
y_{n+1}=(d+g)y_n-d\,gy_{n-1}+\frac{1}{y_n^2}.
$$
For sufficiently long wait times $dg=e^{-(\delta+\gamma)T}$ is small, hence the limiting recurrence is close to
$$
y_{n+1}=(d+g)y_n+\frac{1}{y_n^2},
$$
which with $q_n=(d+g)^{-1/3}y_n^{-1}$ is equivalent to
$$
q_{n+1}=\frac{\lambda q_n}{1+q_n^3},\qquad \lambda=\frac{1}{d+g}.
$$
This is the Maynard Smith-Slatkin population model \cite{Smith1973}, which has the same structure as the right-hand side of the logistic map $q_{n+1}=r q_n(1-q_n)$: A unimodal shape with unique maximum. Thus, in the combined small-target and strongly dissipative regime, the lactation map approaches a one-dimensional smooth unimodal map with a quadratic critical point, providing a natural explanation for the numerically observed period-doubling cascade and apparent chaotic behaviour.

\subsection{Protocol Comparison: Auxiliary results and proofs}
\label{app:protocol_comparison}

\paragraph{Proof of Lem.~\ref{lem:Mday_balance}}

\begin{proof}
Let \(h_*=h(\tau_*)\) and let \(x_*\) denote the amount available
immediately before removal. At a steady state,
\[
M_*=(1-h_*)x_*,
\qquad
m_*=h_*x_*.
\]
Using \(x_*=dm_*+Gp_*\) gives
\[
Gp_*
=
x_*(1-dh_*)
=
M_*\frac{1-dh_*}{1-h_*}
=
\frac{M_*}{r(T,\tau_*)}.
\]
The steady-state production balance is
\[
(1-g)p_*=a_*.
\]
Since \(\Gamma(T)=(1-g)/G\), combining the two relations yields
\[
\frac{a_*}{M_*}
=
\frac{\Gamma(T)}{r(T,\tau_*)}.
\]
At complete emptying \(h_*=0\) and hence \(r(T,\tau_*)=1\).
\end{proof}

\subsection{Daily Milk Output: Auxiliary results and proofs}
\label{app:daily_output}

%-------------------
The following calculations are used in the below proofs. For $r$ as defined in \eqref{eqn:r}, we have before emptying
\begin{equation}
     \label{eqn:partial_tau_r}
     \partial_\tau r(T,\tau)=-h'(\tau)\frac{(1-d(T))}{(1-d(T)h(\tau))^2}>0     
\end{equation}

\begin{equation}
     \label{eqn:partial_T_r}
     \partial_T r(T,\tau)=d'(T)\frac{h(\tau)(1-h(\tau))}{(1-d(T)h(\tau))^2}<0
\end{equation}

\paragraph{Proof of Lem.~\ref{lem:sessionOutput_monotonicity}.}

\begin{proof}
Recall that \(\Gamma(T)\) is strictly increasing in \(T\).

\medskip
For set-duration removal with amount-dependent stimulation, the
steady-state balance is
\[
R(M_*)=\frac{\Gamma(T)}{r(T,\tau)}.
\]
Along a stable positive branch \(R'(M_*)<0\). Using \eqref{eqn:partial_tau_r} and \eqref{eqn:partial_T_r}, one has
\[
\partial_T\!\left(\frac{\Gamma}{r}\right)>0,
\qquad
\partial_\tau\!\left(\frac{\Gamma}{r}\right)<0,
\]
implicit differentiation therefore gives $\partial_T M_*<0$ and $\partial_\tau M_*>0$ before complete emptying. For duration-dependent stimulation,
\[
M_*=\frac{r(T,\tau)a(\tau)}{\Gamma(T)}.
\]
Using \eqref{eqn:partial_tau_r}, \eqref{eqn:partial_T_r}, \(\Gamma'(T)>0\) 
and \(a'(\tau)>0\), we again obtain $\partial_T M_*<0$ and $\partial_\tau M_*>0$.

Finally, for \(\tau\geq\tau_{\mathrm{emp}}\) one has \(r=1\). Hence under
amount-dependent stimulation the steady-state condition
\(R(M_*)=\Gamma(T)\) is independent of \(\tau\), whereas under
duration-dependent stimulation
\[
M_*=\frac{a(\tau)}{\Gamma(T)}
\]
continues to increase with \(\tau\). This completes
the proof.
\end{proof}

%-------------------

\paragraph{Proof of Lem.~\ref{lem:dailyOutput_amtStim_tauMin}.}

\begin{proof}
Fix \(0<T<T_{\mathrm{crit}}\). Along a stable positive steady state,
\[
R(M_*)=\frac{\Gamma(T)}{r(T,\tau)},
\qquad
R'(M_*)<0.
\]
Implicit differentiation with respect to \(\tau\) gives
\[
R'(M_*)\,\partial_\tau M_*
=
-\frac{\Gamma(T)\partial_\tau r}{r^2}.
\]
Defining
\begin{equation}
     \label{eqn:ER_def}
     E_R(M):=\frac{M R'(M)}{R(M)},     
\end{equation}
we obtain
\[
Q(T,\tau)
=
(T+\tau)\frac{\partial_\tau M_*}{M_*}
=
-(T+\tau)\frac{\partial_\tau r}{r}
\frac{1}{E_R(M_*)}.
\]

Before complete emptying, \(r(T,\tau)>0\) and
\(\partial_\tau r(T,\tau)>0\). Moreover, as
\(\tau\downarrow\tau_{\min}(T)\),
\[
\frac{\Gamma(T)}{r(T,\tau)}
\longrightarrow R_{\max}.
\]

For a decreasing-efficiency response, this implies
\(M_*(T,\tau)\to0\). Since \(a(0)=0\), \(a\) is differentiable at zero,
and \(R_{\max}=a'(0)>0\), we have
\[
E_R(M)
=
\frac{M a'(M)}{a(M)}-1
\longrightarrow 0
\qquad\text{as }M\to0^+.
\]

For a unimodal-efficiency response, the stable steady state satisfies
\(M_*>M_{\mathrm{peak}}\) and approaches \(M_{\mathrm{peak}}\) as
\(\tau\downarrow\tau_{\min}(T)\). Since $R'(M_{\mathrm{peak}})=0$, it follows that
\[
E_R(M_*)\longrightarrow0^-.
\]
Thus, we have $E_R(M_*)\longrightarrow0^-$ in both cases, whereas
\[
(T+\tau)\frac{\partial_\tau r}{r}
\]
approaches a finite positive value. Consequently,
\[
Q(T,\tau)\longrightarrow+\infty
\qquad\text{as }\tau\downarrow\tau_{\min}(T).
\]
Hence $Q(T,\tau)>1$ for all \(\tau>\tau_{\min}(T)\) sufficiently close to
\(\tau_{\min}(T)\). By \eqref{eqn:daily_Q_crit}, \(M_{\mathrm{DAY}}(T,\tau)\) is strictly increasing
for \(\tau>\tau_{\min}(T)\) sufficiently close to
\(\tau_{\min}(T)\), as claimed.
\end{proof}

%---------------------------

\paragraph{Proof of Prop.~\ref{prop:dailyOutput_amtStim}.}

\begin{proof}
Fix \(T\in(0,T_{\mathrm{crit}})\) and suppress the dependence on \(T\). For \(\tau_{\min}<\tau<\tau_{\mathrm{emp}}\), define
\[
\sigma(\tau):=(T+\tau)\partial_\tau M_*(T,\tau)-M_*(T,\tau).
\]
Then
\[
\partial_\tau M_{\mathrm{DAY}}(T,\tau)
=\frac{24}{(T+\tau)^2}\sigma(\tau),
\]
while the concavity assumption gives
\[
\sigma'(\tau)
=(T+\tau)\partial_\tau^2M_*(T,\tau)<0.
\]
Thus \(\sigma\), and hence the sign-determining numerator of
\(\partial_\tau M_{\mathrm{DAY}}\), is strictly decreasing before complete
emptying.

\medskip
By Lem.~\ref{lem:dailyOutput_amtStim_tauMin},
\(M_{\mathrm{DAY}}(T,\tau)\) is strictly increasing for \(\tau\) sufficiently
close to \(\tau_{\min}(T)\), so \(\sigma(\tau)>0\) there. At the other
end of the interval, \eqref{eqn:daily_Q_crit} and the definition of
\(Q_A^{\mathrm{pre}}(T)\) imply that the sign of the left limit of
\(\partial_\tau M_{\mathrm{DAY}}\) at \(\tau_{\mathrm{emp}}\) is the sign of
\(Q_A^{\mathrm{pre}}(T)-1\).

\medskip
If \(Q_A^{\mathrm{pre}}(T)<1\), then \(\sigma<0\) for \(\tau\) sufficiently
close to \(\tau_{\mathrm{emp}}\). Since \(\sigma\) is strictly decreasing, it
has exactly one zero \(\tau_{\mathrm{opt}}\in(\tau_{\min},\tau_{\mathrm{emp}})\).
At this point \(M_{\mathrm{DAY}}\) changes from increasing to decreasing, so
\(\tau_{\mathrm{opt}}\) is the unique maximum before complete emptying.
By \eqref{eqn:daily_Q_crit}, it is equivalently characterized by
\(Q(T,\tau_{\mathrm{opt}})=1\).

\medskip
If \(Q_A^{\mathrm{pre}}(T)\geq1\), strict monotonicity of \(\sigma\), together
with \(\sigma>0\) near \(\tau_{\min}\), implies
\(\sigma(\tau)>0\) for all \(\tau_{\min}<\tau<\tau_{\mathrm{emp}}\); in the
case \(Q_A^{\mathrm{pre}}(T)=1\), \(\sigma\) approaches zero only at
\(\tau_{\mathrm{emp}}\). Hence \(M_{\mathrm{DAY}}\) increases up to complete
emptying.

Finally, for \(\tau\geq\tau_{\mathrm{emp}}\), complete emptying has already
been reached and, under amount-dependent stimulation, \(M_*(T,\tau)\)
is independent of \(\tau\). Therefore
\(M_{\mathrm{DAY}}(T,\tau)=24M_*(T,\tau_{\mathrm{emp}})/(T+\tau)\) is strictly
decreasing for \(\tau>\tau_{\mathrm{emp}}\). Thus the unique maximum is
attained at \(\tau_{\mathrm{emp}}\) whenever \(Q_A^{\mathrm{pre}}(T)\geq1\).
\end{proof}

%---------------------------

\paragraph{Proof of Cor.~\ref{cor:dailyOutput_amtStim_T}.}

\begin{proof}
At complete emptying, \(R(M_{\mathrm{emp}}(T))=\Gamma(T)\). Hence defining $E_R(M)$ as in \eqref{eqn:ER_def}, we have
\[
Q_A^{\mathrm{pre}}(T)
=
-\left(1+\frac{T}{\tau_{\mathrm{emp}}}\right)
\frac{1-d(T)}{E_R(M_{\mathrm{emp}}(T))}.
\]

As \(T\to0^+\), we have \(d(T)\to1\) and \(\Gamma(T)\to\gamma<R_{\max}\).
Therefore \(M_{\mathrm{emp}}(T)\) converges to the stable solution of
\(R(M)=\gamma\), for which \(E_R(M)<0\) is finite. Since
\(1-d(T)\to0\), it follows that
\[
Q_A^{\mathrm{pre}}(T)\longrightarrow0
\qquad\text{as }T\to0^+.
\]
Thus \(Q_A^{\mathrm{pre}}(T)<1\) for all sufficiently small \(T>0\).

Next let \(T\to T_{\mathrm{crit}}^-\). Since
\(\Gamma(T)\to R_{\max}\), the stable solution \(M_{\mathrm{emp}}(T)\)
approaches the point at which \(R\) attains its maximum. For a
decreasing-efficiency response this is \(M=0\), and
\(E_R(M_{\mathrm{emp}}(T))\to0^-\). For a unimodal-efficiency response,
\(M_{\mathrm{emp}}(T)\to M_{\mathrm{peak}}\), where \(R'(M_{\mathrm{peak}})=0\), and
again \(E_R(M_{\mathrm{emp}}(T))\to0^-\). Meanwhile,
\(\left(1+T/\tau_{\mathrm{emp}}\right)(1-d(T))\) approaches a finite positive
value. Hence
\[
Q_A^{\mathrm{pre}}(T)\longrightarrow+\infty
\qquad\text{as }T\to T_{\mathrm{crit}}^-.
\]
Therefore \(Q_A^{\mathrm{pre}}(T)>1\) for \(T<T_{\mathrm{crit}}\) sufficiently
close to \(T_{\mathrm{crit}}\).

Finally, \(Q_A^{\mathrm{pre}}(T)\) is continuous for
\(0<T<T_{\mathrm{crit}}\). By the intermediate value theorem, there exists
at least one \(T_{\mathrm{reach}}\in(0,T_{\mathrm{crit}})\) satisfying
\(Q_A^{\mathrm{pre}}(T_{\mathrm{reach}})=1\). The location of the daily-output
maximum then follows directly from
Prop.~\ref{prop:dailyOutput_amtStim}.
\end{proof}
%---------------------------

\paragraph{Proof of Lem.~\ref{lem:Q_durStim}.}

\begin{proof}
For duration-dependent stimulation,
\[
M_*(T,\tau)=\frac{r(T,\tau)a(\tau)}{\Gamma(T)}.
\]
Hence, for fixed \(T\),
\[
\frac{\partial_\tau M_*}{M_*}
=
\frac{\partial_\tau r}{r}+\frac{a'(\tau)}{a(\tau)},
\]
and therefore
\[
Q(T,\tau)
=
\left(1+\frac{T}{\tau}\right)
\left(E_a(\tau)+E_r(T,\tau)\right),
\qquad
E_r(T,\tau):=\frac{\tau\partial_\tau r(T,\tau)}{r(T,\tau)}.
\]

Before complete emptying, the linear removal law gives
\[
r(T,\tau)=\frac{\tau/\tau_\mathrm{emp}}{1-d(T)+d(T)\tau/\tau_\mathrm{emp}},
\]
and hence
\[
E_r(T,\tau)
=
\frac{1-d(T)}
     {1-d(T)+d(T)\tau/\tau_\mathrm{emp}}.
\]
Thus \(E_r(T,\tau)>0\) and is strictly decreasing in \(\tau\). Since
\(E_a(\tau)>0\) is also strictly decreasing, both factors in the expression
for \(Q(T,\tau)\) are positive and strictly decreasing. Hence \(Q(T,\tau)\)
is strictly decreasing for \(0<\tau<\tau_{\mathrm{emp}}\).

For \(\tau>\tau_{\mathrm{emp}}\), one has \(r(T,\tau)=1\) and therefore
\(E_r(T,\tau)=0\). Thus
\(Q(T,\tau)=(1+T/\tau)E_a(\tau)\), which is again strictly decreasing.

As \(\tau\to0^+\), \(E_r(T,\tau)\to1\), while
\(1+T/\tau\to\infty\), and hence \(Q(T,\tau)\to\infty\). As
\(\tau\to\infty\), \(1+T/\tau\to1\) and \(E_a(\tau)\to0\), so
\(Q(T,\tau)\to0\).

Finally, at complete emptying,
\(E_r(T,\tau_{\mathrm{emp}}^-)=1-d(T)\) and
\(E_r(T,\tau_{\mathrm{emp}}^+)=0\). Therefore
\[
Q_D^\mathrm{pre}(T)
=
\left(1+\frac{T}{\tau_{\mathrm{emp}}}\right)
\left[E_a(\tau_{\mathrm{emp}})+1-d(T)\right],
\]
whereas
\[
Q_D^\mathrm{post}(T)
=
\left(1+\frac{T}{\tau_{\mathrm{emp}}}\right)
E_a(\tau_{\mathrm{emp}}).
\]
Since \(0<d(T)<1\) for \(T>0\), it follows immediately that
\(Q_D^\mathrm{post}(T)<Q_D^\mathrm{pre}(T)\).
\end{proof}

%----------

\paragraph{Proof of Prop.~\ref{prop:dailyOutput_durStim}.}

\begin{proof}
Fix \(T>0\). By \eqref{eqn:daily_Q_crit}, the sign of
\(\partial_\tau M_{\mathrm{DAY}}(T,\tau)\) is determined by whether
\(Q(T,\tau)\) is larger or smaller than one. By
Lem.~\ref{lem:Q_durStim}, \(Q(T,\tau)\) is strictly decreasing on each
side of \(\tau_{\mathrm{emp}}\), with
\[
\lim_{\tau\to0^+}Q(T,\tau)=\infty,
\qquad
\lim_{\tau\to\infty}Q(T,\tau)=0,
\]
and one-sided values \(Q_D^{\mathrm{pre}}(T)>Q_D^{\mathrm{post}}(T)\) at complete
emptying.

\medskip
Suppose first that \(Q_D^{\mathrm{pre}}(T)<1\). Since \(Q\) decreases from
infinity to \(Q_D^{\mathrm{pre}}(T)\) on \(0<\tau<\tau_{\mathrm{emp}}\), there is
a unique \(\tau_{\mathrm{opt}}<\tau_{\mathrm{emp}}\) such that
\(Q(T,\tau_{\mathrm{opt}})=1\). Thus \(M_{\mathrm{DAY}}\) increases before
\(\tau_{\mathrm{opt}}\) and decreases afterwards. Since
\(Q_D^{\mathrm{post}}(T)<Q_D^{\mathrm{pre}}(T)<1\), it continues to decrease for
all \(\tau>\tau_{\mathrm{emp}}\). Hence \(\tau_{\mathrm{opt}}\) is the unique
global maximum.

\medskip
Next suppose that
\(Q_D^{\mathrm{post}}(T)\leq1\leq Q_D^{\mathrm{pre}}(T)\). Strict monotonicity of
\(Q\) implies \(Q(T,\tau)>1\) for every
\(0<\tau<\tau_{\mathrm{emp}}\), while \(Q(T,\tau)<1\) for every
\(\tau>\tau_{\mathrm{emp}}\). Hence \(M_{\mathrm{DAY}}\) increases up to complete
emptying and decreases afterwards, so its unique maximum is attained at
\(\tau=\tau_{\mathrm{emp}}\).

\medskip
Finally, suppose that \(Q_D^{\mathrm{post}}(T)>1\). Then also
\(Q_D^{\mathrm{pre}}(T)>1\), so \(M_{\mathrm{DAY}}\) is increasing throughout
\(0<\tau\leq\tau_{\mathrm{emp}}\). For \(\tau>\tau_{\mathrm{emp}}\), \(Q\)
decreases strictly from \(Q_D^{\mathrm{post}}(T)>1\) to zero. Therefore there
is a unique \(\tau_{\mathrm{opt}}>\tau_{\mathrm{emp}}\) satisfying
\(Q(T,\tau_{\mathrm{opt}})=1\), with \(M_{\mathrm{DAY}}\) increasing before and
decreasing after this point. Thus \(\tau_{\mathrm{opt}}\) is the unique
global maximum.

\medskip
This proves all three cases. Whenever the maximum lies away from
complete emptying, its unique location is characterized by
\(Q(T,\tau_{\mathrm{opt}})=1\).
\end{proof}

%----------------

\paragraph{Proof of Cor.~\ref{cor:dailyOutput_durStim_T}.}

\begin{proof}
By Lem.~\ref{lem:Q_durStim},
\[
Q_D^{\mathrm{pre}}(T)
=(1+T/\tau_{\mathrm{emp}})\left[E_a(\tau_{\mathrm{emp}})+1-d(T)\right],
\qquad
Q_D^{\mathrm{post}}(T)
=(1+T/\tau_{\mathrm{emp}})E_a(\tau_{\mathrm{emp}}),
\]
where \(d(T)=e^{-\delta T}\). Since \(T>0\), we immediately have
\(Q_D^{\mathrm{post}}(T)<Q_D^{\mathrm{pre}}(T)\).

Moreover,
\[
\frac{d}{dT}Q_D^{\mathrm{post}}(T)
=E_a(\tau_{\mathrm{emp}})/\tau_{\mathrm{emp}}>0,
\]
while
\[
\frac{d}{dT}Q_D^{\mathrm{pre}}(T)
=
\left[E_a(\tau_{\mathrm{emp}})+1-d(T)\right]/\tau_{\mathrm{emp}}
+(1+T/\tau_{\mathrm{emp}})\delta d(T)>0.
\]
Hence both quantities are strictly increasing with \(T\).

\medskip
Since \(d(T)\to1\) as \(T\to0^+\), both \(Q_D^{\mathrm{pre}}(T)\) and
\(Q_D^{\mathrm{post}}(T)\) converge to \(E_a(\tau_{\mathrm{emp}})\). As
\(T\to\infty\), the factor \(1+T/\tau_{\mathrm{emp}}\) diverges, while
\(E_a(\tau_{\mathrm{emp}})>0\); hence both quantities tend to infinity.

\medskip
Suppose first that \(E_a(\tau_{\mathrm{emp}})<1\). By continuity and strict
monotonicity, there exist unique \(T_{\mathrm{reach}}>0\) and
\(T_{\mathrm{pass}}>0\) satisfying
\(Q_D^{\mathrm{pre}}(T_{\mathrm{reach}})=1\) and
\(Q_D^{\mathrm{post}}(T_{\mathrm{pass}})=1\), respectively. Since
\(Q_D^{\mathrm{post}}(T_{\mathrm{reach}})<Q_D^{\mathrm{pre}}(T_{\mathrm{reach}})=1\),
strict monotonicity of \(Q_D^{\mathrm{post}}\) implies
\(T_{\mathrm{reach}}<T_{\mathrm{pass}}\). The three regimes for the location of
the maximum then follow directly from
Prop.~\ref{prop:dailyOutput_durStim}.

\medskip
Finally, if \(E_a(\tau_{\mathrm{emp}})\geq1\), then for every \(T>0\),
\(Q_D^{\mathrm{post}}(T)=(1+T/\tau_{\mathrm{emp}})E_a(\tau_{\mathrm{emp}})>1\). Hence
Prop.~\ref{prop:dailyOutput_durStim} implies that the maximum lies after
complete emptying for every \(T>0\).
\end{proof}

%-----------------------

\paragraph{Proof of Lemma~\ref{lem:stable_preemptying_max}.}

\begin{proof}
Before complete emptying, \(M_{\mathrm{tar}}=M_*(T,\tau)\) is strictly
increasing with \(\tau\). Hence the maximum at
\(M_{{\mathrm{tar},opt}}\) corresponds to a pre-emptying maximum at a unique
removal duration \(\tau_{\mathrm{opt}}<\tau_{\mathrm{emp}}\).

At such a maximum, \(Q(T,\tau_{\mathrm{opt}})=1\), and therefore
\[
E_a(\tau_{\mathrm{opt}})
=
\frac{\tau_{\mathrm{opt}}}{T+\tau_{\mathrm{opt}}}
-
\tau_{\mathrm{opt}}
\frac{\partial_\tau r(T,\tau_{\mathrm{opt}})}
     {r(T,\tau_{\mathrm{opt}})}
<1.
\]
On the other hand, a target-met fixed point can lose stability only when
\[
E_a(\tau)=E_{\mathrm{crit}}(T,\tau),
\]
with \(E_{\mathrm{crit}}(T,\tau)>1\). Hence the fixed point corresponding to
\(M_{{\mathrm{tar},opt}}\) is locally asymptotically stable.
\end{proof}

%\subsection{Additional Figures}

% ---------------------------------------------------------------- FIGURE ---
%\begin{figure}[t]
%  \centering
%  \includegraphics[width=\textwidth]{fig_dailyOutput_dur}
%  \caption{A,B: Schedule space $(T,\tau)$ for a Michaelis-Menten (decreasing efficiency) response under a linear removal law (A) and an exponential removal law (B). The color indicates the daily milk output $M_\text{DAY}$ at the non-trivial steady state with contour lines for different values of $M_\text{DAY}$. The white-dashed lines shows for a given wait period $T$ the value of $M_\text{tar}$ that maximized the daily output within the target-met region. The used parameters are $\delta=0.03$/h, $\gamma=0.1$/h, $a_\text{max}=37.5$ml/h, $\mu=250$ml, $\nu=1$/h.}
%  \label{fig:dailyOutputDur}
%\end{figure}

\bibliographystyle{plainnat}
\bibliography{references}

\end{document}